\documentclass[twocolumn, a4paper,hyphens]{aa}
\usepackage[varg]{txfonts}
\usepackage{graphics,xcolor}
\usepackage{siunitx,amsmath,amssymb}
\DeclareSIUnit\parsec{pc}

\usepackage{csquotes}
\MakeOuterQuote{"} 

\usepackage{xurl}
\usepackage{hyperref,cleveref}

\usepackage{aasmacros}
\usepackage{textgreek}
\usepackage{array,makecell,booktabs,tabularx}
\usepackage[normalem]{ulem}

\usepackage{natbib}
\bibpunct{(}{)}{;}{a}{}{,}

\newcommand{\nmma}{\textsc{nmma}}
\newcommand{\bilby}{\textsc{Bilby}}
\newcommand{\fiesta}{\textsc{Fiesta}}
\newcommand{\pbilby}{\textsc{Parallel-Bilby}}
\newcommand{\bilpipe}{\textsc{Bilby-Pipe}}
\newcommand{\afgpy}{\textsc{afterglowpy}} 
\newcommand{\diff}{\mathrm{d}}

\newcommand{\mtot}{M_{tot}}
\newcommand{\msun}{M_\odot}
\newcommand{\mtov}{M_{\rm TOV}}
\newcommand{\mdyn}{M_{\rm dyn}}
\newcommand{\mwind}{M_{\rm wind}}
\newcommand{\mdisk}{M_{\rm disk}}
\newcommand{\lambdaT}{\tilde{\Lambda}}
\newcommand{\lsym}{L_{\rm sym}}
\newcommand{\ksym}{K_{\rm sym}}
\newcommand{\ksat}{K_{\rm sat}}
\newcommand{\threensat}{c_{\rm s,3nsat}^2}
\newcommand{\fivensat}{c_{\rm s,5nsat}^2}
\newcommand{\eiso}{E_{\rm iso,0}}
\newcommand{\nsat}{n_{\rm sat}}

\defcitealias{Pang:2023}{P+23}

\crefname{equation}{Eq.}{Eqs.}
\crefname{figure}{Fig.}{Figs.}
\crefname{table}{Table}{Tables}

\begin{document}
\title{\nmma: An extended Bayesian framework for Nuclear Multimessenger Astronomy in the Era of Next-Generation Detectors}
\author{
         H. Rose\inst{\ref{UP}} \corrauth{henrik.rose@uni-potsdam.de}
    \and H. Koehn\inst{\ref{UP}}\email{hauke.koehn@uni-potsdam.de}
    \and T. Wouters\inst{\ref{UU},\ref{Nikhef}}\email{t.r.i.wouters@uu.nl}
    \and P. T. H. Pang\inst{\ref{Nikhef},\ref{UU}}\email{thopang@nikhef.nl}
    \and M. Bulla\inst{\ref{UF},\ref{INFN},\ref{INAF}} \email{bllmtt1@unife.it}
    \and M.~W.~Coughlin\inst{\ref{UMN}} \email{cough052@umn.edu}
    \and T. Dietrich\inst{\ref{UP},\ref{AEI}}\email{tim.dietrich@uni-potsdam.de}
}
\institute{
    Universität Potsdam, Institut für Physik und Astronomie, Karl-Liebknecht-Str. 24/25, 14476 Potsdam, Germany\label{UP} \and
    Institute for Gravitational and Subatomic Physics, Universiteit Utrecht, Princetonplein 1, 3584 CC Utrecht, The Netherlands\label{UU} \and
    Nikhef, Science Park 105, 1098 XG Amsterdam, The Netherlands\label{Nikhef} \and 
    Department of Physics and Earth Science, University of Ferrara, via Saragat 1, I-44122 Ferrara, Italy\label{UF} 
    \and
    INFN, Sezione di Ferrara, via Saragat 1, I-44122 Ferrara, Italy \label{INFN}
    \and
    INAF, Osservatorio Astronomico d’Abruzzo, via Mentore Maggini snc, 64100 Teramo, Italy 
    \label{INAF}
    \and
    School of Physics and Astronomy, U. Minnesota, Minneapolis, MN 55455, USA\label{UMN}
    \and
    Max-Planck-Institut für Gravitationsphysik (Albert-Einstein-Institut), Am Mühlenberg 1, 14476 Potsdam, Germany\label{AEI}
}
\abstract
{ 
    The joint analysis of different (astro-)physical messengers, in particular gravitational-wave data and electromagnetic follow-up observations, allows us to establish, explore and deepen links between different physical fields. 
    As new survey capacities and improved detection methods will lead to a significant increase in the number of multimessenger detections in the upcoming decades, efficient and versatile software frameworks are essential to maximise the scientific outcome of such multimessenger studies.
} { 
    We present a major upgrade to the Nuclear Multimessenger Astronomy (\nmma) framework, incorporating various recent developments in theoretical modelling and machine learning in a modularised and easily extendable Bayesian framework.
    For the first time, this allows direct sampling on nuclear parameters alongside gravitational-wave, kilonova and afterglow parameters. 
} { 
    We combine fast surrogate models for electromagnetic transients with speed-ups from emulators that map nuclear parameters to macroscopic neutron-star properties. 
    Additional acceleration methods for the evaluation of state-of-the-art waveform approximants enable full Bayesian analyses of multimessenger events at the speed required in the era of next-generation detectors.
} { 
    We demonstrate the capabilities of the upgraded \nmma\ framework through a series of representative applications. 
    Re-analysing the 2017 multi-messenger detection of a neutron-star merger, we achieve 20- to 60-fold speed-ups while using more detailed physical models compared to previous studies. 
    Moreover, we demonstrate for a hypothetical future detection how we can simultaneously constrain nuclear parameters and the Hubble parameter with robustly quantified uncertainties.
} {}

\date{Received date /
Accepted date }

\titlerunning{\nmma\ for Next-Generation Detectors}
\maketitle
\nolinenumbers
\section{Introduction}
\label{sec:introduction}
Multimessenger events offer extraordinary opportunities to probe theories, break degeneracies and foster collaboration across physical disciplines:
The supernova SN 1987A, for instance, has sparked extensive research that quickly provided new insights into the supernova mechanism~\citep{Bethe:1990}, stellar evolution~\citep{Arnett:1989}, theories of nucleosynthesis~\citep{Thielemann:1996} or hypothetical particles~\citep{Raffelt:2008}.
Similarly, the binary neutron star (BNS) merger of 17 August 2017, detected through the gravitational-wave (GW) signal GW170817~\citep{Abbott:2017}, followed by the gamma-ray burst GRB~170817A and the kilonova (KN) AT2017gfo~\citep{Abbott:2017b}, led to a better understanding of merger dynamics~\citep{Radice:2020}, the dense-matter equation of state (EoS)~\citep{Lattimer:2021}, KN physics~\citep{Metzger:2019} or the formation of heavy elements~\citep{Cowan:2021}.
Both events remain heavily studied to this day, generating new results from thorough re-analysis of archival data or application of new methods.
While we can gain important insights from a given messenger alone, like tests of general relativity solely based on GWs~\citep{Abbott:2016}, combining information from multiple messengers will usually extend the scope of our analysis, e.g., to constrain neutrino masses~\citep{Koshiba:1992}.
The analysis of BNS mergers this way both benefits from and contributes to the synergy between astrophysical observations, theoretical modelling advances and experimental constraints on nuclear physics~\citep{Shibata:2019,Margutti:2021,Drischler:2021}.

Bayesian methods are ubiquitous in this context to incorporate prior knowledge and uncertainties in the analysis.
Various available tools emphasise different application regimes, e.g.~\citet{Biwer:2019, Breschi:2021,Breschi:2024,Sarin:2024}.
\bilby~\citep{Ashton:2019,Romero_Shaw:2020} stands out as a flexible and user-friendly software for Bayesian parameter inference with an application-agnostic codebase and additional functionality for GW parameter estimation.
The Nuclear Multimessenger Astronomy framework~\citep[\nmma,][]{Pang:2024}, first presented in~\citet[hereafter~\citetalias{Pang:2023}]{Pang:2023}, has extended this basis to enable multimessenger studies that combine GW measurements with observational information from electromagnetic (EM) transients and nuclear physics.
It enables diverse applications, for instance in analysing and classifying EM transients~\citep{Hussenot:2024,Kunert:2024, Barna:2025, Hall:2025, King:2025, Akl:2026, Ducoin:2026}, projecting the number of EM counterparts for future GW observing runs~\citep{Kiendrebeogo:2023}, predicting EM counterparts to a given GW trigger~\citep{Toivonen:2025,Wouters:2025b}, or assessing the nature of compact objects~\citep{Koehn:2024}.

However, the computational costs of Bayesian methods increase with the dimensionality of the parameter space, limiting the potential of multimessenger analyses. 
Without appropriate acceleration methods, Bayesian inference becomes prohibitive for more complex models with many uncertain parameters.
Next-generation GW detectors like Einstein Telescope~\citep[ET,][]{Punturo:2010,Abac:2026} or Cosmic Explorer~\citep{Reitze:2019,Evans:2021} will exacerbate this problem, making more detections than current software infrastructure can handle~\citep{Hu:2025}.
Moreover, their increased sensitivity poses a second challenge that reminds us of the more general problem to gauge the benefits of acceleration methods against an inherent loss of accuracy.
Previous studies have treated nuclear-physics information in \nmma\ through large sets of precomputed EoSs. 
This approach of \citet{Dietrich:2020} avoids the expensive direct exploration of nuclear parameter spaces and allows a sufficient resolution for single-event analyses with present detectors.
The very high precision of next-generation detections across a broad mass range, though, will narrow down the imprint of nuclear information below the implicit resolution limit of precomputed EoSs~\citep{Rose:2023}.

In this work, we address this challenge by a broad overhaul and extensive upgrades to the \nmma\ framework.
We take advantage of the significant progress in machine-learning (ML) methods over recent years to enable efficient joint inference on multimessenger information.
Key changes include the incorporation of new surrogates for nuclear-physics models, improved emulation techniques and enhanced post-processing capabilities.
In particular, we have integrated recently developed EM transient surrogates and have added the capacity to sample from a continuous nuclear parameter space while emulating macroscopic neutron-star (NS) properties at runtime.
This allows us to directly constrain the nuclear parameter space of dense matter from simultaneous Bayesian analysis of GW170817 and its EM counterparts AT2017gfo and GRB~170817A with robust uncertainties.
Moreover, we demonstrate how we can derive tightened constraints on the Hubble parameter from a single favourable BNS merger.
At the same time, we significantly extend the modularity of the code, ensuring quick and simple extensibility to handle new types of data and diverse astrophysical signals.
By aligning the core workflow more closely with \bilby, we allow users to seamlessly connect their research using either tool.

To that end, we review the \nmma-framework and its implementation in \cref{sec:implementation}, before showcasing applications in \cref{sec:applications}.
We conclude with a discussion of identified limitations and further development needs in \cref{sec:conclusion}.
Throughout this paper, we quote all uncertainties at \qty{68}{\percent} confidence level, unless stated otherwise.
We provide magnitudes in the AB system, such that for a given flux density $F_\nu$ and detector response function $e(\nu)$, the apparent magnitude $m$ is given by 
\begin{align}
m = -2.5 \log_{10} \left( \frac{\int F_\nu \frac{e(\nu)}{h\nu} \, \diff\nu}{\qty{3631}{Jy}\int \frac{e(\nu)}{h\nu} \diff\nu} \right).
\end{align}
The \nmma\ codebase and scripts to reproduce the results presented here are available online.\footnote{Codebase: \url{https://github.com/nuclear-multimessenger-astronomy/nmma}; scripts: \url{https://github.com/nuclear-multimessenger-astronomy/nmma\_1.0\_paper}}

\section{Implementation}
\label{sec:implementation}
\subsection{The Nuclear Multimessenger Astronomy Framework}
\nmma\ enables Bayesian analysis of astronomical entities by combining information from various astrophysical messengers, such as GW and EM signals, with nuclear physics models.
The underlying philosophy of Bayesian parameter estimation and model selection is that any information extracted from (uncertain) data should be viewed in light of any prior beliefs we hold.
Bayes' equation expresses this approach mathematically:
\begin{align}
    p(\vec{\theta} | d) = \frac{\mathcal{L}(d | \vec{\theta}) \pi(\vec{\theta})}{Z}. 
\end{align}
It casts the posterior distribution $p$ of model parameters $\vec{\theta}$ in light of data $d$ in relation to a prior distribution $\pi$ and the likelihood $\mathcal{L}$. 
The likelihood measures how well $\vec{\theta}$ can explain the data, usually accounting for uncertainties in the measurement process.
The evidence $ Z = \int \mathcal{L}(d | \vec{\theta}) \pi(\vec{\theta}) \diff\vec{\theta}$
serves \textit{prima facie} as a normalisation constant, but we can attribute it a deeper meaning in the context of model selection:
Among competing models, we usually prefer the one with better balance between low model complexity and high predictive accuracy. 
Since we can treat the size of the prior volume as a complexity measure and the likelihood as a measure of accuracy, we may use the evidence to quantify model preference.
The ratio of evidence for two competing models is known as the Bayes factor $\mathcal{B}^1_2 = Z_1/Z_2$ and quantifies the strength of evidence in favour of one model over another.

In many respects, \nmma\ bases its philosophy and functionality on \bilby~\citep{Ashton:2019,Romero_Shaw:2020}, making it very easy for users to adapt. 
\bilby\ pursues a modular approach to Bayesian inference in a broad sense.
It implements routines for prior handling, likelihood evaluation and parameter conversion, while providing a flexible interface to various sampling algorithms employing both nested sampling~\citep{Skilling:2004} and Markov chain Monte Carlo  methods~\citep{Ashton:2021}.
Additionally, it provides specific tools for GW analysis and hyperparameter studies. 

Previous versions of \nmma\ have extended this architecture with additional functionality to handle EM data alongside nuclear-physics information and to connect model parameters across different messengers through empirical fitting relations.
The EM analysis employs simplified (semi-)analytical expressions or ML surrogates to model KNe, supernovae or GRB afterglows at runtime.
The surrogates are trained on grids of precomputed lightcurves using, e.g., radiation-transport simulations for KNe with \textsc{Possis}~\citep{Bulla:2019} or \textsc{Sedona}~\citep{Kasen:2017}.
\nmma\ includes nuclear information through the impact on the EoS which determines the relation between NS mass $M$ and macroscopic properties like tidal deformability $\Lambda$~\citep{Hinderer:2010,Chatziioannou:2020}, radius $R$ and compactness $C=\frac{GM}{c^2R}$.
While experiments and theoretical ab-initio calculations constrain the EoS up to the regime of nuclear saturation, astrophysical data constrain these macroscopic properties.
$\Lambda$ impacts the late-time GW signal, $C$ is a core model parameters in pulse-profiling approaches pursued by the NICER collaboration~\citep{Miller:2019,Riley:2021} and the radius has an imprint on the expected ejecta masses from which EM models can then predict lightcurves.
We refer to \citetalias{Pang:2023} for further details.

\subsection{Main Novelties}
\label{sec:code_novelties}
Our main improvements for multimessenger studies come from incorporating ML methods in two key areas of \nmma.
First, in the nuclear sector, one would ideally want to sample empirical or theory-based microphysical parameters.
However, the mapping from EoS to macroscopic NS properties takes $\mathcal{O}(\qty{1}{\second})$ to solve the stellar structure equations for an extended range of initial values.
This is prohibitive because seconds-long likelihood evaluations would make the full sampling computationally too expensive and therefore intractable for many repeated inference runs that vary priors, models or sources.
We instead typically aim for evaluation times of at most $\mathcal{O}(\qty{e-2}{\second})$.
\nmma\ has previously solved this by pre-computing large EoS sets that are characteristic of the underlying nuclear parameter space and converting them to macroscopic $M-R-\Lambda$ tables.
Analyses would then draw indices for these sets, weighted in pre-processing by how well the corresponding EoS satisfies observational or experimental constraints; see~\citet{Koehn:2025} for a comprehensive study following this approach.
However, this method bears an inherent resolution limit that will worsen drastically with the increased sensitivity and detection frequency that we expect from next-generation GW detectors~\citep{Rose:2023}.

\nmma\ can now efficiently solve this through emulators that map microphysical parameters to desired macrophysical NS quantities.
Observational and experimental constraints no longer need to be applied in pre-processing, but an EoS-likelihood can evaluate the nuclear data on equal footing with the GW strain and lightcurves.
This way, we no longer need to presuppose the type of merging system, but can infer the nature of the components based on their mass and the EoS samples (see Appendices \ref{app:nuclear} and \ref{app:GW230529}).
Other works have already explored the possibility of using ML surrogates of the TOV solver~\citep{Ferreira:2021,Tiwari:2024,Lalit:2025} and have demonstrated their usefulness in the analysis of GW events~\citep{Thete:2023,Magnall:2025,Somasundaram:2025,Reed:2024,Reed:2026}. 
However, this work is, to the best of our knowledge, the first to apply them consistently in the full multimessenger picture of BNS mergers.

Second, \nmma\ can now directly call \fiesta~\citep{Koehn:2025b}, originally a stand-alone solution to analyse EM data with high-precision surrogates, to predict lightcurves for a broad range of filters.
Especially its GRB afterglow surrogates allow us to drastically reduce the computing time of combined multi-messenger analysis runs. 
GRBs offer a well-known opportunity to break the distance-inclination degeneracy of GW signals~\citep{Nissanke:2010,Usman:2019}.
While the mechanism leading to the prompt emission remains uncertain, the afterglow is well understood and multiple tools are available to model it.
The popular \afgpy~\citep{Ryan:2020,Ryan:2025} is a relatively light-weight and flexible module to compute afterglow lightcurves. 
The recently published \textsc{pyblastafterglow} captures more intricate shock physics~\citep{Nedora:2025}.
\afgpy\ has already been included in \nmma, although it adds considerable computational burden and the runtime of a single forward-model is difficult to predict.
\textsc{pyblastafterglow}'s required computing time is completely prohibitive for stochastic sampling purposes.
\fiesta's surrogates instead provide reliable lightcurve predictions on the timescale of our GW approximant.

We highlight two more developments that deserve special notice:
First, we have harmonised and augmented the use of cosmological relations to let users sample parameters of arbitrary flat \textLambda CDM cosmologies (see \cref{ssec:Future_detection}).
Second, we have substantially restructured the framework's architecture on a more general level to improve modularity and extensibility, adopting many design principles from \bilby.
Users can now treat it as a thin wrapper around \bilby\ and easily adapt familiar concepts to their needs for other classes of multimessenger events. 
We are thus prepared to perform multimessenger analyses of SN events, too, although we highlight a current lack of waveform approximants for the GW signature of core-collapse supernovae.
Appendix~\ref{app:software_upgrades} describes further code and design changes.

\subsection{Performance Comparison}
\label{ssec:benchmark}

\begin{table}[t]
    \caption{
    Model parameters and their priors.
    $\mathcal{N}$ indicates a Gaussian distribution with the given mean and standard deviation.
    $(\log)\,\mathcal{U}[x,y]$ denotes a (log-)uniform distribution on $[x,y]$. 
    Subscripts indicate further specifications: $\mathcal{U}_{\rm CV}$ refers to a comoving volume and $\mathcal{U}_{\rm sph}$ to the surface of a sphere.
    $\sin$ stands for a sine distribution, while $\mathcal{W}$ denotes a weighted categorical prior, see the main text for details.
    $\mathcal{C}$ marks limits on a derived parameter.
    }
    
    \centering
    \begin{tabularx}{\linewidth}{l@{\hspace{2pt}}rl}
    \toprule
    \toprule

    Parameter & Symbol & Prior \\
     \midrule
     \multicolumn{3}{c}{Observational parameters} \\
        luminosity distance [Mpc] & $d_{\rm L}$ & $\mathcal{U}_{\rm CV}[10, 75]$ \\
        inclination [rad] & $\theta_{\rm JN}$ & $\sin[0, \pi]$ \\
        GW polarisation [rad] & $\psi$ & $\mathcal{U}[0, \pi]$ \\
     \midrule
     \multicolumn{3}{c}{GW parameters} \\
        chirp mass~[M$_\odot$] & $\mathcal{M}$ & $\mathcal{U}_{i}[1.18, 1.21]$ \\
        mass ratio & $q$ & $\mathcal{U}_{i}[0.5, 1]$ \\
        Aligned component spins & $\chi_i$ & $\mathcal{U}_{\rm sph}[0, 0.05]$ \\
        \midrule
    \multicolumn{3}{c}{Ejecta parameters} \\
     wind ejection efficiency & $\zeta$ & $\mathcal{U}[0, 1]$ \\
     mass fitting error [M$_\odot$] & $\alpha$ & $\mathcal{N}(0, 0.0004)$ \\
     log dyn. mass~[M$_\odot$] & $\log_{10}(M_{\text{dyn}})$ & $\mathcal{C}[-3, -1.3]$\\
     log wind mass [M$_\odot$] & $\log_{10}(M_{\text{wind}})$ & $\mathcal{C}[-2, -0.9]$\\
     separation angle~[deg] & $\phi_{\text{KN}}$ & $\mathcal{U}[15, 75]$\\

     \midrule
     \multicolumn{3}{c}{GRB parameters} \\
     GRB conversion efficiency & $\epsilon$ & $\log \mathcal{U}[10^{-7}, 0.5]$\\
     log iso.-equiv. energy~[erg] & $\log_{10}(\eiso)$ & $\mathcal{C}[48, 57]$ \\
     jet core angle~[rad] & $\theta_{\rm{c}}$ & $\mathcal{U}[0.01, \pi/5]$\\
     wing factor & $\alpha_{\rm{w}}$ & $\mathcal{U}[0.2, 3.5]$\\
     log ISM density~[cm$^{-3}$] & $\log_{10}(n_{\text{ISM}})$ & $\mathcal{U}[-6, 0]$\\
     $e^-$ spectrum power index & $p$ & $\mathcal{U}[2, 3]$\\
     log $e^-$ energy share & $\log_{10}(\epsilon_e)$ & $\mathcal{U}[-4, 0]$\\
     log magnetic energy share & $\log_{10}(\epsilon_B)$ & $\mathcal{U}[-5, 0]$\\
    \midrule
    \multicolumn{3}{c}{EoS parameters} \\
        EoS index & $i_{\rm EoS}$ & $\mathcal{W}[1, 5000]$\\

     \end{tabularx}
    \label{tab:ref_parameters}
\end{table}

\begin{table*}[ht]
    \centering
    \caption{Overview of models and evaluation methods used in \cref{ssec:benchmark,ssec:mma_bns2017}.
    }
    \begin{tabular}{cccc}
       Messenger & Approach A (=\citetalias{Pang:2023}) & Approach B & Approach C\\
    \toprule
    \toprule
       GW & \makecell{\textsc{IMRPhenomPv2\_NRTidalv2} \\}& \makecell{\textsc{IMRPhenomPv2\_NRTidalv2}\\Multibanding } & \makecell{\textsc{IMRPhenomXAS\_NRTidalv3}\\Multibanding }\\
       \addlinespace
       KN & \makecell{\textsc{Bu2019}\\Gaussian Process Regression } & \makecell{\textsc{Bu2019}\\ Feed-forward Neural Network} & \makecell{\textsc{Bu2026}\\ \fiesta} \\
       \addlinespace
       GRB & Gaussian Jet & \makecell{Gaussian Jet \\ \fiesta }& \makecell{Gaussian Jet \\ \fiesta }\\
       \addlinespace
       EoS & \makecell{pre-weighted table \\ from \citetalias{Pang:2023}} & \makecell{pre-weighted table \\ from \citetalias{Pang:2023}} & \makecell{emulator trained on \\ \citet{Reed:2024}} \\
       \addlinespace
       Priors & \cref{tab:ref_parameters} & \cref{tab:ref_parameters} & \cref{tab:ref_parameters,tab:new_parameters} 
    \end{tabular}
    \label{tab:model_overview}
\end{table*}

\begin{figure*}[ht!]
\centering
\includegraphics[width=0.99\textwidth]{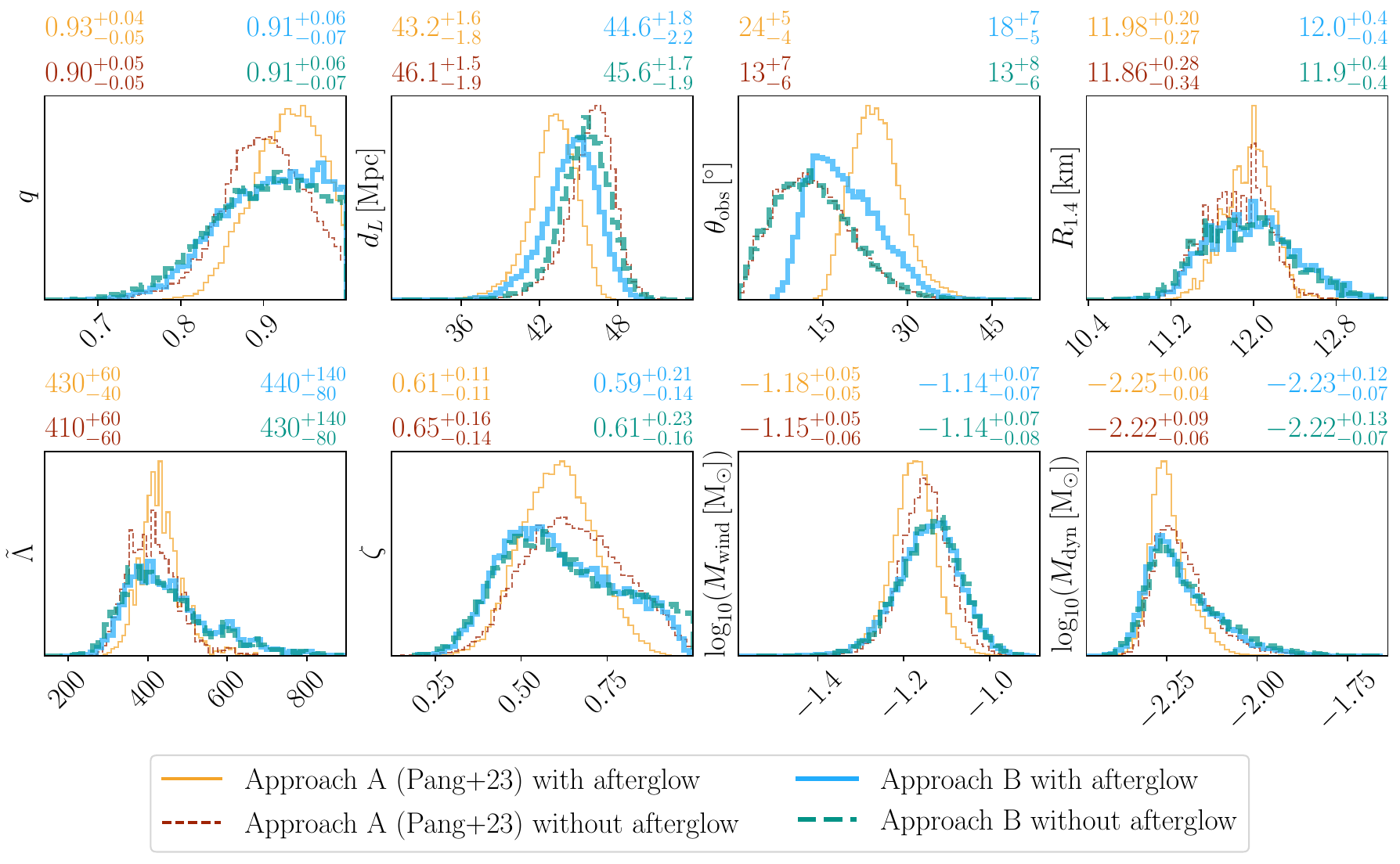}
\caption{
    Posterior comparison for key system parameters of the 2017 BNS merger. 
    Solid (dashed) lines indicate analyses with (without) the GRB~170817A afterglow.
    Thick lines show the results from our re-analysis with the new \nmma\ implementation, while thin lines correspond to the original results from~\citetalias{Pang:2023}.
    Slight discrepancies beyond numerical noise result from a re-fitting of the disk-mass predictions and changes to the computation of the GRB's central isotropic-equivalent energy $\eiso$.
}
\label{fig:performance_reruns_posterior}
\end{figure*}

In order to assess the achieved speed-ups in our new implementation, we compare joint re-analyses of GW170817 and AT2017gfo, including and excluding the afterglow from GRB~170817A, to the results of~\citetalias{Pang:2023}. 
We employ the same data and mostly the same physical models, making small adjustments to the setup only where necessary.
\Cref{tab:ref_parameters} summarises the parameters and prior distributions for our analysis, while \cref{tab:model_overview} gives an overview of the different models and acceleration methods in \citetalias{Pang:2023} (Approach A) and this work (Approach B).

We analyse the glitch-corrected GW strain~\citep{Abbott:2017,Abbott:2021} with the \textsc{IMRPhenomPv2\_NRTidalv2} approximant~\citep{Khan:2016,Dietrich:2019}, but accelerated through multibanding~\citep{Morisaki:2021}. 
The EM data for KN AT2017gfo and the afterglow of GRB~170817A were compiled in~\citet{Coughlin:2018}, based on~\citet{Villar:2017}.
The KN lightcurve modelling applies a surrogate model derived from radiation transport simulations with \textsc{Possis}~\citep{Bulla:2019,Coughlin:2020}, already introduced in~\citet{Dietrich:2020}.
It assumes a three-component ejecta morphology, with isotropically emitted wind ejecta and dynamical ejecta consisting of a red tidal component in the equatorial plane and a blue shocked component in the polar regions, separated by an angle $\phi_{KN}$.
However, the KN surrogate in~\citetalias{Pang:2023} was trained with a Gaussian-process approach that we could not incorporate due to memory constraints and software dependencies.
We have instead used the neural-network surrogate in~\citet{Koehn:2024}.
It was trained on the same simulation grid and produces very similar lightcurves, but cannot yield identical results.
Surrogate models introduce additional systematic uncertainty that the likelihood needs to account for.
We follow \citetalias{Pang:2023} to employ a constant lightcurve modelling uncertainty of $\sigma_{\rm sys}=\qty{1}{mag}$ across all filters.
This systematic uncertainty applies to the GRB afterglow, too, for which we use a \fiesta\ surrogate~\citep{Koehn:2025b} of a Gaussian jet model, trained on \afgpy\ lightcurves~\citep{Ryan:2025}.
As it only works on a more restrictive parameter space, we must tighten our priors on the GRB's core opening angle $\theta_c$ and the electron-spectrum power index $p$.
This should entail no consequences for $p$ because its posterior in~\citetalias{Pang:2023} is still fully enclosed by the tightened prior. 
$\theta_c$, however, was sampled up to the unphysical value $\frac{\pi}{2}$. 
\fiesta\ instead mandates the still conservative limit $\theta_c<\frac{\pi}{5}$.
Moreover, we need to adopt a \fiesta\ reparametrisation that does not sample the GRB's wing opening angle $\theta_w$, but $\alpha_w=\theta_w/\theta_c$. 
We incorporate nuclear information through a set of 5000 tabulated EoSs with the prior weights of~\citetalias{Pang:2023}.
They had combined constraints from chiral effective field theory calculations at low densities with observations of massive pulsars~\citep{Shamohammadi:2023,Fonseca:2021} as well as two NICER measurements~\citep{Miller:2019,Miller:2021}.
\citet{Dietrich:2020,Pang:2023,Koehn:2025b} provide more details on the employed models.

For the purpose of this performance benchmark, we use \textsc{dynesty}~\citep{Speagle:2020} with 2048 live points, too, but deviate from the 256 cores of \citetalias{Pang:2023} due to different system architectures, running our analysis without (with) GRB afterglow on 192 (384) cores.
A lower number of worker processes reduces parallelisation loss~\citep{Handley:2015}, but as the number of workers remains well below the 2048 live points, we can safely treat this effect as secondary (cf. Appendix~\ref{ssec:parallelisation}).
Our runtime is drastically reduced by a factor of 20 (60) to about 1300 (2300) CPUh against the previous 26k (140k) CPUh for the GW+KN(+GRB) analysis, while \cref{fig:performance_reruns_posterior} shows for some important model quantities that the inferred parameters remain broadly consistent.

Apparent discrepancies in the posterior distributions are fully explained by updates to the ejecta treatment.
We here describe their effects and provide a detailed discussion in Appendix~\ref{ssec:bns_ejecta}.
First, our reruns show wider posteriors on the mass ratio up to $q=1$, whereas the older analyses peaked around 0.9.
This relates to a recalibration for estimates of the disk mass $\mdisk$.
In a mass range that is consistent with GW170817, $\mdisk$ no longer monotonically decreases with $q$, but reaches a plateau for symmetric binaries.
Moreover, the new fit leads for the same inputs to the prediction of lower disk masses. 
Since the amount of wind ejecta required to explain the KN remains unchanged, this is countered by a shift towards stiffer EoS and higher wind ejection efficiencies.
Consequently, we see an increased support for higher values of the EoS-mediated observable $\lambdaT$, the \textit{leading-order} (also called \textit{mass-averaged} or \textit{effective}) tidal deformability~\citep{Chatziioannou:2020}, and $R_{\text{1.4}}$, the radius of a \qty{1.4}{\msun} NS.\\
Second, our work and \citetalias{Pang:2023} both find a shift to higher inclinations $\theta_{\rm obs}$ and lower $d_{\rm L}$ when the GRB afterglow is included, although this is less pronounced in our re-analysis.
Note that we identify the system's observational inclination, $\theta_{\rm obs}$, with the angle between the total angular momentum and the line of sight, $\theta_{\rm JN}$, which strictly holds only under the assumption of aligned component spins and rotational symmetry of the EM emission. 
This correlation is naturally expected as $\theta_{\rm JN}$ and $d_{\rm L}$ are degenerate in GW measurements~\citep{Markovic:1993,Cutler:1994}.
GRBs can in principle help to break this degeneracy.
Their highly focused emission is usually only visible from a near-polar viewing angle.
The evolution of GRB~170817A and in particular its afterglow that kept brightening for almost half a year after merger favour a scenario instead where the observation angle lies outside the jet's core region, i.e., $\theta_{\rm obs} > \theta_c$~\citep{Ryan:2020,Nedora:2025}.
At the lower inclinations favoured in GW+KN analysis, this corresponds to higher isotropic-equivalent energies $\eiso\geq\qty{e52}{erg}$ that were previously effectively excluded.

\section{Applications}
\label{sec:applications}
\subsection{Multimessenger Analysis of the 2017 BNS Merger}
\label{ssec:mma_bns2017}

\begin{table}[t]
    \caption{
    Additional parameters for updated models and their priors. 
    $\mathcal{U}_{\rm S}$ refers to the source frame and $\mathcal{U}_{c}$ to the component masses.
    Otherwise, we apply the same notations as in \cref{tab:ref_parameters}.
    }
    
    \centering
    \begin{tabular}{l@{\hspace{-1pt}}r l}
    \toprule
    \toprule

    Parameter & Symbol & Prior \\
    \midrule
        luminosity distance [Mpc] & $d_{\rm L}$ & $\mathcal{U}_{\rm S}[10, 75]$ \\
     \midrule
     \multicolumn{3}{c}{GW parameters} \\
        chirp mass~[M$_\odot$] & $\mathcal{M}$ & $\mathcal{U}_{c}[1.18, 1.21]$ \\
        mass ratio & $q$ & $\mathcal{U}_{c}[0.5, 1]$ \\
    \midrule
    \multicolumn{3}{c}{\textsc{Bu2026} Ejecta parameters} \\
     log dyn. mass~[M$_\odot$] & $\log_{10}(m_{\text{dyn}})$ & $\mathcal{C}[-4, -1.31]$\\
     log wind mass [M$_\odot$] & $\log_{10}(m_{\text{wind}})$ & $\mathcal{C}[-4, -0.56]$\\
     mass-averaged dyn. velocity~[\unit{c}] & $v_{\text{dyn}}$ & $\mathcal{U}[0.12, 0.35]$\\
     mass-averaged dyn. $e^-$ fraction & $Y_{\text{e, dyn}}$ & $\mathcal{U}[0.15, 0.35]$\\
     mass-averaged wind velocity [\unit{c}] & $v_{\text{wind}}$ & $\mathcal{U}[0.05, 0.15]$\\
     mass-averaged wind $e^-$ fraction & $Y_{\text{e, wind}}$ & $\mathcal{U}[0.2, 0.4]$\\
    \midrule
    \multicolumn{3}{c}{Lightcurve nuisance parameters} \\
    \textit{i}-th KN modelling error [mag] & $\sigma_{\text{KN},i}$ & $\mathcal{U}[0.3, 2]$\\
    GRB modelling error [mag] & $\sigma_{\text{GRB}}$ & $\mathcal{U}[0.3, 1]$\\

    \midrule
    \multicolumn{3}{c}{EoS parameters} \\
    sym. energy slope~[MeV] & $L_{\rm sym}$ & $\mathcal{U}[20, 150]$\\
    sym. energy curvature~[MeV] & $K_{\rm sym}$ & $\mathcal{U}[-300, 100]$\\
    incompressibility~[MeV] & $K_{\rm sat}$ & $\mathcal{U}[200, 300]$\\
    $c_s^2$ at 3x saturation density [\unit{c}]& $c_{\rm s,3nsat}^2$ & $\mathcal{U}[0, 1]$\\
    $c_s^2$ at 5x saturation density [\unit{c}]& $c_{\rm s,5nsat}^2$ & $\mathcal{U}[0, 1]$
    \end{tabular}
    \label{tab:new_parameters}
\end{table}

\begin{figure*}[ht]
    \includegraphics[width=0.95\textwidth]{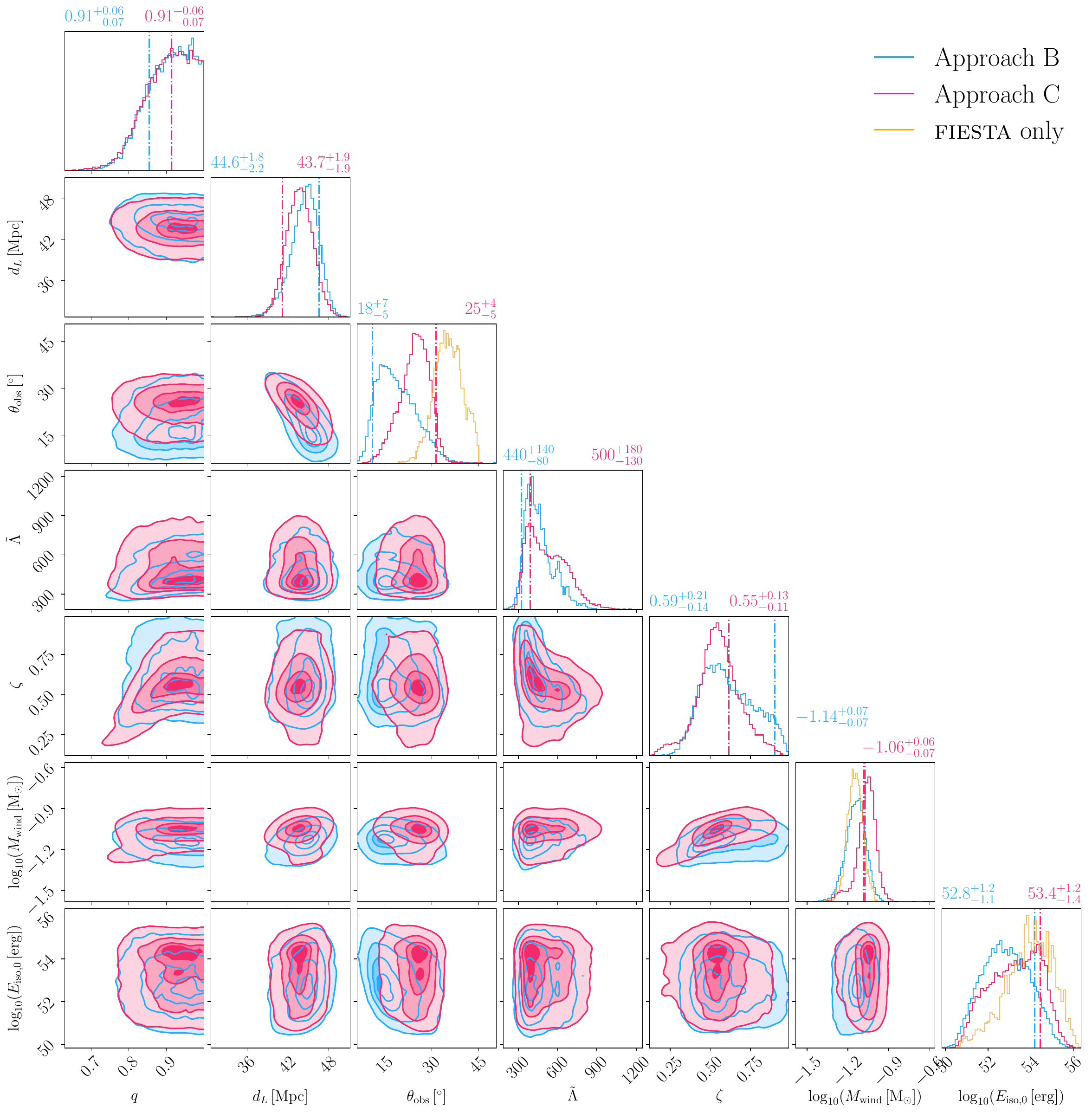}
    \caption{
    Posterior comparison for key system parameters of the 2017 BNS merger, including the GRB afterglow.
    Red contours correspond to the analysis using current models, while thin blue lines use older physical models; see the text for details.
    We additionally show applicable posteriors from an EM-only analysis with \fiesta~\citep{Koehn:2025b}.
    Dashed-dotted vertical lines mark the maximum-posterior solutions for which we show lightcurves in~\cref{fig:lc_comp}
    }
    \label{fig:New_models}
\end{figure*}

\begin{figure*}[ht]
    \centering
    \includegraphics[width=0.95\textwidth]{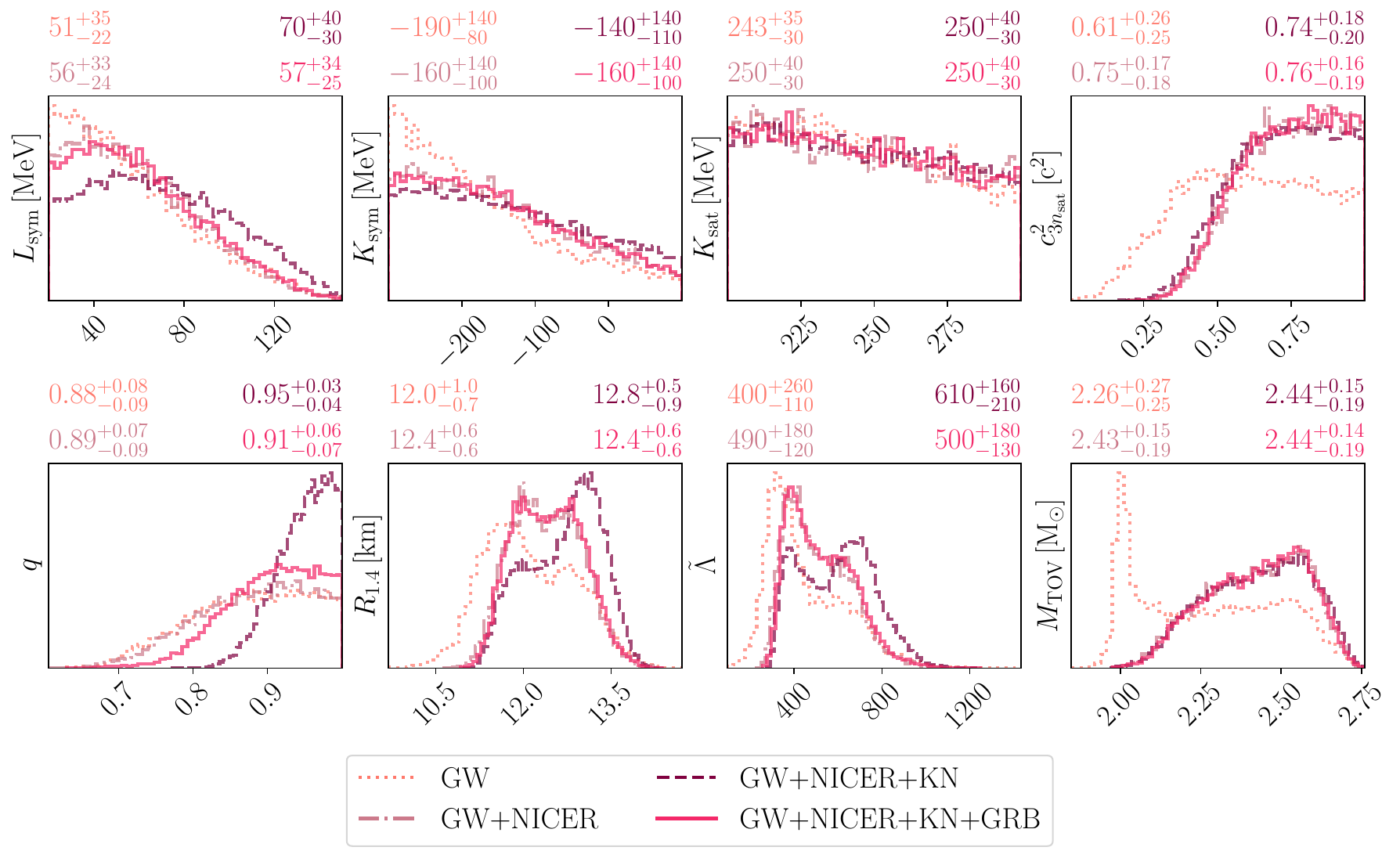} 
    \caption{
    Posterior comparison for microphysical parameters of the 2017 BNS merger when analysing the strain data of GW170817 (dotted) and sequentially including NICER constraints on PSR J0030+0451~\citep{Kini:2026} and J0437-4715~\citep{Choudhury:2024} (dash-dotted), the KN AT2017gfo (dashed) and GRB~170817A (solid).
    We see that GW data alone strongly favour a soft EoS, while the inclusion of NICER constraints stiffens the EoS.
    The KN further intensifies this trend, whereas the GRB afterglow with its broadened distribution of mass ratios again suggests a more moderate stiffening.
    The sharp peak at $\mtov=\qty{2}{\msun}$ in the GW-only case is not physical, but results from our enforcement of a corresponding lower limit in emulator training.
    Because we have excluded EoS with lower $\mtov$ from our training data, we have kept the PSR J1614-2230 mass constraint~\citep{Shamohammadi:2023} to exclude unphysical predictions, leading to an artificial pile-up of EoSs that barely exceed that limit.
    }
    \label{fig:2017_nuclear_params}
\end{figure*}

\begin{figure*}[ht]
    \centering
    \includegraphics[width=\linewidth]{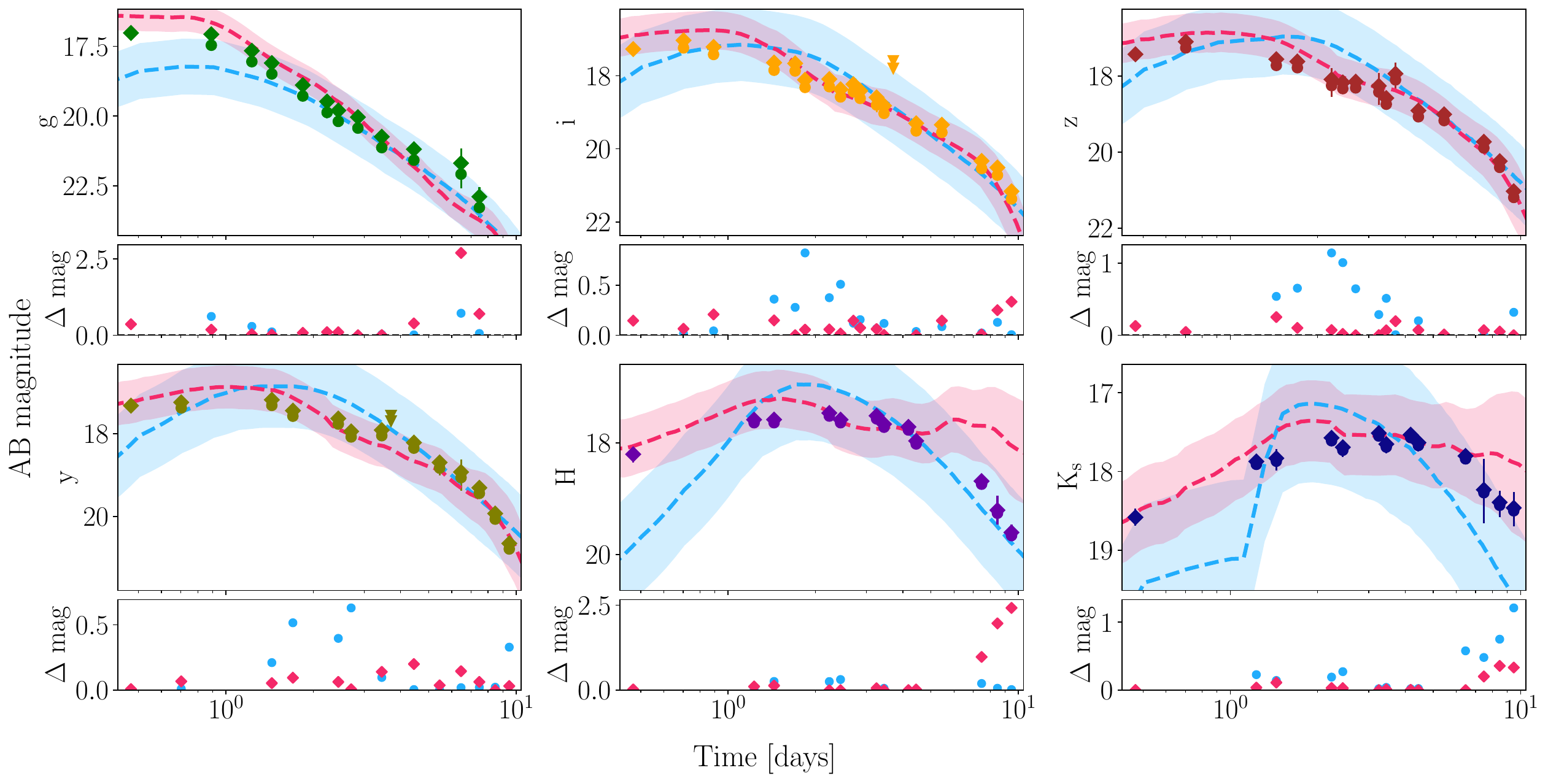}
    \caption{
    Dashed lines mark best-fitting KN lightcurves for the maximum-posterior solution of \cref{fig:New_models}.
    The surrounding bands indicate $\sigma_{\rm sys}$.
    Dots mark the observations compiled in~\citet{Coughlin:2018}, based on~\citet{Villar:2017} and used in \citetalias{Pang:2023} (Approach B, blue lightcurve). 
    Diamonds include an updated reddening-correction~\citep[Approach C, red lightcurve]{Hussenot:2025}.
    Triangles mark upper limits.
    The \textsc{Bu2026} model provides an overall better fit to the data, although it clearly overestimates the late-time infrared flux.
    }
    \label{fig:lc_comp}
\end{figure*}

While these results highlight the consistency of our new implementation with previous findings, they do not leverage various modelling advances over the past few years.
Therefore, we study the impact of using more recent models, summarised as Approach C in \cref{tab:model_overview}. 
Instead of \textsc{IMRPhenomPv2\_NRTidalv2}, we now employ the \textsc{IMRPhenomXAS\_NRTidalv3} approximant~\citep{Pratten:2020,Abac:2024} which is calibrated on a larger parameter space and, despite including dynamical tide corrections, is computationally even more efficient.
Given the expected low spins in BNS systems, we consider precession effects negligible and focus on high evaluation speed.

In the EM sector, we use slightly modified lightcurve data with recent reddening corrections~\citep{Hussenot:2025}.
More importantly, we now use the updated \textsc{Bu2026} KN surrogate model which is based on a major update of \textsc{Possis}~\citep{Bulla:2023} and applies to an extended parameter space over the previously published \textsc{Bu2025}~\citep{Koehn:2025b}.
It no longer separates dynamical ejecta regimes by $\phi_{KN}$, but assumes a smoothly decreasing density of dynamical ejecta towards the poles~\citep{Perego:2017,Radice:2018} and a concurrently increasing electron fraction~\citep{Setzer:2023}.
The mean electron fractions of both dynamical and wind ejecta then parametrise their composition, while their mean velocities control the radial extent~\citep{Kiuchi:2017,Hotokezaka:2018,Kawaguchi:2020}.
Additionally, we follow~\citet{Jhawar:2025} to estimate $\sigma_{\rm sys}$ at runtime.
To that end, we sample $\sigma_{{\rm KN}, i}$ at four sampling times $t_i$ between 0.3 and 10 days after merger and interpolate to $\sigma_{\rm sys}(t_j)$ at the observing times $t_j$ to obtain a time-dependent uncertainty in the KN filters.
We keep the previous surrogate for the afterglow and use a time-independent parameter $\sigma_{\text{GRB}}$ that applies to all its additional filters.

Further, instead of sampling precomputed equations of state, we apply an emulator that uses a feed-forward neural network to predict the macroscopic NS properties from five microphysical EoS parameters.
This uses the same data as~\citet{Reed:2024}, who computed a set of about 270,000 EoSs they obtained from combining a common empirical parametrisation of matter near saturation density (three parameters) with a speed-of-sound extrapolation at higher densities (two parameters).
We provide further details on this parametrisation and our training procedure in Appendix \ref{ssec:emulation_strategy}.
Similar to the prior weighting for precomputed EoSs, we use the following observational and theoretical findings to estimate the likelihood of a given EoS.
First, Shapiro delays found in timing measurements of heavy pulsars provide lower limits on the maximum mass of neutron stars $\mtov$.
We here use measurements for the heavy pulsars J1614-2230 \citep{Shamohammadi:2023} and J0740+6620 \citep{Fonseca:2021}.
Second, we consider $M-R$ estimates from the most extensively studied NICER pulsars J0030+0451~\citep{Kini:2026} and J0740+6620~\citep{Dittmann:2024}.
Appendix~\ref{app:nuclear} contains further details on the likelihood prescriptions.
This way, we use information on the same systems that have also informed the prior weighting of the tabulated EoS in~\citetalias{Pang:2023}.
However, we emphasise that the effective EoS spaces are not comparable because of the fundamentally different construction of the underlying EoSs.
In general, our microphysical priors allow EoSs with much higher radius and tidal deformability than the precomputed EoS sets of~\citet{Pang:2023}.
This generally leads to higher expected disk masses which consequently affect the ejecta and GRB predictions (see Appendix~\ref{ssec:bns_ejecta}).

We follow~\citet{Romero_Shaw:2020} to set $\pi(\mathcal{M})\propto \mathcal{M}$ and $\pi(q)\propto \frac{(1 + q)^{2/5}}{q^{6/5}}$, yielding a uniform two-dimensional probability density for the component masses.
Our new $d_{\rm L}$ prior assumes a uniform source-frame merger rate under the Planck-2018 cosmology~\citep{Planck:2018}. 
It differs from the previous prior that assumed a uniform distribution in a co-moving volume by an additional $1+z$ correction for time dilation in the merger rate, although the impact is negligible within the relatively close distance range of \qtyrange{10}{75}{Mpc}.
See \cref{tab:new_parameters} for an overview of these new parameters in the preferred setup.
Although the additional parameters for the \textsc{Bu2026} KN model with EM nuisance parameters and the EoS emulator increase the dimensionality of the parameter space from 11 (18) to 23 (31) parameters, we obtain similar runtimes of about 1600 (3300) CPUh for the re-analysis without (with) the GRB afterglow.
Figure \ref{fig:New_models} compares posteriors for key parameters under Approaches B and C when including the GRB afterglow, providing highly consistent results.
Changes to the precise values of the parameters reflect the impact of the different model choices. 

Two aspects warrant a deeper discussion.
First, the nuclear model underlying our emulator shifts $\lambdaT$ to higher values and introduces a bimodality seen in agnostic GW analyses~\citep{Abbott:2019b,Narikawa:2019}, too.
Low prior support had eroded the mode at higher $\lambdaT$ for tabulated EoSs.
This stiffening behaviour naturally depends on further model assumptions, too. 
\cref{fig:2017_nuclear_params} compares posteriors on the microphysical parameters and some observables in different setups. 
When ignoring NICER constraints and the EM counterparts, the EoS estimates remain relatively soft and we essentially reproduce the findings of~\citet{Reed:2026} who have trained their emulator on the same EoS data.
In contrast, including NICER and KN data lets the second mode become increasingly more prominent.
Including the GRB afterglow weakens this trend again because it supports a broader mass distribution which can provide the required ejecta masses with slightly softer EoSs.

This touches on the second aspect.
As already discussed in the previous section, the stronger support for very stiff EoSs in the full multimessenger setup corresponds to a further increase in the computed disk mass.
The wind ejecta conversion factor $\zeta$ compensates for this to obtain wind masses that can still account for the observed KN.
The other lightcurve parameters then adjust to obtain appropriate fits. 
In particular, the inclination at $\theta_{\rm obs} = \qty{24(5:4)}{\degree}$ reverts almost exactly to the result of~\citetalias{Pang:2023}, providing a balance between our analysis without the afterglow and the EM-only analysis of~\citet{Koehn:2025} that found $\theta_{\rm obs} = \qty{35(4:5)}{\degree}$.
We show in \cref{fig:lc_comp} the resulting lightcurves in a subset of analysed filters for the maximum-posterior samples of the afterglow-informed runs under approaches B and C.
We see that \textsc{Bu2026} provides an overall better fit to the data.
In fact, the posterior on the estimated fitting errors $\sigma_{\text{KN},i}$ rails against its lower prior bound at \qty{0.3}{mag} for the three earlier sampling times. 
This limit is conservatively set to reflect typical mismatches between \fiesta\ and its underlying training data from full radiation-transport simulations~\citep{Koehn:2025b}.
Only the last error sampling point at 10 days consistently demands a higher modelling error.
A large mismatch between the model and the observed infrared (IR) lightcurves is clearly driving this discrepancy, indicating a need for refined models of the late-time KN evolution.

\subsection{A more detailed view on AT2017gfo}
\label{sec:NICER_fiesta}

Although the posteriors on KN parameters are consistent under the different premises, they do not align with expectations from simulations in numerical relativity (NR). 
Our prior choices reflect the range of typically found values~\citep{Metzger:2019}.
In contrast, the posterior on the wind ejecta's mass-averaged velocity rails against its lower prior bound at \qty{0.05}{c}.
This value has already been found for \textsc{Bu2025}~\citep{Koehn:2025b} and aligns with other lightcurve analyses using independent models, e.g.,~\citep{Siegel:2019,Breschi:2024} as it matches the slow KN fading in the optical bands.
Moreover, we expect the mass-averaged $Y_{\text{e}}$ to be higher for wind than for dynamical ejecta.
As we sample these parameters independently, though, they break this commonly observed correlations in NR simulations~\citep{Nedora:2021,Nedora:2022}.
We instead find $Y_{\text{e, dyn}} =\num{0.34(0.01)}$, far above $Y_{\text{e, wind}} = \num{0.25(0.01)}$.
Low $Y_e$ values generally lead to a stronger r-process that creates more lanthanides.
The high opacity of these elements then causes a redder KN with higher peak magnitudes in the IR bands and a correspondingly faster decline in the visual bands.
Under current modelling assumptions, the observed slow decline in visual bands and sudden post-peak decline in IR bands are therefore hard to reconcile with the expected production of heavy elements.

The versatility of \nmma\ allows us to easily explore these properties in greater detail.
We re-analyse the KN lightcurves with the \textsc{Bu2026} surrogate in different setups, using priors from \cref{tab:new_parameters} with the exception of uniform priors on the ejecta masses.
In order to reduce degeneracies due to the system localisation, we fix the luminosity distance and inclination to the previously found posterior median values at \qty{43.7}{Mpc} and \qty{24.9}{\degree}, respectively.
For these relatively simple problems, we apply the fast \textsc{PyMultinest}~\citep{Buchner:2014} and sample with 2000 live points, cross-checking some results with other samplers, too.

In order to explore the conflict between sudden IR decline and slower optical fading, we first consider how the parameters change in independent analyses of the $grizy$ and $JHK_s$ filters.
The resulting posteriors in \cref{tab:kn_only_results} draw indeed a very complementary picture.
The inferred ejecta properties for IR data fall well within expectations from NR simulations, while the optical data prefer an increase in wind ejecta with a low  $Y_{\text{e, wind}}$ and minor contributions from the dynamical ejecta.
This indicates model limitations in the lanthanide-poor KN regime dominating the optical emission.
A known limitation is the assumption of local thermodynamic equilibrium in radiative transfer calculations which must break down after about a week~\citep{Hotokezaka:2021,Pognan:2022}.
Further, since the \textsc{Bu2026} model assumes an idealised geometry for the expanding ejecta, deviations from that idealisation could likewise cause the ejecta masses to be misestimated~\citep{King:2025}.

\begin{table}[t]
    \renewcommand{\arraystretch}{1.2}
    \caption{
    Posteriors for AT2017gfo when restricting the inference to the filters of either column, excluding those of the other.
    }
    
    \centering
    \begin{tabular}{lrr}
    \toprule
    \toprule
    Parameter & $grizy$ & $JHK_s$\\
    \midrule
    $\log_{10}(m_{\text{dyn}}/\unit{\msun})$ & \num{-2.17(0.10:0.32)} & \num{-1.65(0.20:0.25)}\\
    $\log_{10}(m_{\text{wind}}/\unit{\msun})$ & \num{-0.62(0.04:0.27)} & \num{-1.01(0.13:0.16)}\\
    $v_{\text{dyn}}$ & \num{0.13(0.04:0.01)} & \num{0.21(0.03:0.03)}\\
    $Y_{\text{e, dyn}}$ & \num{0.26(0.04:0.01)} & \num{0.17(0.02:0.01)}\\
    $v_{\text{wind}}$ & \num{0.052(0.003:0.001)}& \num{0.09(0.05:0.03)}\\
    $Y_{\text{e, wind}}$ & \num{0.21(0.01:0.01)} & \num{0.39(0.01:0.01)}
    \end{tabular}
    \label{tab:kn_only_results}
\end{table}

\begin{figure}
    \centering
    \includegraphics[width=0.9\linewidth]{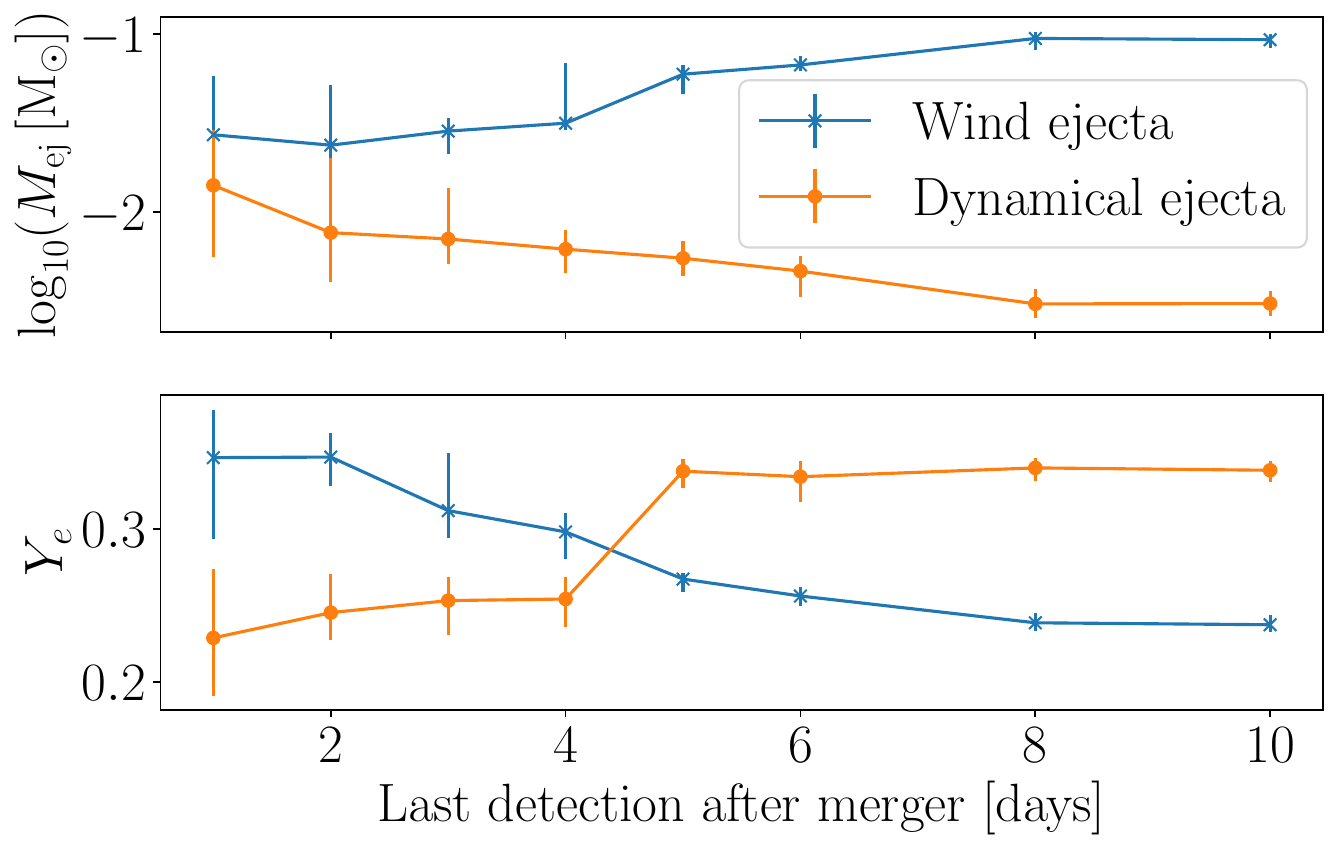}
    \caption{
    Posteriors on the ejecta mass and mean electron fraction of the dynamical and wind ejecta when only considering observations up to $n$ days after merger.
    As we include later observations, the inferred wind ejecta mass grows and increasingly dominates over the dynamical component while its electron fraction decreases, effectively pushing towards a single-component model.
    }
    \label{fig:kn_only_time_comparison}
\end{figure}

Therefore, and given the poor match to IR data at late times, we re-analyse the lightcurves only up to $n$ days after merger.
Figure~\ref{fig:kn_only_time_comparison} shows how $Y_{\text{e, wind}}$ continuously decreases as we include later observations, approaching the regime we would expect for the dynamical ejecta.
The share of dynamical ejecta in turn decreases while its electron fraction jumps towards the upper prior limit when including data after four days.
This balances the relatively slow rise of the IR signal against the higher wind ejecta masses that the \textsc{Bu2026} model needs to describe the late optical data.
As this again drives the posteriors away from the regime expected in NR simulations, we conclude a need for refined models of the late-time KN evolution, especially in the lanthanide-poor regime.

In Appendix \ref{app:extended_AT2017gfo}, we also briefly discuss the impact of enforcing the parameter correlations found in NR simulations as well as including NS population constraints. 
These further illustrate the versatility of \nmma, but provide only minor insights into the limits of KN modelling.

\subsection{Future Detection}
\label{ssec:Future_detection}

\begin{figure*}[t]
    \centering
    \includegraphics[width=0.99\textwidth]{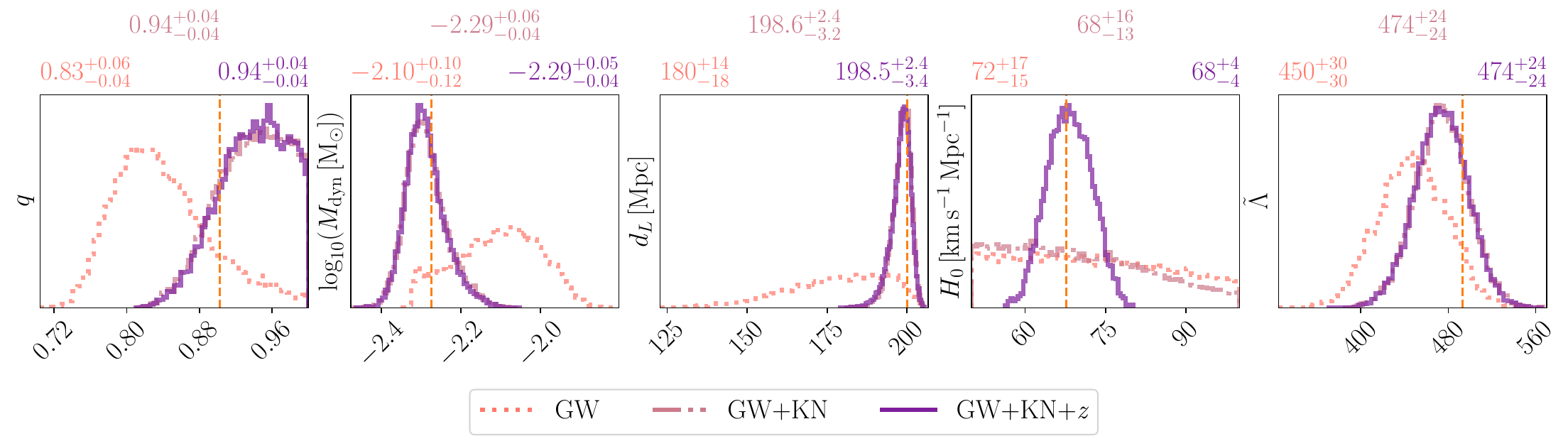}
    \caption{
        Posteriors on important astrophysical parameters with their injection values (dashed orange lines).
        We use dotted (dashed-dotted, solid) lines to indicate analyses using the GW strain (and KN, and redshift).
        We indicate the median and the \qty{68}{\percent} credible intervals on top of each panel.
        }
    \label{fig:future_injection_comparison}
\end{figure*}

To illustrate the usefulness of multimessenger studies as a bridge between microphysics and cosmology, we estimate the Hubble parameter $H_0$ from a hypothetical merger event under increasingly beneficial conditions.
First, we study only the GW signal measured by a triangular ET.
This can provide a basic $d_{\rm L}$ estimate, while the measured $\lambdaT$ implies only a weak constraint on $z$.
Second, we assume the Rubin Observatory~\citep{Izevic:2019} to observe an ensuing KN.
Its lightcurves should tighten our $d_{\rm L}$ estimate, whereas the $z$ measurement should hardly improve when remaining agnostic of the host galaxy.
In a third setup, we assume $z$ to be well measured from the host galaxy or in a very optimistic scenario from robustly identified spectral features~\citep{Domoto:2022,Domoto:2025,Hotokezaka:2022,Hotokezaka:2023,Gillanders:2022,Sneppen:2024}.
We expect the strongest constraints on $H_0$ from this setup.

We inject the binaries at \qty{200}{Mpc} (and, again assuming the Planck-2018 cosmology, with a corresponding redshift of 0.0437), in the sky location of AT2017gfo, such that robust KN observations are still possible.
With a primary at \qty{1.33}{\msun} and a secondary at \qty{1.2}{\msun}, our selected source-frame masses are near the lower plausible end of the expected distributions~\citep{Farrow:2019}.
For the EoS of the injection, we follow estimates that combine experimental and theoretical insights to set $\lsym=\qty{58.9}{MeV}$~\citep{Li:2019} and $\ksat=\qty{240}{MeV}$~\citep{Kumar:2024}.
The other parameters $\ksym=\qty{-290.3}{MeV}$, $\threensat=\qty{0.745}{c^2}$ and $\fivensat=\qty{0.313}{c^2}$ then allow $\mtov$ near \qty{2.3}{\msun} and $R_{1.4}=\qty{11.5}{\kilo\metre}$. 
This EoS is deliberately chosen at the softer end of what our nuclear model can still describe, providing some intuition for possible systematic effects associated with our TOV emulator; cf. Appendix~\ref{app:nuclear}.
Given the discussed limitations of current KN models, we employ ejecta electron fractions and velocities that are more in line with state-of-the-art NR simulations, i.e., $Y_{\text{e, dyn}} = 0.24$, $Y_{\text{e, wind}} = 0.35$, $v_{\text{dyn}} = \qty{0.25}{c}$ and $v_{\text{wind}} = \qty{0.1}{c}$~\citep{Nedora:2022,Neuweiler:2026}.
The remaining parameters are set to their respective values in the maximum-posterior samples of \cref{ssec:mma_bns2017}.
We add Gaussian noise to both the KN lightcurves and the injected GW signal, assuming a mean photometric error of \qty{0.1}{mag} and the ET-D sensitivity curve~\citep{Punturo:2010}, respectively.
This results in an ET network SNR of~\num{130}.

Our analyses then use approach C with the following prior modifications:
We analyse a longer GW signal of \qty{2048}{\second} duration and start waveform evaluation at \qty{10}{Hz}.
We adjust the limits on $\pi(\mathcal{M})$ to $[1.12, 1.15]$ and include $\pi(H_0) [\unit{\kilo\metre\per\second\per\mega\parsec}]= \mathcal{U}[50, 100]$ in each run.
Without redshift information, we additionally sample from $\pi(d_{\rm L})=\mathcal{U}_{\rm SF}[10, 400]$ and compute $z$ as a function of $d_{\rm L}$ in a flat $\Lambda$CDM cosmology.
We obtain it as a variation of Planck-2018 with the sampled $H_0$ using \textsc{astropy}~\citep{astropy:2022}.
When assuming a clearly measured redshift, we compute $d_{\rm L}$ from a $z$ in $\pi(z) = \mathcal{N}(0.0437, 0.003)$ instead, i.e. we assume a Gaussian prior centred on the injection redshift.

In this configuration, inference consumes 1200 CPUh for the GW-only analyses and around 1800 CPUh when including KN information.
Overall, all configurations recover the injected parameters within the $\qty{68}{\percent}$ credible intervals.
Figure~\ref{fig:future_injection_comparison} compares posteriors for some key parameters.
The lightcurve parameters in particular are mostly constrained accurately to a precision of a few percent, while the chirp mass is even measured with $\mathcal{O}(\qty{e-6}{\msun})$ uncertainty.
There are two exceptions to this for degenerate parameters that we have injected near the prior bounds:
In the EM-dark case, we cannot resolve the degeneracy between the inclination $\theta_{\rm obs}=\theta_{\rm JN}$ and $d_{\rm L}$.
Due to the nearly face-on injection, the true luminosity distance receives only marginal posterior support.
Similarly, the injected $K_{\rm sym}=\qty{-290.3}{MeV}$ is very close to its lower prior bound, reconciling the relatively low $R_{1.4}$ with a moderate $\mtov$ and the chosen $L_{\rm sym}$.
Since our nuclear parameters are by construction correlated and the measurement of $\lambdaT$ and the ejecta masses does not have the power to sufficiently resolve the $M-R$ relation from a single event, the soft EoS injection is only captured by upper limits on $\ksat\leq\qty{250}{MeV}$, $\ksym\leq\qty{-230}{MeV}$ and $\lsym\leq\qty{48}{MeV}$.

The noisy early-time lightcurve further contributes to these weak constraints on the injected $M-R$ relation because it favours the reduced dynamical ejecta mass of symmetric binaries. 
The sampled masses and consequently $R$ and $\Lambda$ of both components then appear very similar, reducing our ability to constrain nuclear parameters.
We therefore recover $R_{\text{1.4}}\approx\qtyrange{11.2}{11.6}{\kilo\metre}$ in all configurations with only a moderate shift towards the true $\lambdaT$ injection when we include the lightcurves.

Naturally, the information on $H_0$ is rather weak when not including explicit redshift information, although there is a clear preference for lower values.
This effect is again associated with our very soft injection choice for the EoS.
As we see an acceptable recovery of $d_{\rm L}$ even without the KN, low values of $H_0$ correspond to an underestimated redshift.
Since mass estimates from the GW analysis are redshifted, a low $z$ implies higher source-frame masses.
This allows stiffer EoSs to reach the same observed deformability, providing some counter-balance to our relatively soft injection.

The picture then changes under the strong prior on $z$.
Because the combined information of KN and GW data allow a very accurate $d_{\rm L}$ measurement, we obtain a very informative estimate of $H_0= \qty{68(4)}{\kilo\metre\per\second\per\mega\parsec}$ in excellent agreement with the injected $H_0^{\rm P18} = \qty{67.6}{\kilo\metre\per\second\per\mega\parsec}$.
The level of uncertainty is essentially mandated by the width of the redshift prior.
Although it is not yet competitive, we can expect a drastic reduction from combining multiple events in the era of third-generation detectors~\citep{Abac:2026}.
Moreover, in a multi-detector configuration with other systems like Cosmic Explorer~\citep{Reitze:2019,Evans:2021}, we can expect higher precision in individual measurements of $\lambdaT$ that will further increase the information gain in joint inferences.

\section{Conclusion}
\label{sec:conclusion}
In this article, we have presented significant upgrades and extensions to the Bayesian multimessenger framework \nmma~\citep{Pang:2024}.
By improving the modularisation of its components and aligning the codebase more closely with the popular \bilby\ framework, it allows versatile studies in multimessenger astrophysics broadly construed with present and future detector systems.
Apart from various performance improvements, the introduction of new interfaces, e.g., for compatibility with \fiesta\ EM surrogates, and of emulators for nuclear-physics informed NS properties provides access to new analysis pathways at reasonable computational cost.

We have demonstrated the performance of the new implementation by re-analysing the 2017 BNS merger event with and without including information from the GRB afterglow, showing good consistency with previous results while achieving a 20 to 60-times efficiency gain.
Moreover, we have demonstrated how we can obtain informative constraints on both the Hubble parameter and nuclear parameters from a single event with next-generation detectors. 
Although the derived limits are not competitive with state-of-the-art studies, they open an avenue to strong, robust and independent constraints when analysing multiple events.
In particular, we can expect much stronger nuclear constraints from more asymmetric low-mass mergers.

Various limitations remain that have to be addressed in future works.
In order to achieve a more accurate source classification, it will become necessary to directly incorporate rotation corrections to the NS structure properties~\citep{Koehn:2024}.
Moreover, in the current setup, we use a single waveform approximant to evaluate GW data irrespective of the assumed nature of the binary components.
While we deem this acceptable for the systems studied here, a more generalised approach is desirable that adjusts the waveform approximant to the mass samples under consideration, in particular for studies of potential BH-NS binaries.

At this stage, \nmma\ remains CPU-bound. 
Although important steps are currently being taken to enable likelihood-based inference executed on GPUs~\citep{Edwards:2024,Wong:2023,Wouters:2024,Koehn:2025b,Wouters:2025a,Prathaban:2025} or likelihood-free inference of GWs~\citep{Dax:2025,Hu:2025b} and KNe~\citep{Desai:2025,Brown:2026}, many of the employed models and frameworks still rely on CPU-based implementations.
Our extended framework reflects these developments as it provides interfaces to more recent GPU-centred packages like \fiesta, but keeps support for more extensively tested packages that are trusted by the community, offering an intermediate path as the field moves towards fully GPU-based solutions.

\begin{acknowledgements} 
The authors are very grateful to I. Tews and B. Reed for insightful discussions.
H.R., H.K.\ and T.D. acknowledge funding from the EU Horizon under ERC Starting Grant, no.\ SMArt-101076369. Views and opinions expressed are those of the authors only and do not necessarily reflect those of the European Union or the European Research Council. Neither the European Union nor the granting authority can be held responsible for them.
T.W. is supported by the research program of the Netherlands Organization for Scientific Research (NWO) under grant number OCENW.XL21.XL21.038.
P.T.H.P. is supported by the research program of the Netherlands Organization for Scientific Research (NWO) under grant number VI.Veni.232.021. M. B. acknowledges the Department of Physics and Earth Science of the University of Ferrara for the financial support through the FIRD 2025 grant.  
Analysis presented in this study have been run on the HPC system Jarvis at the University of Potsdam
(DFG-project INST 336/173-1; project number: 502227537).
This research has made use of data or software obtained from the Gravitational Wave Open Science Center (gwosc.org), a service of the LIGO Scientific Collaboration, the Virgo Collaboration, and KAGRA. 
This material is based upon work supported by NSF’s LIGO Laboratory which is a major facility fully funded by the National Science Foundation, as well as the Science and Technology Facilities Council (STFC) of the United Kingdom, the Max-Planck-Society (MPS), and the State of Niedersachsen/Germany for support of the construction of Advanced LIGO and construction and operation of the GEO600 detector. 
Additional support for Advanced LIGO was provided by the Australian Research Council.
Virgo is funded, through the European Gravitational Observatory (EGO), by the French Centre National de Recherche Scientifique (CNRS), the Italian Istituto Nazionale di Fisica Nucleare (INFN) and the Dutch Nikhef, with contributions by institutions from Belgium, Germany, Greece, Hungary, Ireland, Japan, Monaco, Poland, Portugal, Spain. 
KAGRA is supported by Ministry of Education, Culture, Sports, Science and Technology (MEXT), Japan Society for the Promotion of Science (JSPS) in Japan; National Research Foundation (NRF) and Ministry of Science and ICT (MSIT) in Korea; Academia Sinica (AS) and National Science and Technology Council (NSTC) in Taiwan.
\end{acknowledgements}
\bibliographystyle{aa_new} 
\bibliography{_literature}

@ARTICLE{Abac:2026,
       author = {{Abac}, Adrian and {Abramo}, Raul and {Albanesi}, Simone and {Albertini}, Angelica and {Agapito}, Alessandro and {Agathos}, Michalis and {Albertus}, Conrado and {Andersson}, Nils and {Andrade}, Tom{\'a}s and {Andreoni}, Igor and {Angeloni}, Federico and {Antonelli}, Marco and {Antoniadis}, John and {Antonini}, Fabio and {Arca Sedda}, Manuel and {Artale}, M. Celeste and {Ascenzi}, Stefano and {Auclair}, Pierre and {Bachetti}, Matteo and {Badger}, Charles and {Banerjee}, Biswajit and {Barba-Gonz{\'a}lez}, David and {Barta}, D{\'a}niel and {Bartolo}, Nicola and {Bauswein}, Andreas},
        title = "{The Science of the Einstein Telescope}",
      journal = {\jcap},
         year = 2026,
        month = mar,
       volume = {2026},
       number = {3},
          eid = {081},
        pages = {081},
          doi = {10.1088/1475-7516/2026/03/081},
archivePrefix = {arXiv},
       eprint = {2503.12263},
 primaryClass = {gr-qc},
}

@ARTICLE{Abac:2025,
    collaboration = "LIGO Scientific, VIRGO, KAGRA",
       author = {{Abac}, A.~G. and {Abouelfettouh}, I. and {Acernese}, F. and {Ackley}, K. and {Adamcewicz}, C. and {Adhicary}, S. and {Adhikari}, D. and {Adhikari}, N. and {Adhikari}, R.~X. and {Adkins}, V.~K. and {Afroz}, S. and {Agapito}, A. and {Agarwal}, D. and {Agathos}, M. and {Aggarwal}, N. and {Aggarwal}, S. and {Aguiar}, O.~D. and {Ahrend}, I.-L. and {Aiello}, L. and {Ain}, A. and {Ajith}, P. and {Akutsu}, T. and {Albanesi}, S. and {Ali}, W. and {Al-Kershi}, S. and {All{\'e}n{\'e}}, C. and {Allocca}, A. and {Al-Shammari}, S. and {Altin}, P.~A. and {Alvarez-Lopez}, S. and {Amar}, W. and {Amarasinghe}, O. and {Amato}, A. },
        title = "{Open Data from LIGO, Virgo, and KAGRA through the First Part of the Fourth Observing Run}",
      journal = {arXiv e-prints},
         year = 2025,
        month = aug,
          eid = {arXiv:2508.18079},
          doi = {10.48550/arXiv.2508.18079},
archivePrefix = {arXiv},
       eprint = {2508.18079},
 primaryClass = {gr-qc},
}

@ARTICLE{Abac:2024,
       author = {{Abac}, Adrian and {Dietrich}, Tim and {Buonanno}, Alessandra and {Steinhoff}, Jan and {Ujevic}, Maximiliano},
        title = "{New and robust gravitational-waveform model for high-mass-ratio binary neutron star systems with dynamical tidal effects}",
      journal = {\prd},
         year = 2024,
        month = jan,
       volume = {109},
       number = {2},
          eid = {024062},
        pages = {024062},
          doi = {10.1103/PhysRevD.109.024062},
archivePrefix = {arXiv},
       eprint = {2311.07456},
 primaryClass = {gr-qc}
}

@ARTICLE{Abac:2024b,
       author = {{Abac}, A.~G. and {Abbott}, R. and {Abouelfettouh}, I. and {Acernese}, F. and {Ackley}, K. and {Adhicary}, S. and {Adhikari}, N. and {Adhikari}, R.~X. and {Adkins}, V.~K. and {Agarwal}, D. and {Agathos}, M. and {Abchouyeh}, M. Aghaei and {Aguiar}, O.~D. and {Aguilar}, I. and {Aiello}, L. and {Ain}, A. and {Ajith}, P. and {Ak{\c{c}}ay}, S. and {Akutsu}, T. and {Albanesi}, S. and {Alfaidi}, R.~A. and {Al-Jodah}, A. and {All{\'e}n{\'e}}, C. and {Allocca}, A. and {Al-Shammari}, S. and {Altin}, P.~A. and {Alvarez-Lopez}, S. and {Amato}, A. and {Amez-Droz}, L. and {Amorosi}, A. and {Amra}, C. and {Ananyeva}, A. and {Anderson}, S.~B. and {Anderson}, W.~G. and {Andia}, M. and {Ando}, M. and {Andrade}, T. and {Andres}, N. and {Andr{\'e}s-Carcasona}, M. and {Andri{\'c}}, T. and {Anglin}, J. and {Ansoldi}, S. and {Antelis}, J.~M. and {Antier}, S. and {Aoumi}, M. and {Appavuravther}, E.~Z. and {Appert}, S.},
        title = "{Observation of Gravitational Waves from the Coalescence of a 2.5─4.5 M $_{{\ensuremath{\odot}}}$ Compact Object and a Neutron Star}",
      journal = {\apjl},
         year = 2024,
        month = aug,
       volume = {970},
       number = {2},
          eid = {L34},
        pages = {L34},
          doi = {10.3847/2041-8213/ad5beb},
archivePrefix = {arXiv},
       eprint = {2404.04248},
 primaryClass = {astro-ph.HE}
}

@ARTICLE{Abbott:2021,
       author = {{Abbott}, R. and {Abbott}, T.~D. and {Abraham}, S. and {Acernese}, F. and {Ackley}, K. and {Adams}, C. and {Adhikari}, R.~X. and {Adya}, V.~B. and {Affeldt}, C. and {Agathos}, M. and {Agatsuma}, K. and {Aggarwal}, N. and {Aguiar}, O.~D. and {Aich}, A. and {Aiello}, L. and {Ain}, A. and {Ajith}, P. and {Allen}, G. and {Allocca}, A. and {Altin}, P.~A. and {Amato}, A. and {Anand}, S. and {Ananyeva}, A. },
        title = "{Open data from the first and second observing runs of Advanced LIGO and Advanced Virgo}",
      journal = {SoftwareX},
         year = 2021,
        month = jan,
       volume = {13},
          eid = {100658},
        pages = {100658},
          doi = {10.1016/j.softx.2021.100658},
archivePrefix = {arXiv},
       eprint = {1912.11716},
 primaryClass = {gr-qc}
}

@ARTICLE{Abbott:2019b,
       author = {{Abbott}, B.~P. and {Abbott}, R. and {Abbott}, T.~D. and {Acernese}, F. and {Ackley}, K. and {Adams}, C. and {Adams}, T. and {Addesso}, P. and {Adhikari}, R.~X. and {Adya}, V.~B. and {Affeldt}, C. and {Agarwal}, B. and {Agathos}, M. and {Agatsuma}, K. and {Aggarwal}, N. and {Aguiar}, O.~D. and {Aiello}, L. and {Ain}, A. and {Ajith}, P. and {Allen}, B. and {Allen}, G. and {Allocca}, A. and {Aloy}, M.~A. and {Altin}, P.~A. and {Amato}, A. and {Ananyeva}, A. and {Anderson}, S.~B. and {Anderson}, W.~G. and {Angelova}, S.~V. and {Antier}, S.},
        title = "{Properties of the Binary Neutron Star Merger GW170817}",
      journal = {Physical Review X},
         year = 2019,
        month = jan,
       volume = {9},
       number = {1},
          eid = {011001},
        pages = {011001},
          doi = {10.1103/PhysRevX.9.011001},
archivePrefix = {arXiv},
       eprint = {1805.11579},
 primaryClass = {gr-qc}
}

@ARTICLE{Abbott:2017,
       author = {{Abbott}, B.~P. and {Abbott}, R. and {Abbott}, T.~D. and {Acernese}, F. and {Ackley}, K. and {Adams}, C. and {Adams}, T. and {Addesso}, P. and {Adhikari}, R.~X. and {Adya}, V.~B. and {Affeldt}, C. and {Afrough}, M. and {Agarwal}, B. and {Agathos}, M. and {Agatsuma}, K. and {Aggarwal}, N. and {Aguiar}, O.~D. and {Aiello}, L. and {Ain}, A. and {Ajith}, P. and {Allen}, B. and {Allen}, G. and {Allocca}, A. and {Altin}, P.~A. and {Amato}, A. and {Ananyeva}, A. and {Anderson}, S.~B. and {Anderson}, W.~G. and {Angelova}, S.~V. and {Antier}, S.},
        title = "{GW170817: Observation of Gravitational Waves from a Binary Neutron Star Inspiral}",
      journal = {\prl},
         year = 2017,
        month = oct,
       volume = {119},
       number = {16},
          eid = {161101},
        pages = {161101},
          doi = {10.1103/PhysRevLett.119.161101},
archivePrefix = {arXiv},
       eprint = {1710.05832},
 primaryClass = {gr-qc}
}

@ARTICLE{Abbott:2017b,
       author = {{Abbott}, B.~P. and {Abbott}, R. and {Abbott}, T.~D. and {Acernese}, F. and {Ackley}, K. and {Adams}, C. and {Adams}, T. and {Addesso}, P. and {Adhikari}, R.~X. and {Adya}, V.~B. and {Affeldt}, C. and {Afrough}, M. and {Agarwal}, B. and {Agathos}, M. and {Agatsuma}, K. and {Aggarwal}, N. and {Aguiar}, O.~D. and {Aiello}, L. and {Ain}, A. and {Ajith}, P. and {Allen}, B. and {Allen}, G. and {Allocca}, A. },
        title = "{Multi-messenger Observations of a Binary Neutron Star Merger}",
      journal = {\apjl},
         year = 2017,
        month = oct,
       volume = {848},
       number = {2},
          eid = {L12},
        pages = {L12},
          doi = {10.3847/2041-8213/aa91c9},
archivePrefix = {arXiv},
       eprint = {1710.05833},
 primaryClass = {astro-ph.HE}
}

@ARTICLE{Abbott:2016,
       author = {{Abbott}, B.~P. and {Abbott}, R. and {Abbott}, T.~D. and {Abernathy}, M.~R. and {Acernese}, F. and {Ackley}, K. and {Adams}, C. and {Adams}, T. and {Addesso}, P. and {Adhikari}, R.~X. and {Adya}, V.~B. and {Affeldt}, C. and {Agathos}, M. and {Agatsuma}, K. and {Aggarwal}, N. and {Aguiar}, O.~D. and {Aiello}, L. and {Ain}, A. and {Ajith}, P. and {Allen}, B. and {Allocca}, A. and {Altin}, P.~A. and {Anderson}, S.~B. and {Anderson}, W.~G. and {Arai}, K. and {Araya}, M.~C. and {Arceneaux}, C.~C. and {Areeda}, J.~S. and {Arnaud}, N. and {Arun}, K.~G. and {Ascenzi}, S. and {Ashton}, G. and {Ast}, M. and {Aston}, S.~M. and {Astone}, P. and {Aufmuth}, P. and {Aulbert}, C. and {Babak}, S. and {Bacon}, P. and {Bader}, M.~K.~M. and {Baker}, P.~T. and {Baldaccini}, F. and {Ballardin}, G. and {Ballmer}, S.~W. and {Barayoga}, J.~C. and {Barclay}, S.~E. and {Barish}, B.~C. and {Barker}, D. and {Barone}, F. and {Barr}, B. and {Barsotti}, L. and {Barsuglia}, M. and {Barta}, D. and {Bartlett}, J. and {Bartos}, I. and {Bassiri}, R. and {Basti}, A. and {Batch}, J.~C. and {Baune}, C.},
        title = "{Tests of General Relativity with GW150914}",
      journal = {\prl},
         year = 2016,
        month = may,
       volume = {116},
       number = {22},
          eid = {221101},
        pages = {221101},
          doi = {10.1103/PhysRevLett.116.221101},
archivePrefix = {arXiv},
       eprint = {1602.03841},
 primaryClass = {gr-qc},
}

@ARTICLE{Adams:2016,
       author = {{Adams}, T. and {Buskulic}, D. and {Germain}, V. and {Guidi}, G.~M. and {Marion}, F. and {Montani}, M. and {Mours}, B. and {Piergiovanni}, F. and {Wang}, G.},
        title = "{Low-latency analysis pipeline for compact binary coalescences in the advanced gravitational wave detector era}",
      journal = {Classical and Quantum Gravity},
         year = 2016,
        month = sep,
       volume = {33},
       number = {17},
          eid = {175012},
        pages = {175012},
          doi = {10.1088/0264-9381/33/17/175012},
archivePrefix = {arXiv},
       eprint = {1512.02864},
 primaryClass = {gr-qc}
}

@ARTICLE{Agathos:2020,
       author = {{Agathos}, Michalis and {Zappa}, Francesco and {Bernuzzi}, Sebastiano and {Perego}, Albino and {Breschi}, Matteo and {Radice}, David},
        title = "{Inferring prompt black-hole formation in neutron star mergers from gravitational-wave data}",
      journal = {\prd},
         year = 2020,
        month = feb,
       volume = {101},
       number = {4},
          eid = {044006},
        pages = {044006},
          doi = {10.1103/PhysRevD.101.044006},
archivePrefix = {arXiv},
       eprint = {1908.05442},
 primaryClass = {gr-qc}
}

@ARTICLE{Akl:2026,
       author = {{Akl}, D. and {Antier}, S. and {Koehn}, H. and {Pang}, P.~T.~H. and {Geng}, J.~J. and {Gill}, R. and {Abdikamalov}, E. and {Adami}, C. and {Aivazyan}, V. and {Almeida}, L. and {Alshamsi}, S. and {Andrade}, C. and {Andr{\'e}}, Q. and {Angulo-Valdez}, C. and {Atteia}, J.-L. and {Barkaoui}, K. and {Basa}, S. and {Becerra}, R.~L. and {Bendjoya}, P. and {Berdikhan}, D. and {Bernaud}, E. and {Boissier}, S. and {Brunier}, S. and {Burdanov}, A.~Y. and {Butler}, N.~R. and {Chen}, J. and {Colas}, F. and {Corradi}, W. and {Coughlin}, M.~W. and {Darson}, D. and {Dietrich}, T. and {Dornic}, D. and {Douzet}, C. and {Dubois}, C. and {Ducoin}, J.-G. and {du Laz}, T. and {Durroux}, A. and {Dutton}, D. and {Duverne}, P.-A. and {Dux}, F. and {Elhosseiny}, E.~G. and {Esamdin}, A. and {Filippenko}, A.~V. and {Fortin}, F. and {Freeberg}, M. and {Garc{\'\i}a-Garc{\'\i}a}, L. and {Gillon}, M. and {Globus}, N. and {Gokuldass}, P. and {Guessoum}, N. and {Hello}, P. and {Hellot}, R. and {Hendy}, Y.~H.~M. and {Hua}, Y.~L. and {Hussenot-Desenonges}, T. and {Inasaridze}, R. and {Iskandar}, A. and {Jel{\'\i}nek}, M. and {Karpov}, S. and {Klotz}, A. and {Kochiashvili}, N. and {Laskar}, T. and {Le Calloch}, A. and {Lee}, W.~H. and {Leonini}, S. and {Li}, X.~Y. and {Lien}, A. and {Limonta}, C. and {Liu}, J. and {L{\'o}pez-C{\'a}mara}, D. and {Magnani}, F. and {Mao}, J. and {Ma{\v{s}}ek}, M. and {Moreno M{\'e}ndez}, E. and {Menegazzi}, L.~C. and {Mercier}, W. and {Mihov}, B.~M. and {Molham}, M. and {Oates}, S. and {Odeh}, M. and {Peng}, H. and {Pereyra}, M. and {Pillas}, M. and {Pradier}, T. and {Rakotondrainibe}, N.~A. and {Reichart}, D. and {Rivet}, J.-P. and {Romanov}, F.~D. and {S{\'a}nchez-{\'A}lvarez}, F. and {Sasaki}, N. and {Schlekat}, D. and {Schneider}, B. and {Simon}, A. and {Slavcheva-Mihova}, L. and {Strausbaugh}, R. and {Sun}, T.~R. and {Takey}, A. and {Tanasan}, M. and {Turpin}, D. and {de Ugarte Postigo}, A. and {Wang}, L.~T. and {Wang}, X.~F. and {Wang}, Z.~M. and {Watson}, A.~M. and {de Wit}, J. and {Yan}, Y.~S. and {Zheng}, W. and {Z{\'u}{\~n}iga-Fern{\'a}ndez}, S.},
        title = "{Multi-epoch afterglow rebrightenings in GRB 250129A: Evidence for successive shock interactions}",
      journal = {arXiv e-prints},
         year = 2026,
        month = mar,
          eid = {arXiv:2603.08555},
          doi = {10.48550/arXiv.2603.08555},
archivePrefix = {arXiv},
       eprint = {2603.08555},
 primaryClass = {astro-ph.HE}
}

@ARTICLE{Allene:2025,
       author = {{All{\'e}n{\'e}}, Christopher and {Aubin}, Florian and {Bentara}, In{\`e}s and {Buskulic}, Damir and {Guidi}, Gianluca M. and {Juste}, Vincent and {Lethuillier}, Morgan and {Marion}, Fr{\'e}d{\'e}rique and {Mobilia}, Lorenzo and {Mours}, Beno{\^\i}t and {Ouzriat}, Amazigh and {Sainrat}, Thomas and {Sordini}, Viola},
        title = "{The MBTA pipeline for detecting compact binary coalescences in the fourth LIGO-Virgo-KAGRA observing run}",
      journal = {Classical and Quantum Gravity},
         year = 2025,
        month = may,
       volume = {42},
       number = {10},
          eid = {105009},
        pages = {105009},
          doi = {10.1088/1361-6382/add234},
archivePrefix = {arXiv},
       eprint = {2501.04598},
 primaryClass = {gr-qc}
}

@ARTICLE{Arnett:1989,
       author = {{Arnett}, W. David and {Bahcall}, John N. and {Kirshner}, Robert P. and {Woosley}, Stanford E.},
        title = "{Supernova 1987A.}",
      journal = {\araa},
         year = 1989,
        month = jan,
       volume = {27},
        pages = {629-700},
          doi = {10.1146/annurev.aa.27.090189.003213}
}

@ARTICLE{Ashton:2021,
       author = {{Ashton}, G. and {Talbot}, C.},
        title = "{BILBY-MCMC: an MCMC sampler for gravitational-wave inference}",
      journal = {\mnras},
         year = 2021,
        month = oct,
       volume = {507},
       number = {2},
        pages = {2037-2051},
          doi = {10.1093/mnras/stab2236},
archivePrefix = {arXiv},
       eprint = {2106.08730},
 primaryClass = {gr-qc}
}

@ARTICLE{Ashton:2019,
       author = {{Ashton}, Gregory and {H{\"u}bner}, Moritz and {Lasky}, Paul D. and {Talbot}, Colm and {Ackley}, Kendall and {Biscoveanu}, Sylvia and {Chu}, Qi and {Divakarla}, Atul and {Easter}, Paul J. and {Goncharov}, Boris and {Hernandez Vivanco}, Francisco and {Harms}, Jan and {Lower}, Marcus E. and {Meadors}, Grant D. and {Melchor}, Denyz and {Payne}, Ethan and {Pitkin}, Matthew D. and {Powell}, Jade and {Sarin}, Nikhil and {Smith}, Rory J.~E. and {Thrane}, Eric},
        title = "{BILBY: A User-friendly Bayesian Inference Library for Gravitational-wave Astronomy}",
      journal = {\apjs},
         year = 2019,
        month = apr,
       volume = {241},
       number = {2},
          eid = {27},
        pages = {27},
          doi = {10.3847/1538-4365/ab06fc},
archivePrefix = {arXiv},
       eprint = {1811.02042},
 primaryClass = {astro-ph.IM}
}

@ARTICLE{astropy:2022,
       author = {{Astropy Collaboration} and {Price-Whelan}, Adrian M. and {Lim}, Pey Lian and {Earl}, Nicholas and {Starkman}, Nathaniel and {Bradley}, Larry and {Shupe}, David L. and {Patil}, Aarya A. and {Corrales}, Lia and {Brasseur}, C.~E. and {N{\"o}the}, Maximilian and {Donath}, Axel and {Tollerud}, Erik and {Morris}, Brett M. and {Ginsburg}, Adam and {Vaher}, Eero and {Weaver}, Benjamin A. and {Tocknell}, James and {Jamieson}, William and {van Kerkwijk}, Marten H. and {Robitaille}, Thomas P. and {Merry}, Bruce and {Bachetti}, Matteo and {G{\"u}nther}, H. Moritz and {Aldcroft}, Thomas L. and {Alvarado-Montes}, Jaime A. and {Archibald}, Anne M. and {B{\'o}di}, Attila},
        title = "{The Astropy Project: Sustaining and Growing a Community-oriented Open-source Project and the Latest Major Release (v5.0) of the Core Package}",
      journal = {\apj},
         year = 2022,
        month = aug,
       volume = {935},
       number = {2},
          eid = {167},
        pages = {167},
          doi = {10.3847/1538-4357/ac7c74},
archivePrefix = {arXiv},
       eprint = {2206.14220},
 primaryClass = {astro-ph.IM},
}

@ARTICLE{Bailyn:1998,
       author = {{Bailyn}, Charles D. and {Jain}, Raj K. and {Coppi}, Paolo and {Orosz}, Jerome A.},
        title = "{The Mass Distribution of Stellar Black Holes}",
      journal = {\apj},
         year = 1998,
        month = may,
       volume = {499},
       number = {1},
        pages = {367-374},
          doi = {10.1086/305614},
archivePrefix = {arXiv},
       eprint = {astro-ph/9708032},
 primaryClass = {astro-ph}
}

@ARTICLE{Baird:2013,
       author = {{Baird}, Emily and {Fairhurst}, Stephen and {Hannam}, Mark and {Murphy}, Patricia},
        title = "{Degeneracy between mass and spin in black-hole-binary waveforms}",
      journal = {\prd},
         year = 2013,
        month = jan,
       volume = {87},
       number = {2},
          eid = {024035},
        pages = {024035},
          doi = {10.1103/PhysRevD.87.024035},
archivePrefix = {arXiv},
       eprint = {1211.0546},
 primaryClass = {gr-qc}
}

@article{Barbary:2016,
       author = {{Barbary}, Kyle and {Barclay}, Tom and {Biswas}, Rahul and {Craig}, Matt and {Feindt}, Ulrich and {Friesen}, Brian and {Goldstein}, Danny and {Jha}, Saurabh and {Rodney}, Steve and {Sofiatti}, Caroline and {Thomas}, Rollin C. and {Wood-Vasey}, Michael},
        title = "{SNCosmo: Python library for supernova cosmology}",
      journal = {Astrophysics Source Code Library},
         year = 2016,
        month = nov,
archivePrefix = {ascl},
       eprint = {1611.017}
}

@ARTICLE{Bardeen:1972,
       author = {{Bardeen}, James M. and {Press}, William H. and {Teukolsky}, Saul A.},
        title = "{Rotating Black Holes: Locally Nonrotating Frames, Energy Extraction, and Scalar Synchrotron Radiation}",
      journal = {\apj},
         year = 1972,
        month = dec,
       volume = {178},
        pages = {347-370},
          doi = {10.1086/151796}
}

@ARTICLE{Barna:2025,
       author = {{Barna}, Tyler and {Fremling}, Christoffer and {Ahumada}, Tomas and {Andreoni}, Igor and {Banerjee}, Smaranika and {Bloom}, Joshua S. and {Bulla}, Mattia and {Chen}, Tracy X. and {Coughlin}, Michael W. and {Dietrich}, Tim and {Hall}, Xander J. and {Junell}, Alexandra and {Rusholme}, Ben and {Sollerman}, Jesper and {Sravan}, Niharika},
        title = "{IIb or not IIb: A Catalog of ZTF Kilonova Imposters}",
      journal = {\pasp},
         year = 2025,
        month = aug,
       volume = {137},
       number = {8},
          eid = {084105},
        pages = {084105},
          doi = {10.1088/1538-3873/adf578},
archivePrefix = {arXiv},
       eprint = {2506.15900},
 primaryClass = {astro-ph.HE}
}

@ARTICLE{Barr:2024,
       author = {{Barr}, Ewan D. and {Dutta}, Arunima and {Freire}, Paulo C.~C. and {Cadelano}, Mario and {Gautam}, Tasha and {Kramer}, Michael and {Pallanca}, Cristina and {Ransom}, Scott M. and {Ridolfi}, Alessandro and {Stappers}, Benjamin W. and {Tauris}, Thomas M. and {Venkatraman Krishnan}, Vivek and {Wex}, Norbert and {Bailes}, Matthew and {Behrend}, Jan and {Buchner}, Sarah and {Burgay}, Marta and {Chen}, Weiwei and {Champion}, David J. and {Chen}, C.-H. Rosie and {Corongiu}, Alessandro and {Geyer}, Marisa and {Men}, Y.~P. and {Padmanabh}, Prajwal Voraganti and {Possenti}, Andrea},
        title = "{A pulsar in a binary with a compact object in the mass gap between neutron stars and black holes}",
      journal = {Science},
         year = 2024,
        month = jan,
       volume = {383},
       number = {6680},
        pages = {275-279},
          doi = {10.1126/science.adg3005},
archivePrefix = {arXiv},
       eprint = {2401.09872},
 primaryClass = {astro-ph.HE}
}

@ARTICLE{Baumgarte:2000,
       author = {{Baumgarte}, Thomas W. and {Shapiro}, Stuart L. and {Shibata}, Masaru},
        title = "{On the Maximum Mass of Differentially Rotating Neutron Stars}",
      journal = {\apjl},
         year = 2000,
        month = jan,
       volume = {528},
       number = {1},
        pages = {L29-L32},
          doi = {10.1086/312425},
archivePrefix = {arXiv},
       eprint = {astro-ph/9910565},
 primaryClass = {astro-ph}
}

@ARTICLE{Bauswein:2013,
       author = {{Bauswein}, A. and {Baumgarte}, T.~W. and {Janka}, H.-T.},
        title = "{Prompt Merger Collapse and the Maximum Mass of Neutron Stars}",
      journal = {\prl},
         year = 2013,
        month = sep,
       volume = {111},
       number = {13},
          eid = {131101},
        pages = {131101},
          doi = {10.1103/PhysRevLett.111.131101},
archivePrefix = {arXiv},
       eprint = {1307.5191},
 primaryClass = {astro-ph.SR}
}

@ARTICLE{Bethe:1990,
       author = {{Bethe}, H.~A.},
        title = "{Supernova mechanisms}",
      journal = {Reviews of Modern Physics},
         year = 1990,
        month = oct,
       volume = {62},
       number = {4},
        pages = {801-866},
          doi = {10.1103/RevModPhys.62.801}
}

@ARTICLE{Biwer:2019,
       author = {{Biwer}, C.~M. and {Capano}, Collin D. and {De}, Soumi and {Cabero}, Miriam and {Brown}, Duncan A. and {Nitz}, Alexander H. and {Raymond}, V.},
        title = "{PyCBC Inference: A Python-based Parameter Estimation Toolkit for Compact Binary Coalescence Signal}",
      journal = {\pasp},
         year = 2019,
        month = feb,
       volume = {131},
       number = {996},
        pages = {024503},
          doi = {10.1088/1538-3873/aaef0b},
archivePrefix = {arXiv},
       eprint = {1807.10312},
 primaryClass = {astro-ph.IM}
}

@ARTICLE{Blandford:1977,
       author = {{Blandford}, R.~D. and {Znajek}, R.~L.},
        title = "{Electromagnetic extraction of energy from Kerr black holes.}",
      journal = {\mnras},
         year = 1977,
        month = may,
       volume = {179},
        pages = {433-456},
          doi = {10.1093/mnras/179.3.433},
       adsurl = {https://ui.adsabs.harvard.edu/abs/1977MNRAS.179..433B}
}

@ARTICLE{Bogdanovic:2022,
       author = {{Bogdanovi{\'c}}, Tamara and {Miller}, M. Coleman and {Blecha}, Laura},
        title = "{Electromagnetic counterparts to massive black-hole mergers}",
      journal = {Living Reviews in Relativity},
         year = 2022,
        month = dec,
       volume = {25},
       number = {1},
          eid = {3},
        pages = {3},
          doi = {10.1007/s41114-022-00037-8},
archivePrefix = {arXiv},
       eprint = {2109.03262},
 primaryClass = {astro-ph.HE}
}

@ARTICLE{Breschi:2024,
       author = {{Breschi}, Matteo and {Gamba}, Rossella and {Carullo}, Gregorio and {Godzieba}, Daniel and {Bernuzzi}, Sebastiano and {Perego}, Albino and {Radice}, David},
        title = "{Bayesian inference of multi-messenger astrophysical data: Joint and coherent inference of gravitational waves and kilonovae}",
      journal = {\aap},
         year = 2024,
        month = sep,
       volume = {689},
          eid = {A51},
        pages = {A51},
          doi = {10.1051/0004-6361/202449173},
archivePrefix = {arXiv},
       eprint = {2401.03750},
 primaryClass = {astro-ph.HE}
}

@ARTICLE{Breschi:2021,
       author = {{Breschi}, Matteo and {Gamba}, Rossella and {Bernuzzi}, Sebastiano},
        title = "{Bayesian inference of multimessenger astrophysical data: Methods and applications to gravitational waves}",
      journal = {\prd},
         year = 2021,
        month = aug,
       volume = {104},
       number = {4},
          eid = {042001},
        pages = {042001},
          doi = {10.1103/PhysRevD.104.042001},
archivePrefix = {arXiv},
       eprint = {2102.00017},
 primaryClass = {gr-qc},
       adsurl = {https://ui.adsabs.harvard.edu/abs/2021PhRvD.104d2001B}
}

@ARTICLE{Breu:2016,
       author = {{Breu}, Cosima and {Rezzolla}, Luciano},
        title = "{Maximum mass, moment of inertia and compactness of relativistic stars}",
      journal = {\mnras},
         year = 2016,
        month = jun,
       volume = {459},
       number = {1},
        pages = {646-656},
          doi = {10.1093/mnras/stw575},
archivePrefix = {arXiv},
       eprint = {1601.06083},
 primaryClass = {gr-qc}
}

@ARTICLE{Brown:2026,
       author = {{Brown}, Stephanie M. and {Bulla}, Mattia and {Peiris}, Hiranya V. and {Sarin}, Nikhil and {Mortlock}, Daniel and {Thorp}, Stephen and {Jagwani}, Gurjeet and {Rosswog}, Stephan and {Nissanke}, Samaya},
        title = "{Rapid and robust simulation-based inference for kilonovae}",
      journal = {arXiv e-prints},
         year = 2026,
        month = may,
          eid = {arXiv:2605.13983},
          doi = {10.48550/arXiv.2605.13983},
archivePrefix = {arXiv},
       eprint = {2605.13983},
 primaryClass = {astro-ph.IM},
}

@ARTICLE{Buchner:2021,
       author = {{Buchner}, Johannes},
        title = "{UltraNest - a robust, general purpose Bayesian inference engine}",
      journal = {The Journal of Open Source Software},
         year = 2021,
        month = apr,
       volume = {6},
       number = {60},
          eid = {3001},
        pages = {3001},
          doi = {10.21105/joss.03001},
archivePrefix = {arXiv},
       eprint = {2101.09604},
 primaryClass = {stat.CO},
}

@ARTICLE{Buchner:2014,
       author = {{Buchner}, J. and {Georgakakis}, A. and {Nandra}, K. and {Hsu}, L. and {Rangel}, C. and {Brightman}, M. and {Merloni}, A. and {Salvato}, M. and {Donley}, J. and {Kocevski}, D.},
        title = "{X-ray spectral modelling of the AGN obscuring region in the CDFS: Bayesian model selection and catalogue}",
      journal = {\aap},
         year = 2014,
        month = apr,
       volume = {564},
          eid = {A125},
        pages = {A125},
          doi = {10.1051/0004-6361/201322971},
archivePrefix = {arXiv},
       eprint = {1402.0004},
 primaryClass = {astro-ph.HE}
}

@ARTICLE{Bulla:2023,
       author = {{Bulla}, Mattia},
        title = "{The critical role of nuclear heating rates, thermalization efficiencies, and opacities for kilonova modelling and parameter inference}",
      journal = {\mnras},
         year = 2023,
        month = apr,
       volume = {520},
       number = {2},
        pages = {2558-2570},
          doi = {10.1093/mnras/stad232},
archivePrefix = {arXiv},
       eprint = {2211.14348},
 primaryClass = {astro-ph.HE}
}

@ARTICLE{Bulla:2019,
       author = {{Bulla}, M.},
        title = "{POSSIS: predicting spectra, light curves, and polarization for multidimensional models of supernovae and kilonovae}",
      journal = {\mnras},
         year = 2019,
        month = nov,
       volume = {489},
       number = {4},
        pages = {5037-5045},
          doi = {10.1093/mnras/stz2495},
archivePrefix = {arXiv},
       eprint = {1906.04205},
 primaryClass = {astro-ph.HE}
}

@ARTICLE{Burgay:2003,
       author = {{Burgay}, M. and {D'Amico}, N. and {Possenti}, A. and {Manchester}, R.~N. and {Lyne}, A.~G. and {Joshi}, B.~C. and {McLaughlin}, M.~A. and {Kramer}, M. and {Sarkissian}, J.~M. and {Camilo}, F. and {Kalogera}, V. and {Kim}, C. and {Lorimer}, D.~R.},
        title = "{An increased estimate of the merger rate of double neutron stars from observations of a highly relativistic system}",
      journal = {\nat},
         year = 2003,
        month = dec,
       volume = {426},
       number = {6966},
        pages = {531-533},
          doi = {10.1038/nature02124},
archivePrefix = {arXiv},
       eprint = {astro-ph/0312071},
 primaryClass = {astro-ph}
}

@ARTICLE{Cannon:2021,
       author = {{Cannon}, Kipp and {Caudill}, Sarah and {Chan}, Chiwai and {Cousins}, Bryce and {Creighton}, Jolien D.~E. and {Ewing}, Becca and {Fong}, Heather and {Godwin}, Patrick and {Hanna}, Chad and {Hooper}, Shaun and {Huxford}, Rachael and {Magee}, Ryan and {Meacher}, Duncan and {Messick}, Cody and {Morisaki}, Soichiro and {Mukherjee}, Debnandini and {Ohta}, Hiroaki and {Pace}, Alexander and {Privitera}, Stephen and {de Ruiter}, Iris and {Sachdev}, Surabhi and {Singer}, Leo and {Singh}, Divya and {Tapia}, Ron and {Tsukada}, Leo and {Tsuna}, Daichi and {Tsutsui}, Takuya and {Ueno}, Koh and {Viets}, Aaron and {Wade}, Leslie and {Wade}, Madeline},
        title = "{GstLAL: A software framework for gravitational wave discovery}",
      journal = {SoftwareX},
         year = 2021,
        month = jun,
       volume = {14},
          eid = {100680},
          doi = {10.1016/j.softx.2021.100680},
archivePrefix = {arXiv},
       eprint = {2010.05082},
 primaryClass = {astro-ph.IM},
}

@INCOLLECTION{Chakrabarty:2008,
       author = {{Chakrabarty}, D.},
        title = "{The spin distribution of millisecond X-ray pulsars}",
    booktitle = {A Decade of Accreting MilliSecond X-ray Pulsars},
         year = 2008,
       editor = {{Wijnands}, Rudy and {Altamirano}, Diego and {Soleri}, Paolo and {Degenaar}, Nathalie and {Rea}, Nanda and {Casella}, Piergiorgio and {Patruno}, Alessandro and {Linares}, Manuel},
       series = {American Institute of Physics Conference Series},
       volume = {1068},
        month = oct,
    publisher = {AIP},
        pages = {67-74},
          doi = {10.1063/1.3031208},
archivePrefix = {arXiv},
       eprint = {0809.4031},
 primaryClass = {astro-ph}
}

@ARTICLE{Chatziioannou:2020,
       author = {{Chatziioannou}, Katerina},
        title = "{Neutron-star tidal deformability and equation-of-state constraints}",
      journal = {General Relativity and Gravitation},
         year = 2020,
        month = nov,
       volume = {52},
       number = {11},
          eid = {109},
        pages = {109},
          doi = {10.1007/s10714-020-02754-3},
archivePrefix = {arXiv},
       eprint = {2006.03168},
 primaryClass = {gr-qc}
}

@ARTICLE{Choudhury:2024,
       author = {{Choudhury}, Devarshi and {Salmi}, Tuomo and {Vinciguerra}, Serena and {Riley}, Thomas E. and {Kini}, Yves and {Watts}, Anna L. and {Dorsman}, Bas and {Bogdanov}, Slavko and {Guillot}, Sebastien and {Ray}, Paul S. and {Reardon}, Daniel J. and {Remillard}, Ronald A. and {Bilous}, Anna V. and {Huppenkothen}, Daniela and {Lattimer}, James M. and {Rutherford}, Nathan and {Arzoumanian}, Zaven and {Gendreau}, Keith C. and {Morsink}, Sharon M. and {Ho}, Wynn C.~G.},
        title = "{A NICER View of the Nearest and Brightest Millisecond Pulsar: PSR J0437{\textendash}4715}",
      journal = {\apjl},
         year = 2024,
        month = aug,
       volume = {971},
       number = {1},
          eid = {L20},
        pages = {L20},
          doi = {10.3847/2041-8213/ad5a6f},
archivePrefix = {arXiv},
       eprint = {2407.06789},
 primaryClass = {astro-ph.HE}
}

@ARTICLE{Cipolletta:2015,
       author = {{Cipolletta}, F. and {Cherubini}, C. and {Filippi}, S. and {Rueda}, J.~A. and {Ruffini}, R.},
        title = "{Fast rotating neutron stars with realistic nuclear matter equation of state}",
      journal = {\prd},
         year = 2015,
        month = jul,
       volume = {92},
       number = {2},
          eid = {023007},
        pages = {023007},
          doi = {10.1103/PhysRevD.92.023007},
archivePrefix = {arXiv},
       eprint = {1506.05926},
 primaryClass = {astro-ph.SR}
}

@ARTICLE{Combi:2023,
       author = {{Combi}, Luciano and {Siegel}, Daniel M.},
        title = "{Jets from Neutron-Star Merger Remnants and Massive Blue Kilonovae}",
      journal = {\prl},
         year = 2023,
        month = dec,
       volume = {131},
       number = {23},
          eid = {231402},
        pages = {231402},
          doi = {10.1103/PhysRevLett.131.231402},
archivePrefix = {arXiv},
       eprint = {2303.12284},
 primaryClass = {astro-ph.HE},
       adsurl = {https://ui.adsabs.harvard.edu/abs/2023PhRvL.131w1402C}
}

@ARTICLE{Cowan:2021,
       author = {{Cowan}, John J. and {Sneden}, Christopher and {Lawler}, James E. and {Aprahamian}, Ani and {Wiescher}, Michael and {Langanke}, Karlheinz and {Mart{\'\i}nez-Pinedo}, Gabriel and {Thielemann}, Friedrich-Karl},
        title = "{Origin of the heaviest elements: The rapid neutron-capture process}",
      journal = {Reviews of Modern Physics},
         year = 2021,
        month = jan,
       volume = {93},
       number = {1},
          eid = {015002},
        pages = {015002},
          doi = {10.1103/RevModPhys.93.015002},
archivePrefix = {arXiv},
       eprint = {1901.01410},
 primaryClass = {astro-ph.HE}
}

@ARTICLE{Coughlin:2020,
       author = {{Coughlin}, Michael W. and {Dietrich}, Tim and {Antier}, Sarah and {Bulla}, Mattia and {Foucart}, Francois and {Hotokezaka}, Kenta and {Raaijmakers}, Geert and {Hinderer}, Tanja and {Nissanke}, Samaya},
        title = "{Implications of the search for optical counterparts during the first six months of the Advanced LIGO's and Advanced Virgo's third observing run: possible limits on the ejecta mass and binary properties}",
      journal = {\mnras},
         year = 2020,
        month = feb,
       volume = {492},
       number = {1},
        pages = {863-876},
          doi = {10.1093/mnras/stz3457},
archivePrefix = {arXiv},
       eprint = {1910.11246},
 primaryClass = {astro-ph.HE}
}

@ARTICLE{Coughlin:2019,
       author = {{Coughlin}, Michael W. and {Dietrich}, Tim and {Margalit}, Ben and {Metzger}, Brian D.},
        title = "{Multimessenger Bayesian parameter inference of a binary neutron star merger}",
      journal = {\mnras},
         year = 2019,
        month = oct,
       volume = {489},
       number = {1},
        pages = {L91-L96},
          doi = {10.1093/mnrasl/slz133},
archivePrefix = {arXiv},
       eprint = {1812.04803},
 primaryClass = {astro-ph.HE}
}

@ARTICLE{Coughlin:2018,
       author = {{Coughlin}, Michael W. and {Dietrich}, Tim and {Doctor}, Zoheyr and {Kasen}, Daniel and {Coughlin}, Scott and {Jerkstrand}, Anders and {Leloudas}, Giorgos and {McBrien}, Owen and {Metzger}, Brian D. and {O'Shaughnessy}, Richard and {Smartt}, Stephen J.},
        title = "{Constraints on the neutron star equation of state from AT2017gfo using radiative transfer simulations}",
      journal = {\mnras},
         year = 2018,
        month = aug,
       volume = {480},
       number = {3},
        pages = {3871-3878},
          doi = {10.1093/mnras/sty2174},
archivePrefix = {arXiv},
       eprint = {1805.09371},
 primaryClass = {astro-ph.HE}
}

@ARTICLE{Cutler:1994,
       author = {{Cutler}, Curt and {Flanagan}, {\'E}anna E.},
        title = "{Gravitational waves from merging compact binaries: How accurately can one extract the binary's parameters from the inspiral waveform\textbackslash?}",
      journal = {\prd},
         year = 1994,
        month = mar,
       volume = {49},
       number = {6},
        pages = {2658-2697},
          doi = {10.1103/PhysRevD.49.2658},
archivePrefix = {arXiv},
       eprint = {gr-qc/9402014},
 primaryClass = {gr-qc}
}

@ARTICLE{Dal_Canton:2021,
       author = {{Dal Canton}, Tito and {Nitz}, Alexander H. and {Gadre}, Bhooshan and {Cabourn Davies}, Gareth S. and {Villa-Ortega}, Ver{\'o}nica and {Dent}, Thomas and {Harry}, Ian and {Xiao}, Liting},
        title = "{Real-time Search for Compact Binary Mergers in Advanced LIGO and Virgo's Third Observing Run Using PyCBC Live}",
      journal = {\apj},
         year = 2021,
        month = dec,
       volume = {923},
       number = {2},
          eid = {254},
        pages = {254},
          doi = {10.3847/1538-4357/ac2f9a},
archivePrefix = {arXiv},
       eprint = {2008.07494},
 primaryClass = {astro-ph.HE},
}

@ARTICLE{Dax:2025,
       author = {{Dax}, Maximilian and {Green}, Stephen R. and {Gair}, Jonathan and {Gupte}, Nihar and {P{\"u}rrer}, Michael and {Raymond}, Vivien and {Wildberger}, Jonas and {Macke}, Jakob H. and {Buonanno}, Alessandra and {Sch{\"o}lkopf}, Bernhard},
        title = "{Real-time inference for binary neutron star mergers using machine learning}",
      journal = {\nat},
         year = 2025,
        month = mar,
       volume = {639},
       number = {8053},
        pages = {49-53},
          doi = {10.1038/s41586-025-08593-z},
archivePrefix = {arXiv},
       eprint = {2407.09602},
 primaryClass = {gr-qc},
}

@ARTICLE{DES:2024,
       author = {{DES Collaboration} and {Abbott}, T.~M.~C. and {Acevedo}, M. and {Aguena}, M. and {Alarcon}, A. and {Allam}, S. and {Alves}, O. and {Amon}, A. and {Andrade-Oliveira}, F. and {Annis}, J. and {Armstrong}, P. and {Asorey}, J. and {Avila}, S. and {Bacon}, D. and {Bassett}, B.~A. and {Bechtol}, K. and {Bernardinelli}, P.~H. and {Bernstein}, G.~M. and {Bertin}, E. and {Blazek}, J. and {Bocquet}, S. and {Brooks}, D. and {Brout}, D. and {Buckley-Geer}, E. and {Burke}, D.~L. and {Camacho}, H. and {Camilleri}, R. and {Campos}, A. and {Carnero Rosell}, A.},
        title = "{The Dark Energy Survey: Cosmology Results with {\ensuremath{\sim}}1500 New High-redshift Type Ia Supernovae Using the Full 5 yr Data Set}",
      journal = {\apjl},
         year = 2024,
        month = sep,
       volume = {973},
       number = {1},
          eid = {L14},
        pages = {L14},
          doi = {10.3847/2041-8213/ad6f9f},
archivePrefix = {arXiv},
       eprint = {2401.02929},
 primaryClass = {astro-ph.CO}
}

@ARTICLE{Desai:2025,
       author = {{Desai}, M.~M. and {Chatterjee}, D. and {Jhawar}, S. and {Harris}, P. and {Katsavounidis}, E. and {Coughlin}, M.~W.},
        title = "{Rapid parameter estimation for kilonovae using likelihood-free inference}",
      journal = {\mnras},
         year = 2025,
        month = aug,
       volume = {541},
       number = {3},
        pages = {2619-2630},
          doi = {10.1093/mnras/staf1045},
archivePrefix = {arXiv},
       eprint = {2408.06947},
 primaryClass = {astro-ph.IM},
}

@ARTICLE{Dietrich:2020,
       author = {{Dietrich}, Tim and {Coughlin}, Michael W. and {Pang}, Peter T.~H. and {Bulla}, Mattia and {Heinzel}, Jack and {Issa}, Lina and {Tews}, Ingo and {Antier}, Sarah},
        title = "{Multimessenger constraints on the neutron-star equation of state and the Hubble constant}",
      journal = {Science},
         year = 2020,
        month = dec,
       volume = {370},
       number = {6523},
        pages = {1450-1453},
          doi = {10.1126/science.abb4317},
archivePrefix = {arXiv},
       eprint = {2002.11355},
 primaryClass = {astro-ph.HE}
}

@ARTICLE{Dietrich:2019,
       author = {{Dietrich}, Tim and {Samajdar}, Anuradha and {Khan}, Sebastian and {Johnson-McDaniel}, Nathan K. and {Dudi}, Reetika and {Tichy}, Wolfgang},
        title = "{Improving the NRTidal model for binary neutron star systems}",
      journal = {\prd},
         year = 2019,
        month = aug,
       volume = {100},
       number = {4},
          eid = {044003},
        pages = {044003},
          doi = {10.1103/PhysRevD.100.044003},
archivePrefix = {arXiv},
       eprint = {1905.06011},
 primaryClass = {gr-qc}
}

@ARTICLE{Dietrich:2017b,
       author = {{Dietrich}, Tim and {Ujevic}, Maximiliano and {Tichy}, Wolfgang and {Bernuzzi}, Sebastiano and {Br{\"u}gmann}, Bernd},
        title = "{Gravitational waves and mass ejecta from binary neutron star mergers: Effect of the mass ratio}",
      journal = {\prd},
         year = 2017,
        month = jan,
       volume = {95},
       number = {2},
          eid = {024029},
        pages = {024029},
          doi = {10.1103/PhysRevD.95.024029},
archivePrefix = {arXiv},
       eprint = {1607.06636},
 primaryClass = {gr-qc}
}

@ARTICLE{Dittmann:2024,
       author = {{Dittmann}, Alexander J. and {Miller}, M. Coleman and {Lamb}, Frederick K. and {Holt}, Isiah M. and {Chirenti}, Cecilia and {Wolff}, Michael T. and {Bogdanov}, Slavko and {Guillot}, Sebastien and {Ho}, Wynn C.~G. and {Morsink}, Sharon M. and {Arzoumanian}, Zaven and {Gendreau}, Keith C.},
        title = "{A More Precise Measurement of the Radius of PSR J0740+6620 Using Updated NICER Data}",
      journal = {\apj},
         year = 2024,
        month = oct,
       volume = {974},
       number = {2},
          eid = {295},
        pages = {295},
          doi = {10.3847/1538-4357/ad5f1e},
archivePrefix = {arXiv},
       eprint = {2406.14467},
 primaryClass = {astro-ph.HE}
}

@ARTICLE{Domoto:2025,
       author = {{Domoto}, Nanae and {Wanajo}, Shinya and {Tanaka}, Masaomi and {Kato}, Daiji and {Hotokezaka}, Kenta},
        title = "{Thorium in Kilonova Spectra: Exploring the Heaviest Detectable Element}",
      journal = {\apj},
         year = 2025,
        month = jan,
       volume = {978},
       number = {1},
          eid = {99},
        pages = {99},
          doi = {10.3847/1538-4357/ad96b3},
archivePrefix = {arXiv},
       eprint = {2411.16998},
 primaryClass = {astro-ph.HE}
}

@ARTICLE{Domoto:2022,
       author = {{Domoto}, Nanae and {Tanaka}, Masaomi and {Kato}, Daiji and {Kawaguchi}, Kyohei and {Hotokezaka}, Kenta and {Wanajo}, Shinya},
        title = "{Lanthanide Features in Near-infrared Spectra of Kilonovae}",
      journal = {\apj},
         year = 2022,
        month = nov,
       volume = {939},
       number = {1},
          eid = {8},
        pages = {8},
          doi = {10.3847/1538-4357/ac8c36},
archivePrefix = {arXiv},
       eprint = {2206.04232},
 primaryClass = {astro-ph.HE}
}

@ARTICLE{Drischler:2021,
       author = {{Drischler}, C. and {Holt}, J.~W. and {Wellenhofer}, C.},
        title = "{Chiral Effective Field Theory and the High-Density Nuclear Equation of State}",
      journal = {Annual Review of Nuclear and Particle Science},
         year = 2021,
        month = sep,
       volume = {71},
        pages = {403-432},
          doi = {10.1146/annurev-nucl-102419-041903},
archivePrefix = {arXiv},
       eprint = {2101.01709},
 primaryClass = {nucl-th}
}

@ARTICLE{Ducoin:2026,
       author = {{Ducoin}, J.-G. and {Pellouin}, C. and {Aivazyan}, V. and {Akl}, D. and {Alvarez}, F. and {Andrade}, C. and {Angulo}, C. and {Antier}, S. and {Atteia}, J.-L. and {Basa}, S. and {Becerra}, R.~L. and {Benkhaldoun}, Z. and {Bissaldi}, E. and {Breeveld}, A. and {Bruin}, E. de. and {Burns}, E. and {Butler}, N.~R. and {Coughlin}, M.~W. and {Daigne}, F. and {Dietrich}, T. and {Dornic}, D. and {Douzet}, C. and {du Laz}, T. and {Duverne}, P.-A. and {Eggenstein}, H.~B. and {Elhosseiny}, E. and {Esamdin}, A. and {Evans}, P.~A. and {Ag{\"u}{\'\i} Fern{\'a}ndez}, J.~F. and {Ferro}, M. and {Fortin}, F. and {Freeberg}, M. and {Garc{\'\i}a-Garc{\'\i}a}, L. and {Gill}, R. and {Globus}, N. and {Guessoum}, N. and {Hamed}, G.~M. and {Hello}, P. and {Holzmann Airasca}, A. and {Hu}, D.~F. and {Hussenot-Desenonges}, T. and {Inasaridze}, R. and {Iskandar}, A. and {Jiang}, S.~Q. and {Jin}, C.~C. and {Kaeouach}, A. and {Karpov}, S. and {Klingler}, N.~J. and {Klotz}, A. and {Kochiashvili}, N. and {Koehn}, H. and {Kneip}, R. and {Kvernadze}, T. and {Le Calloch}, A. and {Lee}, W.~H. and {Lekic}, A. and {Liang}, Y.~F. and {Limonta}, C. and {Liu}, J. and {L{\'o}pez}, K. Ocelotl. C. and {L{\'o}pez-C{\'a}mara}, D. and {Mabrouk}, R.~H. and {Magnani}, F. and {Mao}, J. and {Ma{\v{s}}ek}, M. and {Moreno M{\'e}ndez}, E. and {Mihov}, B.~M. and {Molham}, M. and {Noysena}, K. and {Odeh}, M. and {Omodei}, N. and {Peng}, H. and {Pereyra}, M. and {Pillas}, M. and {Pillera}, R. and {Pradier}, T. and {Rajabov}, Y. and {Rakotondrainibe}, N.~A. and {Schneider}, B. and {Serrau}, M. and {Slavcheva-Mihova}, L. and {Sokoliuk}, O. and {Sun}, H. and {Takey}, A. and {Tanasan}, M. and {Tinyanont}, K.~S. and {Turpin}, D. and {de Ugarte Postigo}, A. and {Wang}, B.~T. and {Wang}, L.~T. and {Wang}, X.~F. and {Wang}, Z.~M. and {Watson}, A.~M. and {Wu}, H.~Z. and {Wu}, Q.~Y. and {Xu}, J.~J. and {Yan}, Y.~S. and {Yang}, H.~N. and {Yuan}, W. and {Zhao}, H.~S.},
        title = "{GRB 241030A: a bright afterglow challenging forward shock emission}",
      journal = {arXiv e-prints},
         year = 2026,
        month = mar,
          eid = {arXiv:2603.18956},
          doi = {10.48550/arXiv.2603.18956},
archivePrefix = {arXiv},
       eprint = {2603.18956},
 primaryClass = {astro-ph.HE}
}

@ARTICLE{Edwards:2024,
       author = {{Edwards}, Thomas D.~P. and {Wong}, Kaze W.~K. and {Lam}, Kelvin K.~H. and {Coogan}, Adam and {Foreman-Mackey}, Daniel and {Isi}, Maximiliano and {Zimmerman}, Aaron},
        title = "{Differentiable and hardware-accelerated waveforms for gravitational wave data analysis}",
      journal = {\prd},
         year = 2024,
        month = sep,
       volume = {110},
       number = {6},
          eid = {064028},
        pages = {064028},
          doi = {10.1103/PhysRevD.110.064028},
archivePrefix = {arXiv},
       eprint = {2302.05329},
 primaryClass = {astro-ph.IM}
}

@ARTICLE{Evans:2021,
       author = {{Evans}, Matthew and {Adhikari}, Rana X and {Afle}, Chaitanya and {Ballmer}, Stefan W. and {Biscoveanu}, Sylvia and {Borhanian}, Ssohrab and {Brown}, Duncan A. and {Chen}, Yanbei and {Eisenstein}, Robert and {Gruson}, Alexandra and {Gupta}, Anuradha and {Hall}, Evan D. and {Huxford}, Rachael and {Kamai}, Brittany and {Kashyap}, Rahul and {Kissel}, Jeff S. and {Kuns}, Kevin and {Landry}, Philippe and {Lenon}, Amber and {Lovelace}, Geoffrey and {McCuller}, Lee and {Ng}, Ken K.~Y. and {Nitz}, Alexander H. and {Read}, Jocelyn and {Sathyaprakash}, B.~S. and {Shoemaker}, David H. and {Slagmolen}, Bram J.~J. and {Smith}, Joshua R. and {Srivastava}, Varun and {Sun}, Ling and {Vitale}, Salvatore and {Weiss}, Rainer},
        title = "{A Horizon Study for Cosmic Explorer: Science, Observatories, and Community}",
      journal = {arXiv e-prints},
         year = 2021,
        month = sep,
          eid = {arXiv:2109.09882},
          doi = {10.48550/arXiv.2109.09882},
archivePrefix = {arXiv},
       eprint = {2109.09882},
 primaryClass = {astro-ph.IM},
}

@ARTICLE{Ewing:2024,
       author = {{Ewing}, Becca and {Huxford}, Rachael and {Singh}, Divya and {Tsukada}, Leo and {Hanna}, Chad and {Huang}, Yun-Jing and {Joshi}, Prathamesh and {Li}, Alvin K.~Y. and {Magee}, Ryan and {Messick}, Cody and {Pace}, Alex and {Ray}, Anarya and {Sachdev}, Surabhi and {Sakon}, Shio and {Tapia}, Ron and {Adhicary}, Shomik and {Baral}, Pratyusava and {Baylor}, Amanda and {Cannon}, Kipp and {Caudill}, Sarah and {Chaudhary}, Sushant Sharma and {Coughlin}, Michael W. and {Cousins}, Bryce and {Creighton}, Jolien D.~E. and {Essick}, Reed and {Fong}, Heather and {George}, Richard N. and {Godwin}, Patrick and {Harada}, Reiko and {Kennington}, James and {Kuwahara}, Soichiro and {Meacher}, Duncan and {Morisaki}, Soichiro and {Mukherjee}, Debnandini and {Niu}, Wanting and {Posnansky}, Cort and {Toivonen}, Andrew and {Tsutsui}, Takuya and {Ueno}, Koh and {Viets}, Aaron and {Wade}, Leslie and {Wade}, Madeline and {Waratkar}, Gaurav},
        title = "{Performance of the low-latency GstLAL inspiral search towards LIGO, Virgo, and KAGRA's fourth observing run}",
      journal = {\prd},
         year = 2024,
        month = feb,
       volume = {109},
       number = {4},
          eid = {042008},
        pages = {042008},
          doi = {10.1103/PhysRevD.109.042008},
archivePrefix = {arXiv},
       eprint = {2305.05625},
 primaryClass = {gr-qc}
}

@ARTICLE{Farrow:2019,
       author = {{Farrow}, Nicholas and {Zhu}, Xing-Jiang and {Thrane}, Eric},
        title = "{The Mass Distribution of Galactic Double Neutron Stars}",
      journal = {\apj},
         year = 2019,
        month = may,
       volume = {876},
       number = {1},
          eid = {18},
        pages = {18},
          doi = {10.3847/1538-4357/ab12e3},
archivePrefix = {arXiv},
       eprint = {1902.03300},
 primaryClass = {astro-ph.HE}
}

@ARTICLE{Ferreira:2021,
       author = {{Ferreira}, M{\'a}rcio and {Provid{\^e}ncia}, Constan{\c{c}}a},
        title = "{Unveiling the nuclear matter EoS from neutron star properties: a supervised machine learning approach}",
      journal = {\jcap},
         year = 2021,
        month = jul,
       volume = {2021},
       number = {7},
          eid = {011},
        pages = {011},
          doi = {10.1088/1475-7516/2021/07/011},
archivePrefix = {arXiv},
       eprint = {1910.05554},
 primaryClass = {nucl-th},
}

@ARTICLE{Feroz:2019,
       author = {{Feroz}, Farhan and {Hobson}, Michael P. and {Cameron}, Ewan and {Pettitt}, Anthony N.},
        title = "{Importance Nested Sampling and the MultiNest Algorithm}",
      journal = {The Open Journal of Astrophysics},
         year = 2019,
        month = nov,
       volume = {2},
       number = {1},
          eid = {10},
        pages = {10},
          doi = {10.21105/astro.1306.2144},
archivePrefix = {arXiv},
       eprint = {1306.2144},
 primaryClass = {astro-ph.IM}
}

@ARTICLE{Fishbach:2020,
       author = {{Fishbach}, Maya and {Holz}, Daniel E.},
        title = "{Picky Partners: The Pairing of Component Masses in Binary Black Hole Mergers}",
      journal = {\apjl},
         year = 2020,
        month = mar,
       volume = {891},
       number = {1},
          eid = {L27},
        pages = {L27},
          doi = {10.3847/2041-8213/ab7247},
archivePrefix = {arXiv},
       eprint = {1905.12669},
 primaryClass = {astro-ph.HE}
}

@ARTICLE{Fonseca:2021,
       author = {{Fonseca}, E. and {Cromartie}, H.~T. and {Pennucci}, T.~T. and {Ray}, P.~S. and {Kirichenko}, A. Yu. and {Ransom}, S.~M. and {Demorest}, P.~B. and {Stairs}, I.~H. and {Arzoumanian}, Z. and {Guillemot}, L. and {Parthasarathy}, A. and {Kerr}, M. and {Cognard}, I. and {Baker}, P.~T. and {Blumer}, H. and {Brook}, P.~R. and {DeCesar}, M. and {Dolch}, T. and {Dong}, F.~A. and {Ferrara}, E.~C. and {Fiore}, W. and {Garver-Daniels}, N. and {Good}, D.~C. and {Jennings}, R. and {Jones}, M.~L. and {Kaspi}, V.~M. and {Lam}, M.~T. and {Lorimer}, D.~R. and {Luo}, J. and {McEwen}, A. and {McKee}, J.~W. and {McLaughlin}, M.~A. and {McMann}, N. and {Meyers}, B.~W. and {Naidu}, A. and {Ng}, C. and {Nice}, D.~J. and {Pol}, N. and {Radovan}, H.~A. and {Shapiro-Albert}, B. and {Tan}, C.~M. and {Tendulkar}, S.~P. and {Swiggum}, J.~K. and {Wahl}, H.~M. and {Zhu}, W.~W.},
        title = "{Refined Mass and Geometric Measurements of the High-mass PSR J0740+6620}",
      journal = {\apjl},
         year = 2021,
        month = jul,
       volume = {915},
       number = {1},
          eid = {L12},
        pages = {L12},
          doi = {10.3847/2041-8213/ac03b8},
archivePrefix = {arXiv},
       eprint = {2104.00880},
 primaryClass = {astro-ph.HE}
}

@ARTICLE{Foucart:2018,
       author = {{Foucart}, Francois and {Hinderer}, Tanja and {Nissanke}, Samaya},
        title = "{Remnant baryon mass in neutron star-black hole mergers: Predictions for binary neutron star mimickers and rapidly spinning black holes}",
      journal = {\prd},
         year = 2018,
        month = oct,
       volume = {98},
       number = {8},
          eid = {081501},
        pages = {081501},
          doi = {10.1103/PhysRevD.98.081501},
archivePrefix = {arXiv},
       eprint = {1807.00011},
 primaryClass = {astro-ph.HE}
}

@ARTICLE{Gillanders:2022,
       author = {{Gillanders}, J.~H. and {Smartt}, S.~J. and {Sim}, S.~A. and {Bauswein}, A. and {Goriely}, S.},
        title = "{Modelling the spectra of the kilonova AT2017gfo - I. The photospheric epochs}",
      journal = {\mnras},
         year = 2022,
        month = sep,
       volume = {515},
       number = {1},
        pages = {631-651},
          doi = {10.1093/mnras/stac1258},
archivePrefix = {arXiv},
       eprint = {2202.01786},
 primaryClass = {astro-ph.HE}
}

@ARTICLE{Gusakov:2014,
       author = {{Gusakov}, Mikhail E. and {Chugunov}, Andrey I. and {Kantor}, Elena M.},
        title = "{Explaining observations of rapidly rotating neutron stars in low-mass x-ray binaries}",
      journal = {\prd},
         year = 2014,
        month = sep,
       volume = {90},
       number = {6},
          eid = {063001},
        pages = {063001},
          doi = {10.1103/PhysRevD.90.063001},
archivePrefix = {arXiv},
       eprint = {1305.3825},
 primaryClass = {astro-ph.SR}
}

@ARTICLE{Guy:2007,
       author = {{Guy}, J. and {Astier}, P. and {Baumont}, S. and {Hardin}, D. and {Pain}, R. and {Regnault}, N. and {Basa}, S. and {Carlberg}, R.~G. and {Conley}, A. and {Fabbro}, S. and {Fouchez}, D. and {Hook}, I.~M. and {Howell}, D.~A. and {Perrett}, K. and {Pritchet}, C.~J. and {Rich}, J. and {Sullivan}, M. and {Antilogus}, P. and {Aubourg}, E. and {Bazin}, G. and {Bronder}, J. and {Filiol}, M. and {Palanque-Delabrouille}, N. and {Ripoche}, P. and {Ruhlmann-Kleider}, V.},
        title = "{SALT2: using distant supernovae to improve the use of type Ia supernovae as distance indicators}",
      journal = {\aap},
         year = 2007,
        month = apr,
       volume = {466},
       number = {1},
        pages = {11-21},
          doi = {10.1051/0004-6361:20066930},
archivePrefix = {arXiv},
       eprint = {astro-ph/0701828},
 primaryClass = {astro-ph}
}

@ARTICLE{Hall:2025,
       author = {{Hall}, Xander J. and {Busmann}, Malte and {Koehn}, Hauke and {Kunnumkai}, Keerthi and {Palmese}, Antonella and {O'Connor}, Brendan and {Freeburn}, James and {Hu}, Lei and {Gruen}, Daniel and {Dietrich}, Tim and {Bulla}, Mattia and {Coughlin}, Michael W. and {Antier}, Sarah and {Pillas}, Marion and {Price}, Paul A. and {Ahumada}, Tom{\'a}s and {Amsellem}, Ariel and {Andreoni}, Igor and {Augustin}, Jule and {Cabrera}, Tom'as and {Deshpande}, Rasika and {Fab{\`a}-Moreno}, Jennifer and {Gassert}, Julius and {Karpov}, Sergey and {Kasliwal}, Mansi and {Maga{\~n}a Hernandez}, Ignacio and {Mandelbaum}, Rachel and {Fontinele Nunes}, Felipe and {Pang}, Peter T.~H. and {Sommer}, Julian and {Stein}, Robert and {Tabor}, Constantin and {Vega}, Pablo and {Wouters}, Thibeau and {Zuo}, Xiaoxiong},
        title = "{AT2025ulz and S250818k: Investigating early time observations of a subsolar mass gravitational-wave binary neutron star merger candidate}",
      journal = {arXiv e-prints},
         year = 2025,
        month = oct,
          eid = {arXiv:2510.24620},
          doi = {10.48550/arXiv.2510.24620},
archivePrefix = {arXiv},
       eprint = {2510.24620},
 primaryClass = {astro-ph.HE}
}

@ARTICLE{Handley:2015,
       author = {{Handley}, W.~J. and {Hobson}, M.~P. and {Lasenby}, A.~N.},
        title = "{POLYCHORD: next-generation nested sampling}",
      journal = {\mnras},
         year = 2015,
        month = nov,
       volume = {453},
       number = {4},
        pages = {4384-4398},
          doi = {10.1093/mnras/stv1911},
archivePrefix = {arXiv},
       eprint = {1506.00171},
 primaryClass = {astro-ph.IM}
}

@ARTICLE{Hayashi:2025,
       author = {{Hayashi}, Kota and {Kiuchi}, Kenta and {Kyutoku}, Koutarou and {Sekiguchi}, Yuichiro and {Shibata}, Masaru},
        title = "{Jet from Binary Neutron Star Merger with Prompt Black Hole Formation}",
      journal = {\prl},
         year = 2025,
        month = may,
       volume = {134},
       number = {21},
          eid = {211407},
        pages = {211407},
          doi = {10.1103/PhysRevLett.134.211407},
archivePrefix = {arXiv},
       eprint = {2410.10958},
 primaryClass = {astro-ph.HE},
       adsurl = {https://ui.adsabs.harvard.edu/abs/2025PhRvL.134u1407H}
}

@ARTICLE{Hessels:2006,
       author = {{Hessels}, Jason W.~T. and {Ransom}, Scott M. and {Stairs}, Ingrid H. and {Freire}, Paulo C.~C. and {Kaspi}, Victoria M. and {Camilo}, Fernando},
        title = "{A Radio Pulsar Spinning at 716 Hz}",
      journal = {Science},
         year = 2006,
        month = mar,
       volume = {311},
       number = {5769},
        pages = {1901-1904},
          doi = {10.1126/science.1123430},
archivePrefix = {arXiv},
       eprint = {astro-ph/0601337},
 primaryClass = {astro-ph}
}

@ARTICLE{Hinderer:2010,
       author = {{Hinderer}, Tanja and {Lackey}, Benjamin D. and {Lang}, Ryan N. and {Read}, Jocelyn S.},
        title = "{Tidal deformability of neutron stars with realistic equations of state and their gravitational wave signatures in binary inspiral}",
      journal = {\prd},
         year = 2010,
        month = jun,
       volume = {81},
       number = {12},
          eid = {123016},
        pages = {123016},
          doi = {10.1103/PhysRevD.81.123016},
archivePrefix = {arXiv},
       eprint = {0911.3535},
 primaryClass = {astro-ph.HE}
}

@ARTICLE{Hotokezaka:2023,
       author = {{Hotokezaka}, Kenta and {Tanaka}, Masaomi and {Kato}, Daiji and {Gaigalas}, Gediminas},
        title = "{Tellurium emission line in kilonova AT 2017gfo}",
      journal = {\mnras},
         year = 2023,
        month = nov,
       volume = {526},
       number = {1},
        pages = {L155-L159},
          doi = {10.1093/mnrasl/slad128},
archivePrefix = {arXiv},
       eprint = {2307.00988},
 primaryClass = {astro-ph.HE},
}

@ARTICLE{Hotokezaka:2022,
       author = {{Hotokezaka}, Kenta and {Tanaka}, Masaomi and {Kato}, Daiji and {Gaigalas}, Gediminas},
        title = "{Tungsten versus Selenium as a potential source of kilonova nebular emission observed by Spitzer}",
      journal = {\mnras},
         year = 2022,
        month = sep,
       volume = {515},
       number = {1},
        pages = {L89-L93},
          doi = {10.1093/mnrasl/slac071},
archivePrefix = {arXiv},
       eprint = {2204.00737},
 primaryClass = {astro-ph.HE}
}

@ARTICLE{Hotokezaka:2021,
       author = {{Hotokezaka}, Kenta and {Tanaka}, Masaomi and {Kato}, Daiji and {Gaigalas}, Gediminas},
        title = "{Nebular emission from lanthanide-rich ejecta of neutron star merger}",
      journal = {\mnras},
         year = 2021,
        month = oct,
       volume = {506},
       number = {4},
        pages = {5863-5877},
          doi = {10.1093/mnras/stab1975},
archivePrefix = {arXiv},
       eprint = {2102.07879},
 primaryClass = {astro-ph.HE}
}

@ARTICLE{Hotokezaka:2018,
       author = {{Hotokezaka}, Kenta and {Kiuchi}, Kenta and {Shibata}, Masaru and {Nakar}, Ehud and {Piran}, Tsvi},
        title = "{Synchrotron Radiation from the Fast Tail of Dynamical Ejecta of Neutron Star Mergers}",
      journal = {\apj},
         year = 2018,
        month = nov,
       volume = {867},
       number = {2},
          eid = {95},
        pages = {95},
          doi = {10.3847/1538-4357/aadf92},
archivePrefix = {arXiv},
       eprint = {1803.00599},
 primaryClass = {astro-ph.HE}
}

@ARTICLE{Hotokezaka:2011,
       author = {{Hotokezaka}, Kenta and {Kyutoku}, Koutarou and {Okawa}, Hirotada and {Shibata}, Masaru and {Kiuchi}, Kenta},
        title = "{Binary neutron star mergers: Dependence on the nuclear equation of state}",
      journal = {\prd},
         year = 2011,
        month = jun,
       volume = {83},
       number = {12},
          eid = {124008},
        pages = {124008},
          doi = {10.1103/PhysRevD.83.124008},
archivePrefix = {arXiv},
       eprint = {1105.4370},
 primaryClass = {astro-ph.HE}
}

@ARTICLE{Hu:2025,
       author = {{Hu}, Qian and {Veitch}, John},
        title = "{Costs of Bayesian parameter estimation in third-generation gravitational wave detectors: An assessment of current acceleration methods}",
      journal = {\prd},
         year = 2025,
        month = oct,
       volume = {112},
       number = {8},
          eid = {084039},
        pages = {084039},
          doi = {10.1103/dj7k-tk37},
archivePrefix = {arXiv},
       eprint = {2412.02651},
 primaryClass = {gr-qc},
}

@ARTICLE{Hu:2025b,
       author = {{Hu}, Qian and {Irwin}, Jessica and {Sun}, Qi and {Messenger}, Christopher and {Suleiman}, Lami and {Heng}, Ik Siong and {Veitch}, John},
        title = "{Decoding Long-duration Gravitational Waves from Binary Neutron Stars with Machine Learning: Parameter Estimation and Equations of State}",
      journal = {\apjl},
         year = 2025,
        month = jul,
       volume = {987},
       number = {1},
          eid = {L17},
        pages = {L17},
          doi = {10.3847/2041-8213/ade42f},
archivePrefix = {arXiv},
       eprint = {2412.03454},
 primaryClass = {gr-qc},
}

@ARTICLE{Hussenot:2025,
       author = {{Hussenot-Desenonges}, Thomas and {Pillas}, Marion and {Antier}, Sarah and {Hello}, Patrice and {Pang}, Peter T.~H.},
        title = "{Kilonova modelling and parameter inference: Understanding uncertainties and evaluating compatibility between observations and models}",
      journal = {arXiv e-prints},
         year = 2025,
        month = may,
          eid = {arXiv:2505.21392},
          doi = {10.48550/arXiv.2505.21392},
archivePrefix = {arXiv},
       eprint = {2505.21392},
 primaryClass = {astro-ph.HE}
}

@ARTICLE{Hussenot:2024,
       author = {{Hussenot-Desenonges}, T. and {Wouters}, T. and {Guessoum}, N. and {Abdi}, I. and {Abulwfa}, A. and {Adami}, C. and {Ag{\"u}{\'\i} Fern{\'a}ndez}, J.~F. and {Ahumada}, T. and {Aivazyan}, V. and {Akl}, D. and {Anand}, S. and {Andrade}, C.~M. and {Antier}, S. and {Ata}, S.~A. and {D'Avanzo}, P. and {Azzam}, Y.~A. and {Baransky}, A. and {Basa}, S. and {Blazek}, M. and {Bendjoya}, P. and {Beradze}, S. and {Boumis}, P. and {Bremer}, M. and {Brivio}, R. and {Buat}, V. and {Bulla}, M. and {Burkhonov}, O. and {Burns}, E. and {Cenko}, S.~B. and {Coughlin}, M.~W. and {Corradi}, W. and {Daigne}, F. and {Dietrich}, T. and {Dornic}, D. and {Ducoin}, J.-G. and {Duverne}, P.-A. and {Elhosseiny}, E.~G. and {Elnagahy}, F.~I. and {El-Sadek}, M.~A. and {Ferro}, M. and {Le Floc'h}, E. and {Freeberg}, M. and {Fynbo}, J.~P.~U. and {G{\"o}tz}, D. and {Gurbanov}, E. and {Zhu}, Z.},
        title = "{Multiband analyses of the bright GRB 230812B and the associated SN2023pel}",
      journal = {\mnras},
         year = 2024,
        month = may,
       volume = {530},
       number = {1},
        pages = {1-19},
          doi = {10.1093/mnras/stae503},
archivePrefix = {arXiv},
       eprint = {2310.14310},
 primaryClass = {astro-ph.HE},
}

@ARTICLE{Izevic:2019,
       author = {{Ivezi{\'c}}, {\v{Z}}eljko and {Kahn}, Steven M. and {Tyson}, J. Anthony and {Abel}, Bob and {Acosta}, Emily and {Allsman}, Robyn and {Alonso}, David and {AlSayyad}, Yusra and {Anderson}, Scott F. and {Andrew}, John and {Angel}, James Roger P. and {Angeli}, George Z. and {Ansari}, Reza and {Antilogus}, Pierre and {Araujo}, Constanza and {Armstrong}, Robert and {Arndt}, Kirk T. and {Astier}, Pierre and {Aubourg}, {\'E}ric and {Auza}, Nicole and {Axelrod}, Tim S. and {Bard}, Deborah J. and {Barr}, Jeff D. and {Barrau}, Aurelian and {Bartlett}, James G. and {Bauer}, Amanda E. and {Bauman}, Brian J. and {Baumont}, Sylvain and {Bechtol}, Ellen and {Bechtol}, Keith and {Becker}, Andrew C. and {Becla}, Jacek and {Beldica}, Cristina and {Bellavia}, Steve and {Bianco}, Federica B. and {Biswas}, Rahul and {Blanc}, Guillaume and {Blazek}, Jonathan and {Blandford}, Roger D. and {Bloom}, Josh S. and {Bogart}, Joanne},
        title = "{LSST: From Science Drivers to Reference Design and Anticipated Data Products}",
      journal = {\apj},
         year = 2019,
        month = mar,
       volume = {873},
       number = {2},
          eid = {111},
        pages = {111},
          doi = {10.3847/1538-4357/ab042c},
archivePrefix = {arXiv},
       eprint = {0805.2366},
 primaryClass = {astro-ph},
}

@ARTICLE{Jhawar:2025,
       author = {{Jhawar}, Sahil and {Wouters}, Thibeau and {Pang}, Peter T.~H. and {Bulla}, Mattia and {Coughlin}, Michael W. and {Dietrich}, Tim},
        title = "{Data-driven approach for modeling the temporal and spectral evolution of kilonova systematic uncertainties}",
      journal = {\prd},
         year = 2025,
        month = feb,
       volume = {111},
       number = {4},
          eid = {043046},
        pages = {043046},
          doi = {10.1103/PhysRevD.111.043046},
archivePrefix = {arXiv},
       eprint = {2410.21978},
 primaryClass = {astro-ph.HE}
}

@ARTICLE{Kalinani:2026,
       author = {{Kalinani}, Jay V. and {Ciolfi}, Riccardo and {Campanelli}, Manuela and {Giacomazzo}, Bruno and {Pavan}, Andrea and {Wen}, Allen and {Zlochower}, Yosef},
        title = "{Jet─Environment Interaction after Delayed Collapse in Binary Neutron Star Mergers}",
      journal = {\apjl},
         year = 2026,
        month = mar,
       volume = {1000},
       number = {1},
          eid = {L35},
        pages = {L35},
          doi = {10.3847/2041-8213/ae402a},
archivePrefix = {arXiv},
       eprint = {2505.09426},
 primaryClass = {astro-ph.HE},
       adsurl = {https://ui.adsabs.harvard.edu/abs/2026ApJ..1000L..35K}
}

@ARTICLE{Kasen:2017,
       author = {{Kasen}, Daniel and {Metzger}, Brian and {Barnes}, Jennifer and {Quataert}, Eliot and {Ramirez-Ruiz}, Enrico},
        title = "{Origin of the heavy elements in binary neutron-star mergers from a gravitational-wave event}",
      journal = {\nat},
         year = 2017,
        month = nov,
       volume = {551},
       number = {7678},
        pages = {80-84},
          doi = {10.1038/nature24453},
archivePrefix = {arXiv},
       eprint = {1710.05463},
 primaryClass = {astro-ph.HE}
}

@ARTICLE{Kawaguchi:2020,
       author = {{Kawaguchi}, Kyohei and {Shibata}, Masaru and {Tanaka}, Masaomi},
        title = "{Diversity of Kilonova Light Curves}",
      journal = {\apj},
         year = 2020,
        month = feb,
       volume = {889},
       number = {2},
          eid = {171},
        pages = {171},
          doi = {10.3847/1538-4357/ab61f6},
archivePrefix = {arXiv},
       eprint = {1908.05815},
 primaryClass = {astro-ph.HE}
}

@ARTICLE{Kawaguchi:2016,
       author = {{Kawaguchi}, Kyohei and {Kyutoku}, Koutarou and {Shibata}, Masaru and {Tanaka}, Masaomi},
        title = "{Models of Kilonova/Macronova Emission from Black Hole-Neutron Star Mergers}",
      journal = {\apj},
         year = 2016,
        month = jul,
       volume = {825},
       number = {1},
          eid = {52},
        pages = {52},
          doi = {10.3847/0004-637X/825/1/52},
archivePrefix = {arXiv},
       eprint = {1601.07711},
 primaryClass = {astro-ph.HE}
}

@ARTICLE{Kenworthy:2021,
       author = {{Kenworthy}, W.~D. and {Jones}, D.~O. and {Dai}, M. and {Kessler}, R. and {Scolnic}, D. and {Brout}, D. and {Siebert}, M.~R. and {Pierel}, J.~D.~R. and {Dettman}, K.~G. and {Dimitriadis}, G. and {Foley}, R.~J. and {Jha}, S.~W. and {Pan}, Y.-C. and {Riess}, A. and {Rodney}, S. and {Rojas-Bravo}, C.},
        title = "{SALT3: An Improved Type Ia Supernova Model for Measuring Cosmic Distances}",
      journal = {\apj},
         year = 2021,
        month = dec,
       volume = {923},
       number = {2},
          eid = {265},
        pages = {265},
          doi = {10.3847/1538-4357/ac30d8},
archivePrefix = {arXiv},
       eprint = {2104.07795},
 primaryClass = {astro-ph.CO}
}

@ARTICLE{Khan:2016,
       author = {{Khan}, Sebastian and {Husa}, Sascha and {Hannam}, Mark and {Ohme}, Frank and {P{\"u}rrer}, Michael and {Forteza}, Xisco Jim{\'e}nez and {Boh{\'e}}, Alejandro},
        title = "{Frequency-domain gravitational waves from nonprecessing black-hole binaries. II. A phenomenological model for the advanced detector era}",
      journal = {\prd},
         year = 2016,
        month = feb,
       volume = {93},
       number = {4},
          eid = {044007},
        pages = {044007},
          doi = {10.1103/PhysRevD.93.044007},
archivePrefix = {arXiv},
       eprint = {1508.07253},
 primaryClass = {gr-qc}
}

@ARTICLE{Kiendrebeogo:2023,
       author = {{Kiendrebeogo}, R. Weizmann and {Farah}, Amanda M. and {Foley}, Emily M. and {Gray}, Abigail and {Kunert}, Nina and {Puecher}, Anna and {Toivonen}, Andrew and {VandenBerg}, R. Oliver and {Anand}, Shreya and {Ahumada}, Tom{\'a}s and {Karambelkar}, Viraj and {Coughlin}, Michael W. and {Dietrich}, Tim and {Kam}, S. Zacharie and {Pang}, Peter T.~H. and {Singer}, Leo P. and {Sravan}, Niharika},
        title = "{Updated Observing Scenarios and Multimessenger Implications for the International Gravitational-wave Networks O4 and O5}",
      journal = {\apj},
         year = 2023,
        month = dec,
       volume = {958},
       number = {2},
          eid = {158},
        pages = {158},
          doi = {10.3847/1538-4357/acfcb1},
archivePrefix = {arXiv},
       eprint = {2306.09234},
 primaryClass = {astro-ph.HE}
}

@ARTICLE{King:2025,
       author = {{King}, Brendan L. and {De}, Soumi and {Korobkin}, Oleg and {Coughlin}, Michael W. and {Pang}, Peter T.~H. and {Strother}, Terrance T.},
        title = "{Inferring Neutron Star Merger Ejecta Morphology with Kilonovae}",
      journal = {\pasp},
         year = 2025,
        month = oct,
       volume = {137},
       number = {10},
          eid = {104507},
        pages = {104507},
          doi = {10.1088/1538-3873/ae10df},
archivePrefix = {arXiv},
       eprint = {2505.16876},
 primaryClass = {astro-ph.HE}
}

@ARTICLE{Kini:2026,
       author = {{Kini}, Yves and {Mauviard}, Lucien and {Salmi}, Tuomo and {Watts}, Anna L. and {Guillot}, Sebastien and {Dorsman}, Bas and {Choudhury}, Devarshi and {Gonz{\'a}lez-Caniulef}, Denis and {Hoogkamer}, Mariska and {Huppenkothen}, Daniela and {Kazantsev}, Christine and {Kerr}, Matthew and {Nissanke}, Samaya and {Ray}, Paul S. and {Stammler}, Pierre and {Vinciguerra}, Serena},
        title = "{A NICER View of PSR J0030+0451: Updated Constraints from Six Years of NICER Observations}",
      journal = {arXiv e-prints},
         year = 2026,
        month = feb,
          eid = {arXiv:2602.23743},
          doi = {10.48550/arXiv.2602.23743},
archivePrefix = {arXiv},
       eprint = {2602.23743},
 primaryClass = {astro-ph.HE}
}

@ARTICLE{Kiuchi:2018,
       author = {{Kiuchi}, Kenta and {Kyutoku}, Koutarou and {Sekiguchi}, Yuichiro and {Shibata}, Masaru},
        title = "{Global simulations of strongly magnetized remnant massive neutron stars formed in binary neutron star mergers}",
      journal = {\prd},
         year = 2018,
        month = jun,
       volume = {97},
       number = {12},
          eid = {124039},
        pages = {124039},
          doi = {10.1103/PhysRevD.97.124039},
archivePrefix = {arXiv},
       eprint = {1710.01311},
 primaryClass = {astro-ph.HE}
}

@ARTICLE{Kiuchi:2017,
       author = {{Kiuchi}, Kenta and {Kawaguchi}, Kyohei and {Kyutoku}, Koutarou and {Sekiguchi}, Yuichiro and {Shibata}, Masaru and {Taniguchi}, Keisuke},
        title = "{Sub-radian-accuracy gravitational waveforms of coalescing binary neutron stars in numerical relativity}",
      journal = {\prd},
         year = 2017,
        month = oct,
       volume = {96},
       number = {8},
          eid = {084060},
        pages = {084060},
          doi = {10.1103/PhysRevD.96.084060},
archivePrefix = {arXiv},
       eprint = {1708.08926},
 primaryClass = {astro-ph.HE}
}

@ARTICLE{Koehn:2025b,
       author = {{Koehn}, H. and {Wouters}, T. and {Pang}, P.~T.~H. and {Bulla}, M. and {Rose}, H. and {Wichern}, H. and {Dietrich}, T.},
        title = "{Efficient Bayesian analysis of kilonovae and gamma ray burst afterglows with FIESTA}",
      journal = {\aap},
         year = 2025,
        month = dec,
       volume = {704},
          eid = {A55},
        pages = {A55},
          doi = {10.1051/0004-6361/202556626},
archivePrefix = {arXiv},
       eprint = {2507.13807},
 primaryClass = {astro-ph.HE},
}

@ARTICLE{Koehn:2025,
       author = {{Koehn}, Hauke and {Rose}, Henrik and {Pang}, Peter T.~H. and {Somasundaram}, Rahul and {Reed}, Brendan T. and {Tews}, Ingo and {Abac}, Adrian and {Komoltsev}, Oleg and {Kunert}, Nina and {Kurkela}, Aleksi and {Coughlin}, Michael W. and {Healy}, Brian F. and {Dietrich}, Tim},
        title = "{From Existing and New Nuclear and Astrophysical Constraints to Stringent Limits on the Equation of State of Neutron-Rich Dense Matter}",
      journal = {Physical Review X},
         year = 2025,
        month = apr,
       volume = {15},
       number = {2},
          eid = {021014},
        pages = {021014},
          doi = {10.1103/PhysRevX.15.021014},
archivePrefix = {arXiv},
       eprint = {2402.04172},
 primaryClass = {astro-ph.HE}
}

@ARTICLE{Koehn:2024,
       author = {{Koehn}, Hauke and {Wouters}, Thibeau and {Rose}, Henrik and {Pang}, Peter T.~H. and {Somasundaram}, Rahul and {Tews}, Ingo and {Dietrich}, Tim},
        title = "{Classification of compact objects and model comparison using EOS knowledge}",
      journal = {\prd},
         year = 2024,
        month = nov,
       volume = {110},
       number = {10},
          eid = {103015},
        pages = {103015},
          doi = {10.1103/PhysRevD.110.103015},
archivePrefix = {arXiv},
       eprint = {2407.07837},
 primaryClass = {astro-ph.HE}
}

@ARTICLE{Koshiba:1992,
       author = {{Koshiba}, M.},
        title = "{Observational neutrino astrophysics}",
      journal = {\physrep},
         year = 1992,
        month = nov,
       volume = {220},
       number = {5-6},
        pages = {229-381},
          doi = {10.1016/0370-1573(92)90083-C}
}

@misc{Koposov:2025,
       author = {{Koposov}, Sergey and {Speagle}, Josh and {Barbary}, Kyle and {Ashton}, Gregory and {Bennett}, Ed and {Buchner}, Johannes and {Scheffler}, Carl and {Talbot}, Colm and {Cook}, Ben and {Guillochon}, James and {Cubillos}, Patricio and {Asensio Ramos}, Andr{\'e}s and {Dartiailh}, Matthieu and {Ilya} and {Tollerud}, Erik and {Lang}, Dustin and {Johnson}, Ben and {jtmendel} and {Higson}, Edward and {Vandal}, Thomas and {Daylan}, Tansu and {Angus}, Ruth and {patelR} and {Cargile}, Phillip and {Sheehan}, Patrick and {Pitkin}, Matt and {Kirk}, Matthew and {Xu}, Lu and {Leja}, Joel and {joezuntz}},
        title = "{joshspeagle/dynesty: v3.0.0}",
         year = 2025,
        month = oct,
          doi = {10.5281/zenodo.17268284},
      version = {v3.0.0},
    publisher = {Zenodo}
}

@ARTICLE{Krüger:2020,
       author = {{Kr{\"u}ger}, Christian J. and {Foucart}, Francois},
        title = "{Estimates for disk and ejecta masses produced in compact binary mergers}",
      journal = {\prd},
         year = 2020,
        month = may,
       volume = {101},
       number = {10},
          eid = {103002},
        pages = {103002},
          doi = {10.1103/PhysRevD.101.103002},
archivePrefix = {arXiv},
       eprint = {2002.07728},
 primaryClass = {astro-ph.HE}
}

@ARTICLE{Kumar:2024,
       author = {{Kumar}, Rajesh and {Dexheimer}, Veronica and {Jahan}, Johannes and {Noronha}, Jorge and {Noronha-Hostler}, Jacquelyn and {Ratti}, Claudia and {Yunes}, Nico and {Nava Acuna}, Angel Rodrigo and {Alford}, Mark and {Anik}, Mahmudul Hasan and {Chatterjee}, Debarati and {Chatziioannou}, Katerina and {Chen}, Hsin-Yu and {Clevinger}, Alexander and {Conde}, Carlos and {Cruz-Camacho}, Nikolas and {Dore}, Travis and {Drischler}, Christian and {Elfner}, Hannah and {Essick}, Reed and {Friedenberg}, David and {Ghosh}, Suprovo and {Grefa}, Joaquin and {Haas}, Roland and {Haber}, Alexander and {Hammelmann}, Jan and {Harris}, Steven and {Haster}, Carl-Johan and {Hatsuda}, Tetsuo and {Hippert}, Mauricio and {Hirayama}, Renan and {Holt}, Jeremy W. and {Kahangirwe}, Micheal and {Karthein}, Jamie and {Kojo}, Toru and {Landry}, Philippe and {Lin}, Zidu and {Luzum}, Matthew and {Manning}, Timothy Andrew and {Salinas San Martin}, Jordi and {Miller}, Cole and {Most}, Elias Roland and {Mroczek}, Debora and {Muronga}, Azwinndini and {Patino}, Nicolas and {Peterson}, Jeffrey and {Plumberg}, Christopher and {Price}, Damien and {Providencia}, Constanca and {Rougemont}, Romulo and {Roy}, Satyajit and {Shah}, Hitansh and {Shapiro}, Stuart and {Steiner}, Andrew W. and {Strickland}, Michael and {Tan}, Hung and {Togashi}, Hajime and {Portillo Vazquez}, Israel and {Wen}, Pengsheng and {Zhang}, Ziyuan and {Muses Collaboration}},
        title = "{Theoretical and experimental constraints for the equation of state of dense and hot matter}",
      journal = {Living Reviews in Relativity},
         year = 2024,
        month = dec,
       volume = {27},
       number = {1},
          eid = {3},
        pages = {3},
          doi = {10.1007/s41114-024-00049-6},
archivePrefix = {arXiv},
       eprint = {2303.17021},
 primaryClass = {nucl-th}
}

@ARTICLE{Kunert:2024,
       author = {{Kunert}, N. and {Antier}, S. and {Nedora}, V. and {Bulla}, M. and {Pang}, P.~T.~H. and {Anand}, S. and {Coughlin}, M. and {Tews}, I. and {Barnes}, J. and {Hussenot-Desenonges}, T. and {Healy}, B. and {Jegou du Laz}, T. and {Pilloix}, M. and {Kiendrebeogo}, W. and {Dietrich}, T.},
        title = "{Bayesian model selection for GRB 211211A through multiwavelength analyses}",
      journal = {\mnras},
         year = 2024,
        month = jan,
       volume = {527},
       number = {2},
        pages = {3900-3911},
          doi = {10.1093/mnras/stad3463},
archivePrefix = {arXiv},
       eprint = {2301.02049},
 primaryClass = {astro-ph.HE}
}

@ARTICLE{Lalit:2025,
       author = {{Lalit}, Sudhanva and {Semposki}, Alexandra C. and {Maldonado}, Joshua M.},
        title = "{Star Log-extended eMulation: A method for efficient computation of the Tolman-Oppenheimer-Volkoff equations}",
      journal = {Physical Review Research},
         year = 2025,
        month = oct,
       volume = {7},
       number = {4},
          eid = {043037},
        pages = {043037},
          doi = {10.1103/5p3h-b8rf},
archivePrefix = {arXiv},
       eprint = {2411.10556},
 primaryClass = {astro-ph.HE}
}

@ARTICLE{Landry:2021,
       author = {{Landry}, Philippe and {Read}, Jocelyn S.},
        title = "{The Mass Distribution of Neutron Stars in Gravitational-wave Binaries}",
      journal = {\apjl},
         year = 2021,
        month = nov,
       volume = {921},
       number = {2},
          eid = {L25},
        pages = {L25},
          doi = {10.3847/2041-8213/ac2f3e},
archivePrefix = {arXiv},
       eprint = {2107.04559},
 primaryClass = {astro-ph.HE}
}

@ARTICLE{Lattimer:2021,
       author = {{Lattimer}, J.~M.},
        title = "{Neutron Stars and the Nuclear Matter Equation of State}",
      journal = {Annual Review of Nuclear and Particle Science},
         year = 2021,
        month = sep,
       volume = {71},
        pages = {433-464},
          doi = {10.1146/annurev-nucl-102419-124827}
}

@ARTICLE{Lattimer:2001,
       author = {{Lattimer}, J.~M. and {Prakash}, M.},
        title = "{Neutron Star Structure and the Equation of State}",
      journal = {\apj},
         year = 2001,
        month = mar,
       volume = {550},
       number = {1},
        pages = {426-442},
          doi = {10.1086/319702},
archivePrefix = {arXiv},
       eprint = {astro-ph/0002232},
 primaryClass = {astro-ph},
}

@ARTICLE{Li:2019,
       author = {{Li}, Bao-An and {Krastev}, Plamen G. and {Wen}, De-Hua and {Zhang}, Nai-Bo},
        title = "{Towards understanding astrophysical effects of nuclear symmetry energy}",
      journal = {European Physical Journal A},
         year = 2019,
        month = jul,
       volume = {55},
       number = {7},
          eid = {117},
        pages = {117},
          doi = {10.1140/epja/i2019-12780-8},
archivePrefix = {arXiv},
       eprint = {1905.13175},
 primaryClass = {nucl-th}
}

@ARTICLE{Lo:2011,
       author = {{Lo}, Ka-Wai and {Lin}, Lap-Ming},
        title = "{The Spin Parameter of Uniformly Rotating Compact Stars}",
      journal = {\apj},
         year = 2011,
        month = feb,
       volume = {728},
       number = {1},
          eid = {12},
        pages = {12},
          doi = {10.1088/0004-637X/728/1/12},
archivePrefix = {arXiv},
       eprint = {1011.3563},
 primaryClass = {astro-ph.HE}
}

@ARTICLE{Lund:2025,
       author = {{Lund}, Kelsey A. and {Somasundaram}, Rahul and {McLaughlin}, Gail C. and {Miller}, Jonah M. and {Mumpower}, Matthew R. and {Tews}, Ingo},
        title = "{Kilonova Emissions from Neutron Star Merger Remnants: Implications for the Nuclear Equation of State}",
      journal = {\apj},
         year = 2025,
        month = jul,
       volume = {987},
       number = {1},
          eid = {56},
        pages = {56},
          doi = {10.3847/1538-4357/add148},
archivePrefix = {arXiv},
       eprint = {2408.07686},
 primaryClass = {astro-ph.HE},
}

@ARTICLE{Magnall:2025,
       author = {{Magnall}, Spencer J. and {Goode}, Simon R. and {Sarin}, Nikhil and {Lasky}, Paul D.},
        title = "{Directly inferring cosmology and the neutron-star equation of state from gravitational-wave mergers}",
      journal = {\mnras},
         year = 2025,
        month = nov,
       volume = {543},
       number = {4},
        pages = {3673-3683},
          doi = {10.1093/mnras/staf1613},
archivePrefix = {arXiv},
       eprint = {2410.07754},
 primaryClass = {gr-qc}
}

@ARTICLE{Margutti:2021,
       author = {{Margutti}, Raffaella and {Chornock}, Ryan},
        title = "{First Multimessenger Observations of a Neutron Star Merger}",
      journal = {\araa},
         year = 2021,
        month = sep,
       volume = {59},
        pages = {155-202},
          doi = {10.1146/annurev-astro-112420-030742},
archivePrefix = {arXiv},
       eprint = {2012.04810},
 primaryClass = {astro-ph.HE}
}

@ARTICLE{Markovic:1993,
       author = {{Markovi{\'c}}, Dragoljub},
        title = "{Possibility of determining cosmological parameters from measurements of gravitational waves emitted by coalescing, compact binaries}",
      journal = {\prd},
         year = 1993,
        month = nov,
       volume = {48},
       number = {10},
        pages = {4738-4756},
          doi = {10.1103/PhysRevD.48.4738}
}

@ARTICLE{Mauviard:2025,
       author = {{Mauviard}, Lucien and {Guillot}, Sebastien and {Salmi}, Tuomo and {Choudhury}, Devarshi and {Dorsman}, Bas and {Gonz{\'a}lez-Caniulef}, Denis and {Hoogkamer}, Mariska and {Huppenkothen}, Daniela and {Kazantsev}, Christine and {Kini}, Yves and {Olive}, Jean-Francois and {Stammler}, Pierre and {Watts}, Anna L. and {Mendes}, Melissa and {Rutherford}, Nathan and {Schwenk}, Achim and {Svensson}, Isak and {Bogdanov}, Slavko and {Kerr}, Matthew and {Ray}, Paul S. and {Guillemot}, Lucas and {Cognard}, Isma{\"e}l and {Theureau}, Gilles},
        title = "{A NICER view of the 1.4 solar-mass edge-on pulsar PSR J0614--3329}",
      journal = {arXiv e-prints},
         year = 2025,
        month = jun,
          eid = {arXiv:2506.14883},
          doi = {10.48550/arXiv.2506.14883},
archivePrefix = {arXiv},
       eprint = {2506.14883},
 primaryClass = {astro-ph.HE}
}

@ARTICLE{Metzger:2019,
       author = {{Metzger}, Brian D.},
        title = "{Kilonovae}",
      journal = {Living Reviews in Relativity},
         year = 2019,
        month = dec,
       volume = {23},
       number = {1},
          eid = {1},
        pages = {1},
          doi = {10.1007/s41114-019-0024-0},
archivePrefix = {arXiv},
       eprint = {1910.01617},
 primaryClass = {astro-ph.HE}
}

@ARTICLE{Miller:2019,
       author = {{Miller}, M.~C. and {Lamb}, F.~K. and {Dittmann}, A.~J. and {Bogdanov}, S. and {Arzoumanian}, Z. and {Gendreau}, K.~C. and {Guillot}, S. and {Harding}, A.~K. and {Ho}, W.~C.~G. and {Lattimer}, J.~M. and {Ludlam}, R.~M. and {Mahmoodifar}, S. and {Morsink}, S.~M. and {Ray}, P.~S. and {Strohmayer}, T.~E. and {Wood}, K.~S. and {Enoto}, T. and {Foster}, R. and {Okajima}, T. and {Prigozhin}, G. and {Soong}, Y.},
        title = "{PSR J0030+0451 Mass and Radius from NICER Data and Implications for the Properties of Neutron Star Matter}",
      journal = {\apjl},
         year = 2019,
        month = dec,
       volume = {887},
       number = {1},
          eid = {L24},
        pages = {L24},
          doi = {10.3847/2041-8213/ab50c5},
archivePrefix = {arXiv},
       eprint = {1912.05705},
 primaryClass = {astro-ph.HE}
}

@ARTICLE{Miller:2021,
       author = {{Miller}, M.~C. and {Lamb}, F.~K. and {Dittmann}, A.~J. and {Bogdanov}, S. and {Arzoumanian}, Z. and {Gendreau}, K.~C. and {Guillot}, S. and {Ho}, W.~C.~G. and {Lattimer}, J.~M. and {Loewenstein}, M. and {Morsink}, S.~M. and {Ray}, P.~S. and {Wolff}, M.~T. and {Baker}, C.~L. and {Cazeau}, T. and {Manthripragada}, S. and {Markwardt}, C.~B. and {Okajima}, T. and {Pollard}, S. and {Cognard}, I. and {Cromartie}, H.~T. and {Fonseca}, E. and {Guillemot}, L. and {Kerr}, M. and {Parthasarathy}, A. and {Pennucci}, T.~T. and {Ransom}, S. and {Stairs}, I.},
        title = "{The Radius of PSR J0740+6620 from NICER and XMM-Newton Data}",
      journal = {\apjl},
         year = 2021,
        month = sep,
       volume = {918},
       number = {2},
          eid = {L28},
        pages = {L28},
          doi = {10.3847/2041-8213/ac089b},
archivePrefix = {arXiv},
       eprint = {2105.06979},
 primaryClass = {astro-ph.HE}
}

@ARTICLE{Morisaki:2021,
       author = {{Morisaki}, Soichiro},
        title = "{Accelerating parameter estimation of gravitational waves from compact binary coalescence using adaptive frequency resolutions}",
      journal = {\prd},
         year = 2021,
        month = aug,
       volume = {104},
       number = {4},
          eid = {044062},
        pages = {044062},
          doi = {10.1103/PhysRevD.104.044062},
archivePrefix = {arXiv},
       eprint = {2104.07813},
 primaryClass = {gr-qc}
}

@ARTICLE{Narikawa:2019,
       author = {{Narikawa}, Tatsuya and {Uchikata}, Nami and {Kawaguchi}, Kyohei and {Kiuchi}, Kenta and {Kyutoku}, Koutarou and {Shibata}, Masaru and {Tagoshi}, Hideyuki},
        title = "{Discrepancy in tidal deformability of GW170817 between the Advanced LIGO twin detectors}",
      journal = {Physical Review Research},
         year = 2019,
        month = oct,
       volume = {1},
       number = {3},
          eid = {033055},
        pages = {033055},
          doi = {10.1103/PhysRevResearch.1.033055},
archivePrefix = {arXiv},
       eprint = {1812.06100},
 primaryClass = {astro-ph.HE}
}

@ARTICLE{Nedora:2025,
       author = {{Nedora}, Vsevolod and {Crosato Menegazzi}, Ludovica and {Peretti}, Enrico and {Dietrich}, Tim and {Shibata}, Masaru},
        title = "{Multiphysics framework for fast modelling of gamma-ray burst afterglows}",
      journal = {\mnras},
         year = 2025,
        month = apr,
       volume = {538},
       number = {3},
        pages = {2089-2115},
          doi = {10.1093/mnras/staf302},
archivePrefix = {arXiv},
       eprint = {2409.16852},
 primaryClass = {astro-ph.HE},
}

@ARTICLE{Nedora:2022,
       author = {{Nedora}, Vsevolod and {Schianchi}, Federico and {Bernuzzi}, Sebastiano and {Radice}, David and {Daszuta}, Boris and {Endrizzi}, Andrea and {Perego}, Albino and {Prakash}, Aviral and {Zappa}, Francesco},
        title = "{Mapping dynamical ejecta and disk masses from numerical relativity simulations of neutron star mergers}",
      journal = {Classical and Quantum Gravity},
         year = 2022,
        month = jan,
       volume = {39},
       number = {1},
          eid = {015008},
        pages = {015008},
          doi = {10.1088/1361-6382/ac35a8},
archivePrefix = {arXiv},
       eprint = {2011.11110},
 primaryClass = {astro-ph.HE}
}

@ARTICLE{Nedora:2021,
       author = {{Nedora}, Vsevolod and {Bernuzzi}, Sebastiano and {Radice}, David and {Daszuta}, Boris and {Endrizzi}, Andrea and {Perego}, Albino and {Prakash}, Aviral and {Safarzadeh}, Mohammadtaher and {Schianchi}, Federico and {Logoteta}, Domenico},
        title = "{Numerical Relativity Simulations of the Neutron Star Merger GW170817: Long-term Remnant Evolutions, Winds, Remnant Disks, and Nucleosynthesis}",
      journal = {\apj},
         year = 2021,
        month = jan,
       volume = {906},
       number = {2},
          eid = {98},
        pages = {98},
          doi = {10.3847/1538-4357/abc9be},
archivePrefix = {arXiv},
       eprint = {2008.04333},
 primaryClass = {astro-ph.HE}
}

@ARTICLE{Nelson:2020,
       author = {{Nelson}, Benjamin E. and {Ford}, Eric B. and {Buchner}, Johannes and {Cloutier}, Ryan and {D{\'\i}az}, Rodrigo F. and {Faria}, Jo{\~a}o P. and {Hara}, Nathan C. and {Rajpaul}, Vinesh M. and {Rukdee}, Surangkhana},
        title = "{Quantifying the Bayesian Evidence for a Planet in Radial Velocity Data}",
      journal = {\aj},
         year = 2020,
        month = feb,
       volume = {159},
       number = {2},
          eid = {73},
        pages = {73},
          doi = {10.3847/1538-3881/ab5190},
archivePrefix = {arXiv},
       eprint = {1806.04683},
 primaryClass = {astro-ph.EP}
}

@ARTICLE{Neuweiler:2026,
       author = {{Neuweiler}, Anna and {Gieg}, Henrique and {Rose}, Henrik and {Koehn}, Hauke and {Markin}, Ivan and {Schianchi}, Federico and {Brodie}, Liam and {Haber}, Alexander and {Nedora}, Vsevolod and {Bulla}, Mattia and {Dietrich}, Tim},
        title = "{General-relativistic radiation magnetohydrodynamics simulations of binary neutron star mergers: The influence of spin on the multimessenger picture}",
      journal = {\prd},
         year = 2026,
        month = feb,
       volume = {113},
       number = {4},
          eid = {043038},
        pages = {043038},
          doi = {10.1103/mxlf-8sbm},
archivePrefix = {arXiv},
       eprint = {2510.14850},
 primaryClass = {astro-ph.HE}
}

@ARTICLE{Nissanke:2010,
       author = {{Nissanke}, Samaya and {Holz}, Daniel E. and {Hughes}, Scott A. and {Dalal}, Neal and {Sievers}, Jonathan L.},
        title = "{Exploring Short Gamma-ray Bursts as Gravitational-wave Standard Sirens}",
      journal = {\apj},
         year = 2010,
        month = dec,
       volume = {725},
       number = {1},
        pages = {496-514},
          doi = {10.1088/0004-637X/725/1/496},
archivePrefix = {arXiv},
       eprint = {0904.1017},
 primaryClass = {astro-ph.CO}
}

@article{Pang:2024,
       author = {{Pang}, Peter T.~H. and {Dietrich}, Tim and {Coughlin}, Michael W. and {Bulla}, Mattia and {Tews}, Ingo and {Almualla}, Mouza and {Barna}, Tyler and {Kiendrebeogo}, Ramodgwend{\'e} Weizmann and {Kunert}, Nina and {Mansingh}, Gargi and {Reed}, Brandon and {Sravan}, Niharika and {Toivonen}, Andrew and {Antier}, Sarah and {VandenBerg}, Robert O. and {Heinzel}, Jack and {Nedora}, Vsevolod and {Salehi}, Pouyan and {Sharma}, Ritwik and {Somasundaram}, Rahul and {Van Den Broeck}, Chris},
        title = "{NMMA: Nuclear Multi Messenger Astronomy framework}",
      journal = {Astrophysics Source Code Library},
         year = 2024,
        month = feb,
archivePrefix = "ascl",
       eprint = {2402.001}
}

@ARTICLE{Pang:2023,
       author = {{Pang}, Peter T.~H. and {Dietrich}, Tim and {Coughlin}, Michael W. and {Bulla}, Mattia and {Tews}, Ingo and {Almualla}, Mouza and {Barna}, Tyler and {Kiendrebeogo}, Ramodgwend{\'e} Weizmann and {Kunert}, Nina and {Mansingh}, Gargi and {Reed}, Brandon and {Sravan}, Niharika and {Toivonen}, Andrew and {Antier}, Sarah and {VandenBerg}, Robert O. and {Heinzel}, Jack and {Nedora}, Vsevolod and {Salehi}, Pouyan and {Sharma}, Ritwik and {Somasundaram}, Rahul and {Van Den Broeck}, Chris},
        title = "{An updated nuclear-physics and multi-messenger astrophysics framework for binary neutron star mergers}",
      journal = {Nature Communications},
         year = 2023,
        month = dec,
       volume = {14},
          eid = {8352},
        pages = {8352},
          doi = {10.1038/s41467-023-43932-6},
archivePrefix = {arXiv},
       eprint = {2205.08513},
 primaryClass = {astro-ph.HE}
}

@ARTICLE{Perego:2017,
       author = {{Perego}, Albino and {Radice}, David and {Bernuzzi}, Sebastiano},
        title = "{AT 2017gfo: An Anisotropic and Three-component Kilonova Counterpart of GW170817}",
      journal = {\apjl},
         year = 2017,
        month = dec,
       volume = {850},
       number = {2},
          eid = {L37},
        pages = {L37},
          doi = {10.3847/2041-8213/aa9ab9},
archivePrefix = {arXiv},
       eprint = {1711.03982},
 primaryClass = {astro-ph.HE}
}

@ARTICLE{Planck:2018,
       author = {{Planck Collaboration} and {Aghanim}, N. and {Akrami}, Y. and {Ashdown}, M. and {Aumont}, J. and {Baccigalupi}, C. and {Ballardini}, M. and {Banday}, A.~J. and {Barreiro}, R.~B. and {Bartolo}, N. and {Basak}, S. and {Battye}, R. and {Benabed}, K. and {Bernard}, J.-P. and {Bersanelli}, M. and {Bielewicz}, P. and {Bock}, J.~J. and {Bond}, J.~R. and {Borrill}, J. and {Bouchet}, F.~R. and {Boulanger}, F. and {Bucher}, M. and {Burigana}, C. and {Butler}, R.~C. and {Calabrese}, E. and {Cardoso}, J.-F. and {Carron}, J. and {Challinor}, A. and {Chiang}, H.~C. and {Chluba}, J. and {Colombo}, L.~P.~L. and {Combet}, C. and {Contreras}, D. and {Crill}, B.~P. and {Cuttaia}, F. and {de Bernardis}, P. and {de Zotti}, G.},
        title = "{Planck 2018 results. VI. Cosmological parameters}",
      journal = {\aap},
         year = 2020,
        month = sep,
       volume = {641},
          eid = {A6},
        pages = {A6},
          doi = {10.1051/0004-6361/201833910},
archivePrefix = {arXiv},
       eprint = {1807.06209},
 primaryClass = {astro-ph.CO}
}

@ARTICLE{Pognan:2022,
       author = {{Pognan}, Quentin and {Jerkstrand}, Anders and {Grumer}, Jon},
        title = "{NLTE effects on kilonova expansion opacities}",
      journal = {\mnras},
         year = 2022,
        month = jul,
       volume = {513},
       number = {4},
        pages = {5174-5197},
          doi = {10.1093/mnras/stac1253},
archivePrefix = {arXiv},
       eprint = {2202.09245},
 primaryClass = {astro-ph.HE}
}

@ARTICLE{Prathaban:2025,
       author = {{Prathaban}, Metha and {Yallup}, David and {Alvey}, James and {Yang}, Ming and {Templeton}, Will and {Handley}, Will},
        title = "{Gravitational-wave inference at GPU speed: A bilby-like nested sampling kernel within blackjax-ns}",
      journal = {arXiv e-prints},
         year = 2025,
        month = sep,
          eid = {arXiv:2509.04336},
          doi = {10.48550/arXiv.2509.04336},
archivePrefix = {arXiv},
       eprint = {2509.04336},
 primaryClass = {gr-qc}
}

@ARTICLE{Pratten:2020,
       author = {{Pratten}, Geraint and {Husa}, Sascha and {Garc{\'\i}a-Quir{\'o}s}, Cecilio and {Colleoni}, Marta and {Ramos-Buades}, Antoni and {Estell{\'e}s}, H{\'e}ctor and {Jaume}, Rafel},
        title = "{Setting the cornerstone for a family of models for gravitational waves from compact binaries: The dominant harmonic for nonprecessing quasicircular black holes}",
      journal = {\prd},
         year = 2020,
        month = sep,
       volume = {102},
       number = {6},
          eid = {064001},
        pages = {064001},
          doi = {10.1103/PhysRevD.102.064001},
archivePrefix = {arXiv},
       eprint = {2001.11412},
 primaryClass = {gr-qc}
}

@ARTICLE{Price:2017,
       author = {{Price-Whelan}, Adrian M. and {Foreman-Mackey}, Daniel},
        title = "{schwimmbad: A uniform interface to parallel processing pools in Python}",
      journal = {The Journal of Open Source Software},
         year = 2017,
        month = sep,
       volume = {2},
          eid = {357},
        pages = {357},
          doi = {10.21105/joss.00357}
}

@ARTICLE{Punturo:2010,
       author = {{Punturo}, M. and {Abernathy}, M. and {Acernese}, F. and {Allen}, B. and {Andersson}, N. and {Arun}, K. and {Barone}, F. and {Barr}, B. and {Barsuglia}, M. and {Beker}, M. and {Beveridge}, N. and {Birindelli}, S. and {Bose}, S. and {Bosi}, L. and {Braccini}, S. and {Bradaschia}, C. and {Bulik}, T. and {Calloni}, E. and {Cella}, G. and {Chassande Mottin}, E. and {Chelkowski}, S. and {Chincarini}, A. and {Clark}, J. and {Coccia}, E. and {Colacino}, C. and {Colas}, J. and {Cumming}, A. and {Cunningham}, L. and {Cuoco}, E. },
        title = "{The Einstein Telescope: a third-generation gravitational wave observatory}",
      journal = {Classical and Quantum Gravity},
         year = 2010,
        month = oct,
       volume = {27},
       number = {19},
          eid = {194002},
        pages = {194002},
          doi = {10.1088/0264-9381/27/19/194002}
}

@ARTICLE{Ramachandran:2017,
       author = {{Ramachandran}, Prajit and {Zoph}, Barret and {Le}, Quoc V.},
        title = "{Searching for Activation Functions}",
      journal = {arXiv e-prints},
         year = 2017,
        month = oct,
          eid = {arXiv:1710.05941},
          doi = {10.48550/arXiv.1710.05941},
archivePrefix = {arXiv},
       eprint = {1710.05941},
 primaryClass = {cs.NE}
}

@ARTICLE{Radice:2020,
       author = {{Radice}, David and {Bernuzzi}, Sebastiano and {Perego}, Albino},
        title = "{The Dynamics of Binary Neutron Star Mergers and GW170817}",
      journal = {Annual Review of Nuclear and Particle Science},
         year = 2020,
        month = oct,
       volume = {70},
        pages = {95-119},
          doi = {10.1146/annurev-nucl-013120-114541},
archivePrefix = {arXiv},
       eprint = {2002.03863},
 primaryClass = {astro-ph.HE}
}

@ARTICLE{Radice:2018,
       author = {{Radice}, David and {Perego}, Albino and {Hotokezaka}, Kenta and {Fromm}, Steven A. and {Bernuzzi}, Sebastiano and {Roberts}, Luke F.},
        title = "{Binary Neutron Star Mergers: Mass Ejection, Electromagnetic Counterparts, and Nucleosynthesis}",
      journal = {\apj},
         year = 2018,
        month = dec,
       volume = {869},
       number = {2},
          eid = {130},
        pages = {130},
          doi = {10.3847/1538-4357/aaf054},
archivePrefix = {arXiv},
       eprint = {1809.11161},
 primaryClass = {astro-ph.HE}
}

@INCOLLECTION{Raffelt:2008,
       author = {{Raffelt}, Georg G.},
        title = "{Astrophysical Axion Bounds}",
    booktitle = {Axions},
    publisher = {Springer Heildelberg},
         year = 2008,
       editor = {{Kuster}, Markus and {Raffelt}, Georg and {Beltr{\'a}n}, Berta},
       volume = {741},
        pages = {51},
          doi = {10.1007/978-3-540-73518-2_3}
}

@ARTICLE{Reed:2024,
       author = {{Reed}, Brendan T. and {Somasundaram}, Rahul and {De}, Soumi and {Armstrong}, Cassandra L. and {Giuliani}, Pablo and {Capano}, Collin and {Brown}, Duncan A. and {Tews}, Ingo},
        title = "{Toward Accelerated Nuclear-physics Parameter Estimation from Binary Neutron Star Mergers: Emulators for the Tolman─Oppenheimer─Volkoff Equations}",
      journal = {\apj},
         year = 2024,
        month = oct,
       volume = {974},
       number = {2},
          eid = {285},
        pages = {285},
          doi = {10.3847/1538-4357/ad737c},
archivePrefix = {arXiv},
       eprint = {2405.20558},
 primaryClass = {astro-ph.HE},
       adsurl = {https://ui.adsabs.harvard.edu/abs/2024ApJ...974..285R}
}

@ARTICLE{Reed:2026,
       author = {{Reed}, Brendan T. and {Armstrong}, Cassandra L. and {Somasundaram}, Rahul and {Brown}, Duncan A. and {Capano}, Collin and {De}, Soumi and {Tews}, Ingo},
        title = "{Direct inference of nuclear equation-of-state parameters from gravitational-wave observations}",
      journal = {Classical and Quantum Gravity},
         year = 2026,
        month = mar,
       volume = {43},
       number = {5},
          eid = {055015},
        pages = {055015},
          doi = {10.1088/1361-6382/ae49d9},
archivePrefix = {arXiv},
       eprint = {2506.15984},
 primaryClass = {astro-ph.HE},
}

@INPROCEEDINGS{Reitze:2019,
       author = {{Reitze}, David and {Adhikari}, Rana X. and {Ballmer}, Stefan and {Barish}, Barry and {Barsotti}, Lisa and {Billingsley}, GariLynn and {Brown}, Duncan A. and {Chen}, Yanbei and {Coyne}, Dennis and {Eisenstein}, Robert and {Evans}, Matthew and {Fritschel}, Peter and {Hall}, Evan D. and {Lazzarini}, Albert and {Lovelace}, Geoffrey and {Read}, Jocelyn and {Sathyaprakash}, B.~S. and {Shoemaker}, David and {Smith}, Joshua and {Torrie}, Calum and {Vitale}, Salvatore and {Weiss}, Rainer and {Wipf}, Christopher and {Zucker}, Michael},
        title = "{Cosmic Explorer: The U.S. Contribution to Gravitational-Wave Astronomy beyond LIGO}",
    booktitle = {Bulletin of the American Astronomical Society},
         year = 2019,
       volume = {51},
        month = sep,
          eid = {35},
        pages = {35},
          doi = {10.48550/arXiv.1907.04833},
archivePrefix = {arXiv},
       eprint = {1907.04833},
 primaryClass = {astro-ph.IM},
       adsurl = {https://ui.adsabs.harvard.edu/abs/2019BAAS...51g..35R}
}

@ARTICLE{Rezzolla:2018,
       author = {{Rezzolla}, Luciano and {Most}, Elias R. and {Weih}, Lukas R.},
        title = "{Using Gravitational-wave Observations and Quasi-universal Relations to Constrain the Maximum Mass of Neutron Stars}",
      journal = {\apjl},
         year = 2018,
        month = jan,
       volume = {852},
       number = {2},
          eid = {L25},
        pages = {L25},
          doi = {10.3847/2041-8213/aaa401},
archivePrefix = {arXiv},
       eprint = {1711.00314},
 primaryClass = {astro-ph.HE}
}

@ARTICLE{Rigault:2025,
       author = {{Rigault}, M. and {Smith}, M. and {Goobar}, A. and {Maguire}, K. and {Dimitriadis}, G. and {Johansson}, J. and {Nordin}, J. and {Burgaz}, U. and {Dhawan}, S. and {Sollerman}, J. and {Regnault}, N. and {Kowalski}, M. and {Nugent}, P. and {Andreoni}, I. and {Amenouche}, M. and {Aubert}, M. and {Barjou-Delayre}, C. and {Bautista}, J. and {Bellm}, E. and {Betoule}, M. },
        title = "{ZTF SN Ia DR2: Overview}",
      journal = {\aap},
         year = 2025,
        month = feb,
       volume = {694},
          eid = {A1},
        pages = {A1},
          doi = {10.1051/0004-6361/202450388},
archivePrefix = {arXiv},
       eprint = {2409.04346},
 primaryClass = {astro-ph.CO}
}

@ARTICLE{Rigault:2025b,
       author = {{Rigault}, M. and {Smith}, M. and {Regnault}, N. and {Kenworthy}, W.~D. and {Maguire}, K. and {Goobar}, A. and {Dimitriadis}, G. and {Johansson}, J. and {Amenouche}, M. and {Aubert}, M. and {Barjou-Delayre}, C. and {Bellm}, E.~C. and {Burgaz}, U. and {Carreres}, B. and {Copin}, Y. and {Deckers}, M. and {de Jaeger}, T. and {Dhawan}, S. and {Feinstein}, F. and {Fouchez}, D. and {Galbany}, L. and {Ginolin}, M. and {Graham}, M.~J. and {Kim}, Y.-L. and {Kowalski}, M. and {Kuhn}, D. and {Kulkarni}, S.~R. and {M{\"u}ller-Bravo}, T.~E. and {Nordin}, J. and {Popovic}, B. and {Purdum}, J. and {Rosnet}, P. and {Rosselli}, D. and {Racine}, B. and {Ruppin}, F. and {Sollerman}, J. and {Terwel}, J.~H. and {Townsend}, A.},
        title = "{ZTF SN Ia DR2: Study of Type Ia supernova light-curve fits}",
      journal = {\aap},
         year = 2025,
        month = feb,
       volume = {694},
          eid = {A2},
        pages = {A2},
          doi = {10.1051/0004-6361/202450377},
archivePrefix = {arXiv},
       eprint = {2406.02073},
 primaryClass = {astro-ph.CO},
}

@ARTICLE{Riley:2021,
       author = {{Riley}, Thomas E. and {Watts}, Anna L. and {Ray}, Paul S. and {Bogdanov}, Slavko and {Guillot}, Sebastien and {Morsink}, Sharon M. and {Bilous}, Anna V. and {Arzoumanian}, Zaven and {Choudhury}, Devarshi and {Deneva}, Julia S. and {Gendreau}, Keith C. and {Harding}, Alice K. and {Ho}, Wynn C.~G. and {Lattimer}, James M. and {Loewenstein}, Michael and {Ludlam}, Renee M. and {Markwardt}, Craig B. and {Okajima}, Takashi and {Prescod-Weinstein}, Chanda and {Remillard}, Ronald A. and {Wolff}, Michael T. and {Fonseca}, Emmanuel and {Cromartie}, H. Thankful and {Kerr}, Matthew and {Pennucci}, Timothy T. and {Parthasarathy}, Aditya and {Ransom}, Scott and {Stairs}, Ingrid and {Guillemot}, Lucas and {Cognard}, Ismael},
        title = "{A NICER View of the Massive Pulsar PSR J0740+6620 Informed by Radio Timing and XMM-Newton Spectroscopy}",
      journal = {\apjl},
         year = 2021,
        month = sep,
       volume = {918},
       number = {2},
          eid = {L27},
        pages = {L27},
          doi = {10.3847/2041-8213/ac0a81},
archivePrefix = {arXiv},
       eprint = {2105.06980},
 primaryClass = {astro-ph.HE}
}

@article{Romero_Shaw:2020,
       author = {{Romero-Shaw}, I.~M. and {Talbot}, C. and {Biscoveanu}, S. and {D'Emilio}, V. and {Ashton}, G. and {Berry}, C.~P.~L. and {Coughlin}, S. and {Galaudage}, S. and {Hoy}, C. and {H{\"u}bner}, M. and {Phukon}, K.~S. and {Pitkin}, M. and {Rizzo}, M. and {Sarin}, N. and {Smith}, R. and {Stevenson}, S. and {Vajpeyi}, A. and {Ar{\`e}ne}, M. and {Athar}, K. and {Banagiri}, S. and {Bose}, N. and {Carney}, M. and {Chatziioannou}, K. and {Clark}, J.~A. and {Colleoni}, M. and {Cotesta}, R. and {Edelman}, B. and {Estell{\'e}s}, H. and {Garc{\'\i}a-Quir{\'o}s}, C. and {Ghosh}, Abhirup and {Green}, R. and {Haster}, C.-J. and {Husa}, S. and {Keitel}, D. and {Kim}, A.~X. and {Hernandez-Vivanco}, F. and {Maga{\~n}a Hernandez}, I. and {Karathanasis}, C. and {Lasky}, P.~D. and {De Lillo}, N. and {Lower}, M.~E. and {Macleod}, D. and {Mateu-Lucena}, M. and {Miller}, A. and {Millhouse}, M. and {Morisaki}, S. and {Oh}, S.~H. and {Ossokine}, S. and {Payne}, E. and {Powell}, J. and {Pratten}, G. and {P{\"u}rrer}, M. and {Ramos-Buades}, A. and {Raymond}, V. and {Thrane}, E. and {Veitch}, J. and {Williams}, D. and {Williams}, M.~J. and {Xiao}, L.},
        title = "{Bayesian inference for compact binary coalescences with BILBY: validation and application to the first LIGO-Virgo gravitational-wave transient catalogue}",
      journal = {\mnras},
         year = 2020,
        month = dec,
       volume = {499},
       number = {3},
        pages = {3295-3319},
          doi = {10.1093/mnras/staa2850},
archivePrefix = {arXiv},
       eprint = {2006.00714},
 primaryClass = {astro-ph.IM}
}

@ARTICLE{Rose:2023,
       author = {{Rose}, Henrik and {Kunert}, Nina and {Dietrich}, Tim and {Pang}, Peter T.~H. and {Smith}, Rory and {Van Den Broeck}, Chris and {Gandolfi}, Stefano and {Tews}, Ingo},
        title = "{Revealing the strength of three-nucleon interactions with the proposed Einstein Telescope}",
      journal = {\prc},
         year = 2023,
        month = aug,
       volume = {108},
       number = {2},
          eid = {025811},
        pages = {025811},
          doi = {10.1103/PhysRevC.108.025811},
archivePrefix = {arXiv},
       eprint = {2303.11201},
 primaryClass = {astro-ph.HE}
}

@ARTICLE{Ryan:2025,
       author = {{Ryan}, Geoffrey and {van Eerten}, Hendrik and {Piro}, Luigi and {Troja}, Eleonora and {O'Connor}, Brendan and {Ricci}, Roberto},
        title = "{afterglowpy: Compute and fit GRB afterglows}",
      journal = {Astrophysics Source Code Library},
         year = 2025,
        month = may,
       eprint = {2505.014},
archivePrefix = {ascl}
}

@ARTICLE{Ryan:2020,
       author = {{Ryan}, Geoffrey and {van Eerten}, Hendrik and {Piro}, Luigi and {Troja}, Eleonora},
        title = "{Gamma-Ray Burst Afterglows in the Multimessenger Era: Numerical Models and Closure Relations}",
      journal = {\apj},
         year = 2020,
        month = jun,
       volume = {896},
       number = {2},
          eid = {166},
        pages = {166},
          doi = {10.3847/1538-4357/ab93cf},
archivePrefix = {arXiv},
       eprint = {1909.11691},
 primaryClass = {astro-ph.HE}
}

@ARTICLE{Saji:2025,
       author = {{Saji}, J. and {Dainotti}, M.~G. and {Bhardwaj}, S. and {Janiuk}, A.},
        title = "{Exploring jet structure and dynamics in short gamma-ray bursts: A case study on GRB 090510}",
      journal = {\aap},
         year = 2025,
        month = oct,
       volume = {702},
          eid = {A13},
        pages = {A13},
          doi = {10.1051/0004-6361/202554147},
archivePrefix = {arXiv},
       eprint = {2507.14938},
 primaryClass = {astro-ph.HE},
       adsurl = {https://ui.adsabs.harvard.edu/abs/2025A&A...702A..13S}
}

@ARTICLE{Salafia:2021,
       author = {{Salafia}, O.~S. and {Giacomazzo}, B.},
        title = "{Accretion-to-jet energy conversion efficiency in GW170817}",
      journal = {\aap},
         year = 2021,
        month = jan,
       volume = {645},
          eid = {A93},
        pages = {A93},
          doi = {10.1051/0004-6361/202038590},
archivePrefix = {arXiv},
       eprint = {2006.07376},
 primaryClass = {astro-ph.HE},
       adsurl = {https://ui.adsabs.harvard.edu/abs/2021A&A...645A..93S}
}

@ARTICLE{Salafia:2022,
       author = {{Salafia}, Om Sharan and {Ghirlanda}, Giancarlo},
        title = "{The Structure of Gamma Ray Burst Jets}",
      journal = {Galaxies},
         year = 2022,
        month = aug,
       volume = {10},
       number = {5},
          eid = {93},
        pages = {93},
          doi = {10.3390/galaxies10050093},
archivePrefix = {arXiv},
       eprint = {2206.11088},
 primaryClass = {astro-ph.HE},
       adsurl = {https://ui.adsabs.harvard.edu/abs/2022Galax..10...93S}
}

@ARTICLE{Sarin:2024,
       author = {{Sarin}, Nikhil and {H{\"u}bner}, Moritz and {Omand}, Conor M.~B. and {Setzer}, Christian N. and {Schulze}, Steve and {Adhikari}, Naresh and {Sagu{\'e}s-Carracedo}, Ana and {Galaudage}, Shanika and {Wallace}, Wendy F. and {Lamb}, Gavin P. and {Lin}, En-Tzu},
        title = "{REDBACK: a Bayesian inference software package for electromagnetic transients}",
      journal = {\mnras},
         year = 2024,
        month = jun,
       volume = {531},
       number = {1},
        pages = {1203-1227},
          doi = {10.1093/mnras/stae1238},
archivePrefix = {arXiv},
       eprint = {2308.12806},
 primaryClass = {astro-ph.HE}
}

@ARTICLE{Setzer:2023,
       author = {{Setzer}, Christian N. and {Peiris}, Hiranya V. and {Korobkin}, Oleg and {Rosswog}, Stephan},
        title = "{Modelling populations of kilonovae}",
      journal = {\mnras},
         year = 2023,
        month = apr,
       volume = {520},
       number = {2},
        pages = {2829-2842},
          doi = {10.1093/mnras/stad257},
archivePrefix = {arXiv},
       eprint = {2205.12286},
 primaryClass = {astro-ph.HE}
}

@ARTICLE{Shamohammadi:2023,
       author = {{Shamohammadi}, M. and {Bailes}, M. and {Freire}, P.~C.~C. and {Parthasarathy}, A. and {Reardon}, D.~J. and {Shannon}, R.~M. and {Venkatraman Krishnan}, V. and {Bernadich}, M.~C. i. and {Cameron}, A.~D. and {Champion}, D.~J. and {Corongiu}, A. and {Flynn}, C. and {Geyer}, M. and {Kramer}, M. and {Miles}, M.~T. and {Possenti}, A. and {Spiewak}, R.},
        title = "{Searches for Shapiro delay in seven binary pulsars using the MeerKAT telescope}",
      journal = {\mnras},
         year = 2023,
        month = apr,
       volume = {520},
       number = {2},
        pages = {1789-1806},
          doi = {10.1093/mnras/stac3719},
archivePrefix = {arXiv},
       eprint = {2212.04051},
 primaryClass = {astro-ph.HE}
}

@ARTICLE{Shibata:2019,
       author = {{Shibata}, Masaru and {Hotokezaka}, Kenta},
        title = "{Merger and Mass Ejection of Neutron Star Binaries}",
      journal = {Annual Review of Nuclear and Particle Science},
         year = 2019,
        month = oct,
       volume = {69},
        pages = {41-64},
          doi = {10.1146/annurev-nucl-101918-023625},
archivePrefix = {arXiv},
       eprint = {1908.02350},
 primaryClass = {astro-ph.HE}
}

@ARTICLE{Siegel:2019,
       author = {{Siegel}, Daniel M.},
        title = "{GW170817 -the first observed neutron star merger and its kilonova: Implications for the astrophysical site of the r-process}",
      journal = {European Physical Journal A},
         year = 2019,
        month = nov,
       volume = {55},
       number = {11},
          eid = {203},
        pages = {203},
          doi = {10.1140/epja/i2019-12888-9},
archivePrefix = {arXiv},
       eprint = {1901.09044},
 primaryClass = {astro-ph.HE}
}

@ARTICLE{Siegel:2013,
       author = {{Siegel}, Daniel M. and {Ciolfi}, Riccardo and {Harte}, Abraham I. and {Rezzolla}, Luciano},
        title = "{Magnetorotational instability in relativistic hypermassive neutron stars}",
      journal = {\prd},
         year = 2013,
        month = jun,
       volume = {87},
       number = {12},
          eid = {121302},
        pages = {121302},
          doi = {10.1103/PhysRevD.87.121302},
archivePrefix = {arXiv},
       eprint = {1302.4368},
 primaryClass = {gr-qc}
}

@incollection{Skilling:2004,
       author = {{Skilling}, John},
        title = "{Nested Sampling}",
    booktitle = {Bayesian Inference and Maximum Entropy Methods in Science and Engineering},
         year = 2004,
       editor = {{Fischer}, Rainer and {Preuss}, Roland and {Toussaint}, Udo Von},
       volume = {735},
        month = nov,
    publisher = {AIP},
        pages = {395-405},
          doi = {10.1063/1.1835238}
}

@ARTICLE{Smith:2020,
       author = {{Smith}, Rory J.~E. and {Ashton}, Gregory and {Vajpeyi}, Avi and {Talbot}, Colm},
        title = "{Massively parallel Bayesian inference for transient gravitational-wave astronomy}",
      journal = {\mnras},
         year = 2020,
        month = nov,
       volume = {498},
       number = {3},
        pages = {4492-4502},
          doi = {10.1093/mnras/staa2483},
archivePrefix = {arXiv},
       eprint = {1909.11873},
 primaryClass = {gr-qc}
}

\appendix
\nolinenumbers
\crefname{section}{Appendix}{Appendices}
\Crefname{section}{Appendix}{Appendices}

\section{Implementation Details}
\label{app:software_upgrades}
Development was driven overall by considerations of reusability, tighter interlocking with \bilby\ and the parallelisation requirements (\Cref{ssec:parallelisation}) of full multimessenger analyses.
The \nmma\ package now comprises a new \texttt{core} module (\Cref{sssec:core}) for shared functionalities.
They find usage in dedicated modules for messenger-specific applications (\Cref{ssec:em_module,ssec:eos_module,ssec:gw_module}).
They have seen different levels of refactoring over previous versions in order to streamline their interplay for multimessenger studies in the \texttt{joint} module (\Cref{ssec:joint_module}).

\subsection{Parallelisation}
\label{ssec:parallelisation}
Bayesian inference typically requires the evaluation of millions of test samples.
In nested sampling, the parameter space is explored using a suite of "live points" that are ranked by their likelihood.
In each iteration step, the likelihood function is evaluated at random points in the parameter space until the sampler finds a new point with higher likelihood than the current lowest-ranked live point.
The number of required test points can grows rapidly with the dimensionality of the underlying parameter space, making parallelisation mandatory in order to keep analyses tractable.
Parallelised nested sampling usually employs a master-worker architecture, where the workers concurrently suggest replacements for the lowest-ranked live point to the master process.
This leverages the fact that despite nested sampling being inherently sequential, the search for live points is approximately independent and can therefore run in parallel.
"Approximately independent" refers to the fact that a suggested live point only improves upon the lowest-ranked live point at the time the search started.
The margin for acceptance, however, increases as the master sequentially evaluates the returned suggestions.
If the likelihood of a returned point falls below this new margin, the master rejects it and the worker's efforts were effectively wasted.
The parallelisation loss depends on the ratio between the number of workers and live points.
It remains on an acceptable level as long as $n_{\rm workers} \lesssim n_{\rm live}$~\citep{Handley:2015}.

The available parallelisation capacities of \bilby\ depend on the sampler in use.
Its preferred nested sampling framework is the popular and robust \textsc{dynesty}~\citep{Speagle:2020}.
While \textsc{PyMultinest}~\citep{Feroz:2019,Buchner:2014} or \textsc{Ultranest}~\citep{Buchner:2021} only support MPI-based parallelisation, the \bilby-implemented \textsc{dynesty} interface can only be parallelised through \textsc{python}'s \textsc{multiprocessing} module on a single compute node.
However, some of the emulator-based models available in \nmma, for instance its \fiesta\ interface, can suffer from deadlocks in multiprocessing.
Additionally, the resource demands of full multimessenger analyses still warrant usage of parallel computing capacities across nodes, especially in time-critical situations.

\citet{Smith:2020} have developed \pbilby\ as an MPI-parallelised interface that allows using \bilby\ with \textsc{dynesty} on cluster-scale, inspiring the original implementation of \nmma.
However, it was introduced when the multiprocessing capability had not yet been implemented in \bilby. 
This, together with the introduction of more efficient waveform approximants, has alleviated the need for \pbilby\ in GW-only analyses.
Development has consequently stalled, limiting \nmma's ability to adapt to and incorporate software advances.
We have therefore reimplemented the MPI-based parallelisation approach of \pbilby with its master-worker architecture that employs \textsc{schwimmbad}'s MPI interface~\citep{Price:2017}.
While the master process wraps \bilby's \textsc{dynesty}-implementation, \texttt{Worker} instances independently initialise the likelihood and the underlying emulators to avoid communication overhead and racing conditions. 
The new version of \nmma\ hence adapts utilities of \pbilby without depending on it.
This allows us to move forward to \bilby's latest version and in extension to exploit significant performance improvements in version 3.0~\citep{Koposov:2025} of our default sampler \textsc{dynesty}.

\subsection{Ejecta Treatment}
\label{ssec:bns_ejecta}
The prime application of \nmma\ remains the Bayesian analysis of compact object mergers, combining GW and EM data alongside nuclear physics information.
The most important intersection of the different messengers occurs by phenomenological relations for ejecta properties.
Models for the EM counterpart take the dynamical ($\mdyn$) and wind ejecta ($\mwind$), among others, as input.
The amount of ejecta produced by a merger in turn depends on properties of the binary.
While most other changes to the code focused on improved performance and extended functionalities, we have made some changes to the ejecta treatment that can have significant impact on inference results.
We therefore elaborate on the relations we use as well as intuitions underlying them and we detail the modifications introduced relative to~\citetalias{Pang:2023}.

The code assumes that compact objects have a disjoint mass spectrum and are of stellar origin, i.e., it does not account for primordial black holes or potential EM counterparts of massive BH mergers~\citep{Bogdanovic:2022}.
When both objects are BHs, we do not expect any ejecta and hence no transient.

For the dynamical ejecta, \nmma\ applies empirical relations that were suggested by \citet{Krüger:2020}.
If only the secondary is an NS and the primary is a BH, they express $\mdyn$ in units of the NS's baryonic mass as \begin{align}
    \frac{\mdyn}{M_{\rm NS}^b}= a_1\, q^{n_1} \frac{1-2C_{\rm NS}}{C_{\rm NS}} - a_2\, q^{n_2}\, r_{\rm ISCO}(\chi_{\rm BH}) + a_4 \label{eq:m_dyn_bhns}
\end{align}
with the coefficients $a_1 =0.007116 $, $a_2=0.001436$, $a_4= -0.02762$, $n_1=-0.2497$ and $n_2 =-1.352$.
We remark that the \nmma\  convention $q=\frac{M_2}{M_1}\leq1$ requires inverting the signs of $n_1$ and $n_2$ against \citet{Krüger:2020}.
The functional form modifies an ansatz of~\citet{Kawaguchi:2016} to satisfy the physical expectation of a reduced $\mdyn$ from very compact NSs.
In this expression, $r_{\rm ISCO}$ denotes the innermost stable circular orbit around the BH in units of its mass and depends only on $\chi_{\rm BH}$, the dimensionless BH spin aligned with the orbital angular momentum~\citep{Bardeen:1972}.
The baryonic mass, sometimes also called rest mass~\citep{Lattimer:2021}, is the hypothetical mass content of the neutron star if it was spread out infinitely.
The measurable gravitational mass $M$ is reduced due to the binding energy of its constituents.
We follow \citet{Lattimer:2001} to conveniently approximate their ratio as \begin{align}
    \frac{M^b}{M} = 1 + \frac{6}{5}\frac{C}{2-C}. \label{eq:bar_to_grav_mass}
\end{align}

When dealing with a BNS instead, \citet{Krüger:2020} provide the relation \begin{align}
    \frac{\mdyn}{\qty{e-3}{\msun}} = \left( \frac{a}{C_1} + b \left(\frac{M_2}{M_1}\right)^n + cC_1 \right) M_1 + ( 1\longleftrightarrow 2) \label{eq:m_dyn_bns}
\end{align}
with the optimal coefficients $a= -9.3335$, $b=114.17$, $c = -337.56$, $n = 1.5465$, again found from fitting to NR simulations.
This relation naturally encompasses our expectations that more asymmetric systems should eject more mass dynamically, while an increased compactness reduces the amount of ejecta. 
In case either expression for $\mdyn$ returns negative values, we assume the absence of a dynamical ejecta component. 

In absence of feasible models of the wind ejection mechanism, we introduce a fudge factor $\zeta$ to obtain $\mwind=\zeta \, \mdisk$.
For the disk mass $\mdisk$ itself, there are various phenomenological relations available~\citep{Radice:2018,Foucart:2018,Krüger:2020,Dietrich:2020,Lund:2025}.
To model the massive disk in a BH-NS merger, we adopt the prescription of~\citet{Foucart:2018}, according to which \begin{align}
    \frac{M_{\rm ej}}{M_{\rm NS}^b} &= \left( a \frac{1-2C_{\rm NS}}{\eta^{1/3}} - b\ r_{\rm ISCO}(\chi_{\rm BH}) \frac{C_{\rm NS}}{\eta} + c \right) ^d
    \label{eq:mej_bhns}
\end{align}
describes the total disk mass in units of the baryonic NS mass.
In this expression, $\eta = q/(1+q)^2$ is the symmetric mass ratio and the optimal coefficients are $a=0.406$, $b=0.139$, $c=0.255$ and $d=1.761$.
While $c$ and $d$ account for unmodelled non-linearities, the first and second term satisfy our intuition that significant ejecta are only expected as tidal disruption occurs well before the NS approaches $r_{\rm ISCO}$ and that less compact NSs are easier to disrupt.
We again use \cref{eq:bar_to_grav_mass} to express the estimate in terms of our assumed gravitational mass and then combine this result with \cref{eq:m_dyn_bhns} to set $\mdisk = M_{\rm ej} - \mdyn$.
If \cref{eq:mej_bhns} yields negative results, we again set all ejecta masses to 0.

For $\mdisk$ in BNS mergers, \nmma\ varies the approach of~\citet{Dietrich:2020} which sets \begin{align}
    \log_{10}\frac{\mdisk}{\msun} = a(q) \left(b(q) \tanh\left(\frac{c-(M_1+M_2)/M_{pc}}{d} \right) +1 \right) \label{eq:disk_mass}
\end{align}
above a minimum disk mass of \qty{e-3}{\msun}.
This functional, inspired by~\citet{Radice:2018,Coughlin:2019}, allows a smooth transition between the regime of massive disks around \qty{e-1}{\msun}, seen in simulations of low-mass mergers, and the residual disk masses found in simulations where the remnant experiences prompt collapse.
It differs from \citet{Coughlin:2019} by introducing a dependence on $q$ in the parameters $a$ and $b$ as \begin{align}
    a(q) &= a_0 + a_1\,\hat{\beta}(q) \quad \text{ and}\\
    b(q) &= b_0 + b_1\,\hat{\beta}(q)\quad \text{ with}\\
    \hat{\beta}(q) &= \frac{1}{2}\tanh(\beta (q- q_0)).
\end{align}
This prescription reproduces the high $\mdisk$ seen in asymmetric binaries, but increases the number of fit parameters.
The best fit additionally depends on the threshold $M_{pc}$ where we expect prompt collapse.
It naturally relates to the EoS through $\mtov$ and is commonly parametrised as $M_{pc} = k \, \mtov$ with $k\sim 1.3-1.7$~\citep{Bauswein:2013,Agathos:2020,Metzger:2019}.
We can parametrise $k$ in dependence of the EoS, too.
\citet{Dietrich:2020} followed \citet{Agathos:2020} to employ \begin{align}
    k = 2.392-3.29 C_{\rm TOV}. \label{eq:k_thresh_old}
\end{align}
Here, $C_{\rm TOV}$ is the compactness of an NS at the TOV point.
This prescription was also used in \citetalias{Pang:2023}, although, because of a version mismatch, it claimed to have used an alternative method to determine $k$.
This alternative method uses a dependence on $C^*= \frac{G \mtov}{c^2 R_{\text{1.6}}}$, where $R_{\text{1.6}}$ is the radius of a \qty{1.6}{\msun} NS.
From \citet{Bauswein:2013}, one can then determine
\begin{align}
    k = 2.380- 3.606C^*. \label{eq:k_thresh_new}
\end{align}
Only in this work we actually use this latter expression to fit the free parameters in \cref{eq:disk_mass} to the same compilation of NR simulations~\citep{Hotokezaka:2011,Dietrich:2017b,Radice:2018,Kiuchi:2018}, resulting in $a_0=-1.725$, $a_1=-2.337$, $b_0=-0.564$, $b_1= -0.437$, $c=0.958$, $d=0.057$, $\beta=5.879$ and $q_0=0.886$.

Moreover, the disk mass estimates impact the GRB energies predicted in \nmma.
Given the currently insufficient understanding of the GRB mechanism, \nmma\ makes the minimal assumption that the disk serves as the principal energy reservoir of short GRBs~\citep{Blandford:1977, Salafia:2021, Combi:2023, Hayashi:2025, Kalinani:2026}, implying that the total kinetic energy of the jet is a fraction $\epsilon$ of the remaining disk mass (asides from a contribution that goes into the prompt gamma-ray emission)
\begin{align}
    E_{\rm jet} = \epsilon \, (1-\zeta) \mdisk.
    \label{eq:jet_energy_from_disk}
\end{align} 
This kinetic energy is anisotropic, but axisymmetric and thus follows an angular distribution
\begin{align}
    E_{\rm jet} = \int \frac{\diff E}{\diff\Omega}(\theta, \varphi)\ \diff\Omega = 2\pi \int \frac{\diff E}{\diff \Omega}(\theta)\ \sin(\theta) \diff\theta .
\end{align}
Several profiles $\diff E/\diff\Omega$ have been suggested in the literature from hydrodynamic simulations~\citep{Ryan:2020, Salafia:2022, Saji:2025}.
Popular choices that are implemented in \afgpy\ assume $\diff E/\diff\Omega = \varepsilon_c\ w(\theta)$ with
\begin{align}
    w(\theta) &= H(\theta_c-\theta) \qquad &\text{(tophat)}, \\
    w(\theta) &= \exp \left(\frac{-\theta^2}{2\theta_c^2}\right)\ H(\theta_w-\theta) \qquad \quad &\text{(Gaussian)}, \label{eq:gaussian_grb}\\
    w(\theta) &= \left(1 + \frac{\theta^2}{b\theta_c^2}\right)^{-b/2}\ H(\theta_w-\theta) \qquad \quad &\text{(power-law)} . \label{eq:powerlaw_grb}
\end{align} 
Here, $H(x)$ is the Heaviside step function, $\theta_c$ is the jet core opening angle as the characteristic angular scale and $\theta_w$ is the wing angle of the jet that serves as its boundary.
\Cref{eq:powerlaw_grb} introduces an additional parameter for the power-law index $b$.
The most important input parameter for most afterglow models is the central isotropic energy-equivalent
\begin{align}
    \eiso = 4\pi \frac{\diff E}{\diff\Omega}_{|\theta=0} = 4\pi \varepsilon_c.
\end{align}
Therefore, in \nmma, we use \cref{eq:jet_energy_from_disk} to then solve 
\begin{align}
    E_{\rm jet} = \frac{\eiso}{2} \int_0^{\theta_w} w(\theta) \, \sin(\theta) \diff\theta
    \label{eq:eiso_conversion}
\end{align}
for $\eiso$.
In previous versions, \nmma\ imprecisely used 
\begin{align}
    \eiso= \epsilon\, (1-\zeta)\mdisk \label{eq:old_E0}
\end{align} to absorb these connections in a conversion factor $\epsilon$.
This choice was problematic, though, because for small values of $\theta_c$ the integral in \cref{eq:eiso_conversion} remains (very) small and $\eiso$ can consequently far exceed $E_{\rm jet}$.
Since $\epsilon$ was and is supposed to represent an energy conversion factor, its value is bounded between 0 and 1.
High, but physically permissible values of $\eiso$ hence became suppressed by the old prescription.
Now, $\epsilon$ accurately represents the energy conversion from disk to kinetic jet energy.
We note that within the code, due to \afgpy\ nomenclature, $\eiso$ is actually named $\mathtt{E0}$.

\subsection{The \texttt{core} module}
\label{sssec:core}
We have introduced a \texttt{core} module as the shared backbone of any messenger-specific applications. 
Most importantly, it defines the \texttt{NMMALikelihood}-class from which other likelihoods in \nmma\ inherit.
While we mostly align with the \bilby\ design philosophy, we deviate from it by no longer holding priors and likelihoods structurally separate.
In \bilby, they only interact through the respective sampling algorithm which draws samples from the prior, eventually checks them for known constraints and then evaluates the likelihood at these points.
This implicitly assumes that constraints apply only to some equivalent parameter set that is computationally cheap to obtain and logically connected to the sampling parameters.
For example, in the context of GW analysis, it is indeed beneficial to sample the more precisely measurable parameters chirp mass $\mathcal{M}$ and mass ratio $q$ instead of the component masses, while enforcing constraints on the latter keeps them within physical bounds.
These assumptions do not hold in \nmma: 
When we use emulators to map nuclear parameters to macroscopic NS properties, they introduce significant computational burden comparable to the likelihood call itself, warranting only a single evaluation.
Additionally, we treat \texttt{Constraint}-objects in a \texttt{PriorDict} simply as a convenient interface to easily adjust bounds on arbitrary parameters, even if they might logically rather relate to the likelihood than the sampling parameters.
Consequently, the \texttt{NMMALikelihood}-class handles parameter conversion and constraint evaluation and defines a common interface for messenger-specific likelihood-objects.

To initialise an \texttt{NMMALikelihood}, we require a sub-model with a \texttt{log\_likelihood}-method and a \bilby-style dictionary of priors.
In the GW-sector, for instance, the sub-model is a \bilby-\texttt{GravitationalWaveTransient}, while the \texttt{EMTransient}-class provides methods to generate lightcurves. 
Moreover, methods to visualise inference results or to do additional parameter conversion in the parent likelihood frequently wrap around corresponding methods in the sub-model.

Additionally, the \texttt{core} module contains utilities to handle various parameter conversion routines, read data files and parse arguments from the command line or \texttt{yaml} configuration files.
These are particularly useful to set up multiple similar analyses that differ only in a few aspects, e.g. the data files or prior ranges.

\subsection{The \texttt{em} module}
\label{ssec:em_module}
The \texttt{em} module contains all functionality required to read, process and model transients across the electromagnetic spectrum. 
It has seen the largest overhaul compared to previous versions of \nmma.
Although backward compatibility has been largely maintained, we note that various defaults and argument names have changed.
The \texttt{EMTransient}-class provides a common interface to handle lightcurve data and to evaluate model lightcurves by comparing a synthetic lightcurve to (real or injected) observations, taking observational and modelling uncertainties into account.
For a given set of parameters, we obtain these lightcurves for a set of sampling times by evaluating either simplified (semi-)analytical expressions or ML surrogates that were trained in advance on grids of precomputed lightcurves. 
The currently supported transient types are KNe, supernovae, GRB afterglows and combinations thereof, partly wrapping around respective implementations in \fiesta~\citep{Koehn:2025b}, \afgpy~\citep{Ryan:2020,Ryan:2025} and \textsc{sncosmo}~\citep{Barbary:2016}, but the modularised architecture makes it easy to add new models or transient types.

To that end, we provide a convenient interface through the \texttt{LightCurveModelContainer}-class to which users only need to add a method that computes lightcurves in AB magnitudes. 
Previously, these needed to be absolute magnitudes, whereas \nmma\ now preferentially processes lightcurves in apparent magnitude.
This relates to the presumably most important modification over previous code versions:
In order to compare a synthetic lightcurve to observations, we interpolate the model lightcurve to the observing times in the respective filters.
Previously, the sampling times for model lightcurves were set in the observer's frame. 
The code would compute absolute magnitudes at redshift-correct sampling times and then transform them to apparent magnitudes.
However, every lightcurve model is only valid within a characteristic time range.
It requires special care on the user's side to not evaluate the model outside its validity range.
We now set sampling times by default to the model's range of validity in the source frame and return apparent magnitudes at the corresponding observer-frame times.
Consequently, the exact values of model lightcurves will differ slightly from previous versions, especially for high-redshift sources.

Another noteworthy change concerns the treatment of systematic uncertainties. 
\nmma\ can employ constant or variable, time-dependent or time-independent systematic errors $\sigma_{\rm sys}$ to account for modelling uncertainties~\citep{Jhawar:2025}. 
We have introduced a \texttt{SystematicsHandler}-class to subsume all relevant functionality in a unified way with a simple interface to set up corresponding priors.

\subsection{The \texttt{eos} module}
\label{ssec:eos_module}
Significant code expansions handling nuclear parameters comprise the \texttt{eos} module.
We have implemented a new \texttt{EoSGenerator}-class to provide an easily customisable interface to appropriately pre-trained emulators that map nuclear parameters to macroscopic NS properties.
At its core, any subclass only needs to define a method \texttt{emulate\_macro\_eos} to correctly call the underlying emulator and an \texttt{adjust\_format}-method that transforms the emulator output to a table-like format of equal-length arrays for $M-R-\Lambda$.
For demonstration purposes of this work, we have trained a minimal example following~\citep{Reed:2024} that we discuss in~\cref{ssec:emulation_strategy}.

Calls to the new \texttt{EoSConverter}-class then use either these generators, sets of tabulated EoSs as previously implemented or simple quasi-universal relations to compute characteristic radius, mass or \textLambda\ values.
An \texttt{EquationofStateLikelihood}, employing a combination of various mass(-radius) measurements as its sub-model, provides means to evaluate the likelihood of a given EoS or pre-determine weights for tabulated EoS sets.
Users can easily extend it to include new types of constraints based on theoretical, experimental or observational constraints.

\subsection{The \texttt{gw} module}
\label{ssec:gw_module}
Contrary to the various extensions for other messengers, we have cut many previous specifications to handle GW information.
The \texttt{gw} module has instead become a thin wrapper around \bilby's GW functionality to more easily adapt to future developments by the larger GW community.
This allows us to handle more efficient methods that were previously not available due to dependency constraints, including relative binning~\citep{Zackay:2018} and multibanding~\citep{Morisaki:2021}.

\subsection{The \texttt{joint} module}
\label{ssec:joint_module}
The \texttt{joint} module combines these functionalities to explore the full multimessenger picture.
While multimessenger events in the GW era remain rare, detailed injection studies offer means to assess the capabilities of future detectors and to optimise search strategies.
For this purpose, \nmma\ adapts the injection handling of \bilpipe~\citep{Romero_Shaw:2020} in the \texttt{NMMAInjectionCreator}-class, sampling from a \bilby-\texttt{PriorDict} to generate a large set of injection parameters.
This allows users to easily create or extend populations of signals across all implemented messengers, including various options to apply quality cuts, for instance for GW signals to meet a given SNR threshold or for EM transients that should exceed a telescope's limiting magnitude to be clearly detectable.

Finally, we have replaced the rather rigid likelihood classes for joint analyses of specific astrophysical sources with a generic \texttt{MultiMessengerLikelihood} that can handle injected or real data from arbitrary messengers.
To simplify usage on clusters where jobs often start only after long wait times, we detach the error-prone setup process and handle it with a re-implementation of \bilpipe's \texttt{DataGenerationInput}.


\section{TOV Emulation Strategy}
\label{ssec:emulation_strategy}
In this work, we use an emulator based on~\citet{Reed:2024} who employed a frequently used decomposition of the energy per nucleon $E$ into a symmetric part $\mathcal{E}$ for matter containing equal amounts of protons and neutrons and a symmetry energy term $S$. 
When the number density $n$ is close to the nuclear saturation density $n_{\rm sat}$, we can apply a threefold Taylor expansion, here truncated at second order, in terms of a density parameter $x = \frac{n-n_{\rm sat}}{3n_{\rm sat}}$ and the asymmetry parameter $\delta = \frac{n_n - n_p}{n}$, where $n_n$ and $n_p$ are the neutron and proton number densities, respectively:
\begin{align}
E(n, \delta) &= \mathcal{E}(n) + S(n) \delta^2 + \ldots \label{eq:energy_decomposition}\\
\mathcal{E}(n) &= E_{\rm sat} + \frac{1}{2} K_{\rm sat} x^2 + \ldots \label{eq:symmetric_energy}\\
S(n) &= S_{\rm sat} + L_{\rm sym} x + \frac{1}{2} K_{\rm sym} x^2 + \ldots \label{eq:symmetry_energy}
\end{align}
As the strong force is isospin-symmetric, all odd powers of $\delta$ must vanish in~\cref{eq:energy_decomposition}.
The empirical fact that nuclear saturation occurs requires the linear term in $x$ to vanish in~\cref{eq:symmetric_energy}, too, while further odd powers appear in a higher-order expansion.
Hence, we only have a linear term in the Taylor expansion of~\cref{eq:symmetry_energy}, the slope parameter $L_{\rm sym}$.
At higher densities, the Taylor expansion breaks down and we need to introduce additional parameters to capture the behaviour of the EoS.
The approach of \citet{Reed:2024} is to introduce the parameters $c_{\rm s,3nsat}^2$ and $c_{\rm s,5nsat}^2$ that characterise the speed of sound squared at \qty{3}{\nsat} and \qty{5}{\nsat}, respectively.
An EoS is then computed assuming that \cref{eq:energy_decomposition,eq:symmetric_energy,eq:symmetry_energy} hold up to \qty{2}{\nsat} and that the speed of sound was constant between \qty{3}{\nsat} and \qty{4}{\nsat} as well as above \qty{5}{\nsat}, while linearly interpolating $c_s^2$ in the ranges \qtyrange{2}{3}{\nsat} and \qtyrange{4}{5}{\nsat}, respectively. 
Fixing $E_{\rm sat}=\qty{-16}{\mega\electronvolt}$, $E_{\rm sym}= \qty{32}{MeV}$ and $n_{\rm sat} = \qty{0.16}{\per\cubic\femto\meter}$ in pure neutron matter ($\delta=1$), \citet{Reed:2024} have computed about 270\,000 EoSs varying the remaining five parameters within the ranges given in \cref{tab:new_parameters} and calculated the corresponding $M-R-\Lambda$ relations.
The training data includes about 70\,000 EoS with $\mtov<\qty{2}{\msun}$ which we reject in pre-processing as we find this to drastically improve the prediction accuracy for realistic EoSs.
However, we then need to reject corresponding predictions in applications.
In this work, we achieve this by enforcing lower limits on $\mtov$. 

In contrast to the emulator of \citet{Reed:2024}, we use a more flexible architecture with variable mass grids:
They had used a fixed grid of 30 masses between 1 and 2 $\msun$ on which they could predict $\Lambda$.
Our mass grids are characterised by a 3-tuple $(N_{\rm low}, N_{\rm high}, M_{\rm split})$.
It comprises $N_{\rm low}$ equally spaced masses up to $M_{\rm split}$, to which we attach $N_{\rm high}$ masses with a different spacing up to the respective EoS's $\mtov$.
The emulator in turn predicts an array of length $2(N_{\rm low} + N_{\rm high})+1$, comprising values for $R$ and $\log \Lambda$ as well as $\mtov$.
The \texttt{adjust\_format}-method of our \texttt{EoSGenerator} then reconstructs for any prediction the corresponding mass grid from $(N_{\rm low}, N_{\rm high}, M_{\rm split})$ and $\mtov$ and maps $\log \Lambda$ to $\Lambda$.

We here use a simple feed-forward neural network and reserve a dedicated analysis of more sophisticated approaches to future work.
We rescale all input parameters to mean zero and unit variance and adopt $N_{\rm low}=25$, $N_{\rm high}=15$ with $M_{\rm split} = \qty{1.8}{\msun}$ for our training, which avoids an additional interpolation error at low masses due to different base lengths, while still flexibly capturing a variable upper mass end.
Our neural network architecture consists of four hidden layers of 256 neurons each with a Swish activation function~\citep{Ramachandran:2017} and a mean-squared-error loss function.

\section{Example Applications}
\label{sec:examples}

\begin{table}[t]
    \caption{
    Additional model parameters and their priors for the example runs.
    We apply the same notations as in \cref{tab:ref_parameters}.
    }
    
    \centering
    \begin{tabular}{lrl}
    \toprule
    \toprule

    Parameter & Symbol & Prior \\

    \midrule
    \multicolumn{3}{c}{SALT2/3 model parameters} \\
    amplitude& $x_0$ & $\log \mathcal{U}[10^{-4}, 10^{1}]$\\
    shape parameter & $x_1$ & $\mathcal{U}[-5, 5]$\\
    color parameter & $c$ & $\mathcal{U}[-0.5, 2.]$\\
    peak time [\unit{d}] & $t_0$ & $\mathcal{U}[5, 20]$\\
    modelling error [mag] & $\sigma_{\text{sys}}$ & $\mathcal{U}[0.1, 1]$\\

    \midrule
    \multicolumn{3}{c}{PSR J0514-4002E timing parameters} \\
     cosine of inclination & $\cos i$ & $\mathcal{U}[0.0, 1.0]$\\
     binary mass function [\unit{\msun}] & $f$ & 
     \makecell[l]{$
        \begin{aligned}
            \mathcal{N}(&0.41672, \\[-4pt]
                        &0.00022)\\
        \end{aligned}
        $}\\
     total mass [\unit{\msun}] & $M_{tot}$ & 
     \makecell[l]{$
        \begin{aligned}
            \mathcal{N}(&3.8870, \\[-4pt]
                        &0.0045)\\
        \end{aligned}
        $}\\
     pulsar mass [\unit{\msun}] & $M_{\rm {PSR}}$ & $\mathcal{U}[1.17, 2.0]$\\
    
    \midrule
    \multicolumn{3}{c}{GW230529 waveform parameters} \\
        primary spin magnitude & $a_1$ & $\mathcal{U}[0, 0.99]$ \\
        secondary spin magnitude & $a_2$ & $\mathcal{U}[0, 0.05]$ \\
        component tilt [rad] & $\theta_i$ & $\sin[0, \pi]$ \\
        relative spin azimuth [rad] & $\Delta\phi$ & $\mathcal{U}[0, 2\pi]$ \\
        spin phase [rad] & $\phi_{JL}$ & $\mathcal{U}[0, 2\pi]$ \\
    
    \end{tabular}
    \label{tab:example_parameters}
\end{table}

To illustrate the capabilities of \nmma\ for broad use cases, we discuss further applications of increasing computational complexity. 
\Cref{tab:example_parameters} summarises additional parameters and their respective priors.
Most examples presented here use \textsc{PyMultinest}~\citep{Feroz:2019,Buchner:2014}.
We emphasise that \nmma\ adopts the \bilby-sampler API, so users can choose from a wide set of alternative samplers. 
While \textsc{PyMultinest} can introduce biases on complex parameter spaces~\citep{Nelson:2020}, this is negligible for the well-behaved examples discussed here.
Nevertheless, we have checked the results with \textsc{dynesty}~\citep{Speagle:2020} and other samplers, too. 
Scripts to run these applications are available online.\footnote{\url{https://github.com/nuclear-multimessenger-astronomy/nmma_1.0_paper}}

\subsection{Model Selection for Supernova Lightcurves}
\label{ex:supernova}

\begin{figure}[t]
    \centering
    \includegraphics[width=0.95\linewidth]{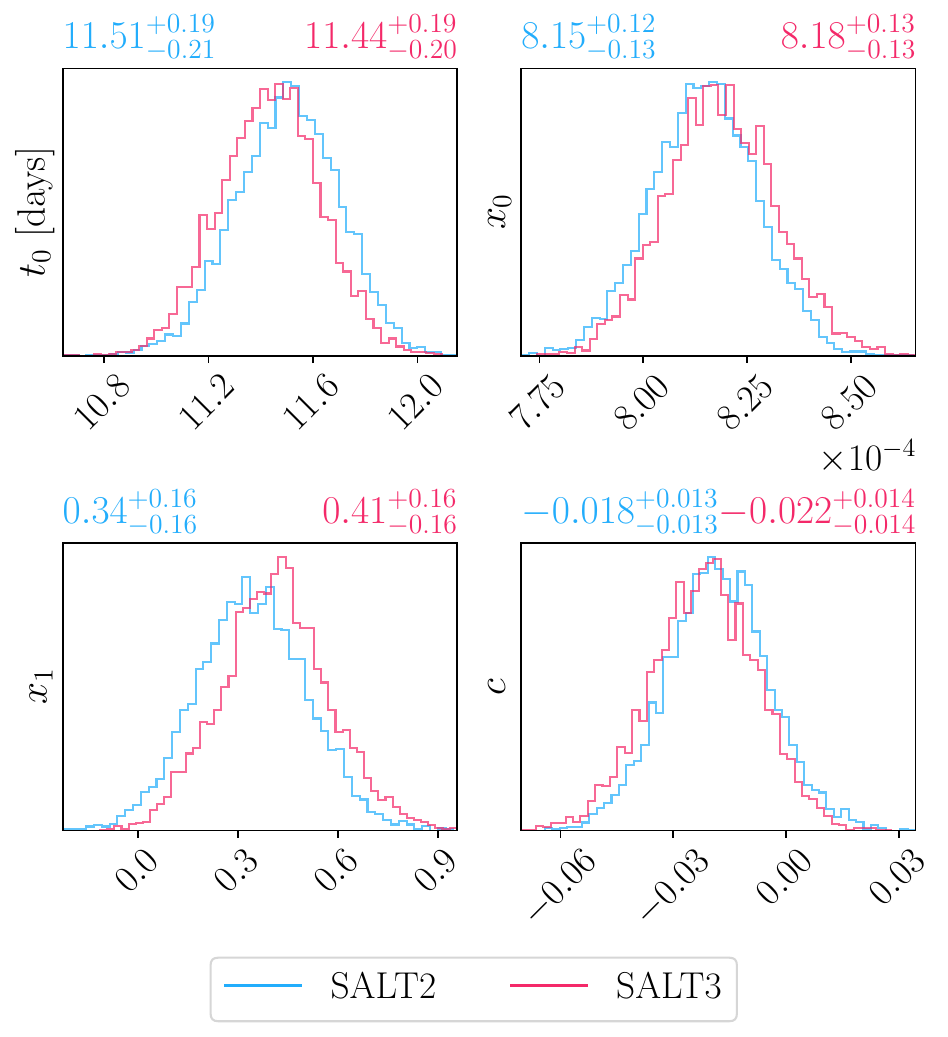}

    \caption{Model posteriors for SALT2/3 on SN 2020adlp/ZTF20acyvbhj. Note that \nmma\ uses here the first detection as the (arbitrary) reference time for the B-band peak time $t_0$.}
    \label{fig:sn_example}
\end{figure}

As an introductory application for lightcurve studies, we consider a SN model-selection problem.
\nmma\ can call any supernova model in \textsc{sncosmo}~\citep{Barbary:2016} and thus allows easy comparisons between such models.
Since \nmma\ saves the inference outcome as \bilby\ result objects, users may conveniently proceed with the latter's capacities for hyper-parameter inference to derive stringent cosmological conclusions.

We here consider a relatively well-covered thermonuclear (Type Ia) SN detection by the Zwicky Transient Facility~\citep{Rigault:2025}, SN 2020adlp/ZTF20acyvbhj, under the two popular phenomenological models SALT2~\citep{Guy:2007} and SALT3~\citep{Kenworthy:2021}.
They both parametrise the "broader-brighter" and "redder-dimmer" relations of Type Ia supernovae through a shape parameter $x_1$ and a colour parameter $c$ (alongside an overall amplitude parameter $x_0$), offering an improved lightcurve normalisation required for strong cosmological constraints~\citep[e.g.][]{DES:2024}.
SALT3 uses extended training data with a modified training algorithm, leading to reduced uncertainties and a better separation of its parameters~\citep{Kenworthy:2021}.
However, \citet{Rigault:2025b} report worse fits particularly before and near peak brightness.

For a given set of parameters $\vec{\theta} = (x_0, x_1, c, t_0)$, where $t_0$ denotes the B-band peak time, \nmma\ computes a synthetic lightcurve $m^{\star}(t, \vec{\theta})$ and compares it to observations $m$ at sample times $t_j$ using its standard EM likelihood
\begin{align}
    \ln\mathcal{L}(\vec{\theta}|d) &= - \frac{1}{2} \sum_{t_j}\frac{(m(t_j) - m^{\star}(t_j, \vec{\!\theta}\,))^2}{\sigma(t_j)^2}
    + \ln (2\pi\sigma(t_j)^2)
\end{align}
with
\begin{align}
    \sigma(t_j)^2 &= \sigma_{\text{obs}}(t_j)^2 + \sigma_{\text{sys}}(t_j)^2 .
    \label{eq:em_likelihood}
\end{align}
In this expression, $\sigma_{\text{obs}}$ is the observational uncertainty, whereas $\sigma_{\text{sys}}$ describes the uncertainty of the model.
As in Sec.~\ref{ssec:mma_bns2017}, we freely sample $\sigma_{\text{sys}}$ to obtain data-driven estimates for the modelling uncertainties~\citep{Jhawar:2025}. 
\Cref{fig:sn_example} shows posteriors under either model, ignoring additional information from the available spectra for simplicity.
Both provide very good fits to the lightcurve data.
While the estimates for the colour correction $c$ agree excellently with the findings of~\citet{Rigault:2025}, our posteriors on the stretch $x_1\approx\num{0.4}$ remain below their $x_1=\num{1.05(0.08)}$, although the precise values vary considerably depending on the exact time frame of the analysis.
Overall, we see slightly better fits under SALT2, leading to a Bayes factor of $\ln B^{\rm SALT3}_{\rm SALT2} \approx -1.6$.
We remark that contrary to the general trend that \citet{Rigault:2025b} observe in their complete dataset, this results from the inclusion of late-time observations of the rebrightening about 20 days after peak. 
When only studying the early-time lightcurves around the brightness peak, SALT3 provides the better model for this supernova.

\subsection{Compact Object Parameter Estimation}
\label{ex:psr_companion}
\begin{figure}[t]
    \centering
    \includegraphics[width=\linewidth]{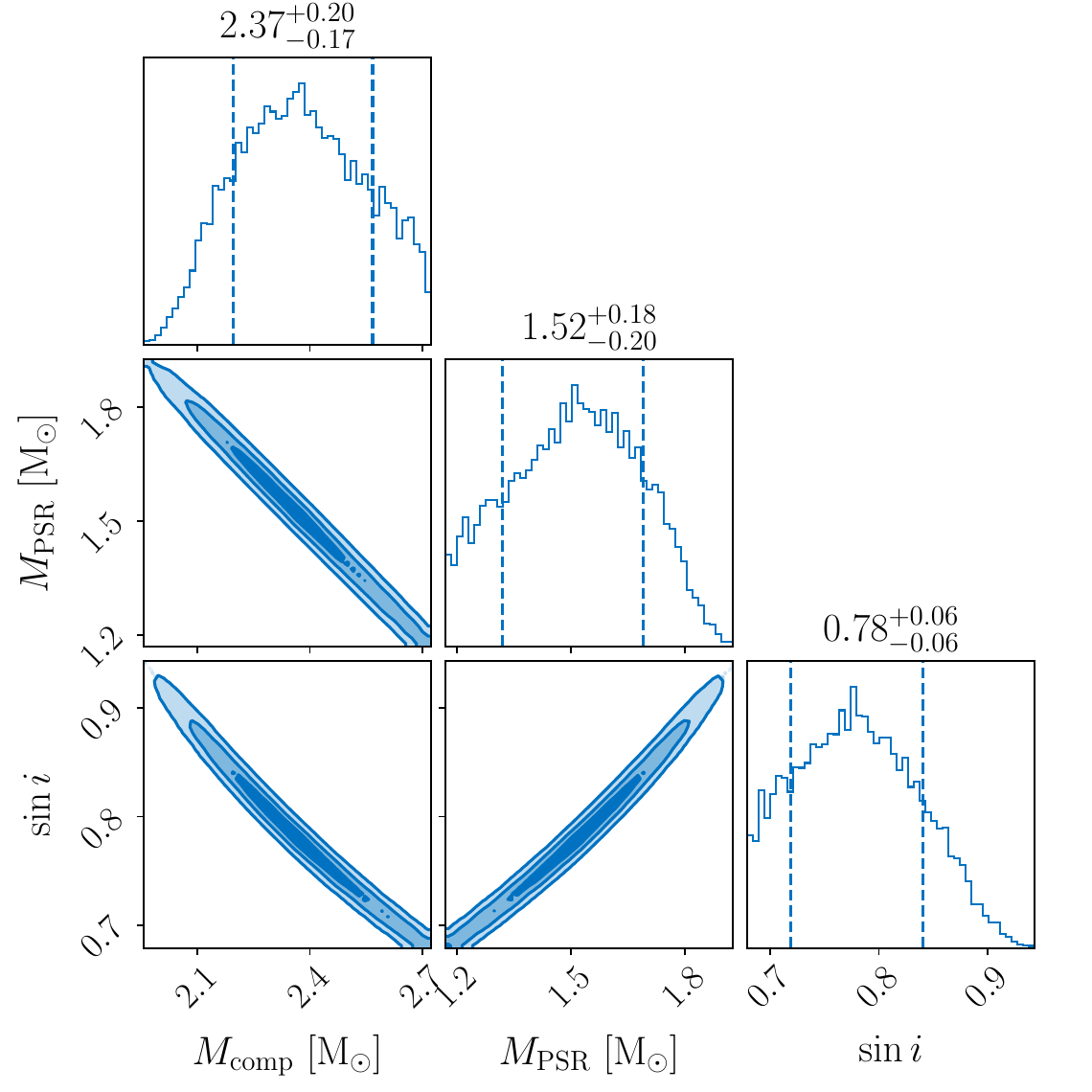}
    \caption{Posteriors for the masses of PSR J0514-4002E and its companion as well as the inclination angle of the system. The companion mass is constrained to be above \qty{2.1}{\msun} at \qty{95}{\percent} confidence, marginally consistent with a massive NS.
    }
    \label{fig:psr_example}
\end{figure}

On the surface, users may treat \nmma\ as a wrapper around \bilby\ with its familiar concepts. 
For such a simple use case, we reconsider the heavy companion of PSR J0514-4002E.
\citet{Barr:2024} had used pulsar timing methods to measure orbital parameters from which they then constrained its companion's mass to \qtyrange{2.09}{2.71}{\msun} (\qty{95}{\percent} level), right in the lower mass gap.
This term refers to the portion of the mass spectrum of compact objects between roughly \qty{2}{\msun} and \qty{5}{\msun}.
Mass measurements based on EM detections of NSs near $\mtov$ and low-mass BH are strongly deficient in this region~\citep{Bailyn:1998,Zevin:2020}, indicating a heavily suppressed formation.

We can easily approximate the~\citet{Barr:2024} mass estimate by considering a pulsar-timing sub-model that evaluates their reported values for the post-Keplerian parameters $\gamma_E$, the Einstein delay, and $h_3$, a parameter related to the Shapiro delay, 
\begin{align}
    \gamma_E &= e \left(\frac{P G^2}{2\pi c^6} \right)^{1/3} \frac{M_c(\mtot+M_c)}{\mtot^{4/3}} = \qty{0.0111(84)}{\second}\, \quad\text{ and }\\
    h_3 &= \frac{G M_c}{c^3} \left(\frac{\sin i}{1+\cos i} \right)^3= \qty{- 0.02(91)}{\micro\second}
\end{align}
as a multivariate Gaussian in \nmma.
Here, $e$ is the eccentricity of the system, $P$ is the orbital period, $M_c$ is the companion mass and $\mtot$ is the total mass of the system.
Neglecting the very small uncertainties on $P$ and $e$, we can then use the more reliably measured binary mass function $f$ and total mass $M_{tot}$ (as found by the advance of periastron) with their respective uncertainties as Gaussian priors.
A uniform prior on $\cos i$ further expresses prior ignorance on the orbital inclination and allows us to obtain the mass posteriors given in \cref{fig:psr_example}, naturally matching the results of \citet{Barr:2024}.

\subsection{Joint Inference with Equation of State Information}
\label{app:nuclear}

\begin{figure}
    \centering
    \includegraphics[width=0.45\textwidth]{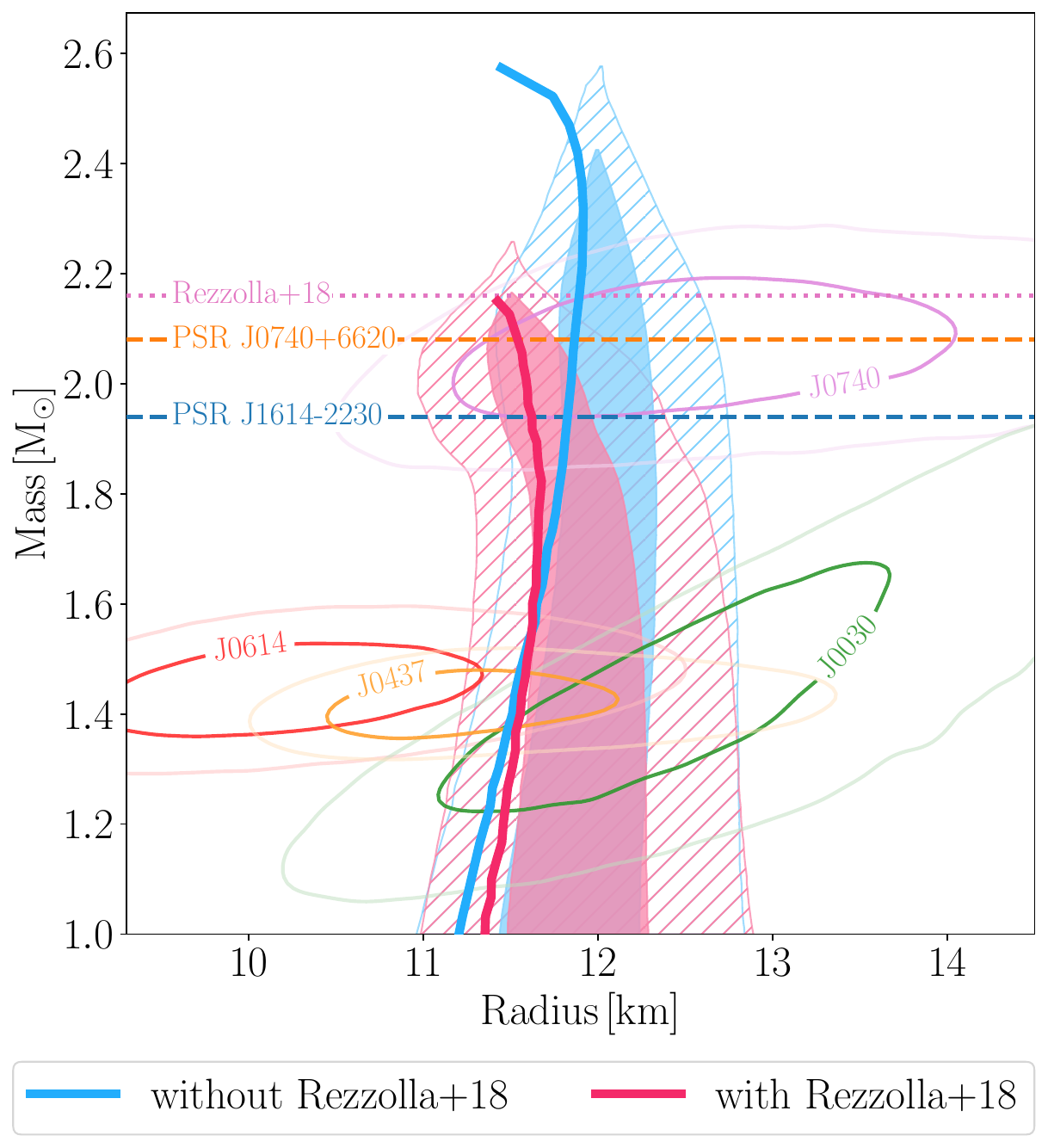}
    \caption{ 
    Effective $M-R$ posteriors from different observational constraints, applying agnostic (blue) and experimentally informed nuclear priors (red).
    Solid lines show the mass-radius curves with highest posterior support. 
    The filled (hatched) contours surrounding them show the radius spread with \qty{50}{\percent} (\qty{90}{\percent}) posterior support.
    The applied constraints are visualised as dashed (dotted) lines to indicate lower (upper) bounds on the maximum mass from heavy pulsar observations, while empty contours show mass-radius constraints from NICER measurements, again at the \qty{50}{\percent} and \qty{90}{\percent} levels.
    For the tighter prior, we additionally apply an upper mass limit from~\citet{Rezzolla:2018}. 
    }
    \label{fig:eos_example}
\end{figure}

\nmma's added value shows when we use the emulator of~\cref{ssec:emulation_strategy} to simultaneously perform inference on nuclear properties, too, assessing how multiple astrophysical NS measurements constrain the nuclear parameter space and thus allowing a more quantitative assessment of PSR J0514-4002E's companion.
We here take three types of data with corresponding likelihood expressions into account.
Assuming Gaussian uncertainties $\sigma$ on a given lower or upper mass limit $d=M_{\rm est}$, we can ascribe a set of nuclear parameters $\vec{\theta}$ the likelihood \begin{align}
    \mathcal{L}(d|\vec{\theta}) = \frac{1}{2} \pm \frac{1}{2}{\rm erf}\left(\frac{\mtov(\vec{\theta})-M_{\rm est}}{\sqrt{2\sigma^2}} \right),
\end{align}
where erf denotes the Gauss error function and the + (-) sign corresponds to a lower (upper) limit.
In other words, the likelihood of $\mtov$ is just the (inverse) cumulative distribution function if we take the reported measurement uncertainties to characterise a normal distribution.
For mass-radius measurements based on NICER data, we express the likelihood of a given measurement posterior $d=P(M, R)$ as \begin{align}
    \mathcal{L}(d|\vec{\theta}) = \int_0^{\mtov(\vec{\theta})} P(M, R(M, \vec{\theta})) \diff M.
\end{align}
We here take a more extended set of NICER data for the pulsars J0030+0451 \citep{Kini:2026}, J0437-4715 \citep{Choudhury:2024}, J0614-3329 \citep{Mauviard:2025} and J0740+6620 \citep{Dittmann:2024} into account.

\Cref{fig:eos_example} shows the best-fitting EoSs when our likelihood includes/excludes the $\mtov$ limit of~\citep{Rezzolla:2018}.
In particular the NICER data on J0614-3329 pull them to relatively low radii around \qty{11.5}{\kilo\metre} at $\qty{1.4}{\msun}$ while the full posteriors are centred on \qty{12}{km}, reflecting the impact of our priors. 
Without the upper $\mtov$ limit we see a stronger support for stiffer EoS that becomes primarily noticeable in a shift of $\threensat$ from \qty{0.55(0.08:0.09)}{c^2} to \qty{0.78(0.14:0.18)}{c^2}, whereas the other microphysical parameters remain largely unaffected.
Correspondingly, the effective $M-R$ posteriors remain indistinguishable below \qty{1.6}{\msun}.
The difference at high masses in contrast strongly impacts the resulting $\mtov$ estimates at \qty{2.17(0.08:0.07)}{\msun} and \qty{2.41(0.14:0.17)}{\msun}, respectively.
As we jointly analyse these constraints with PSR J0514-4002E, we find a probability for the companion to be an NS of \qty{11.9}{\percent} with the upper limit and \qty{56.2}{\percent} otherwise.

\subsection{Additional AT2017gfo analyses}
\label{app:extended_AT2017gfo}
In addition to the discussion in Sec.~\ref{sec:NICER_fiesta}, we also study how inference changes when we enforce the velocity and electron fraction ratios seen in numerical modelling.
To that end, we only sample the dynamical ejecta parameters independently with modified priors based on \citet{Nedora:2022}, who found narrow distributions of mean velocities and electron fractions around \qty{0.2(0.03)}{c} and \num{0.2(0.04)}, respectively.
These values are in good agreement with our findings in recent state-of-the-art simulations, too~\citep{Neuweiler:2026}.
We then alternatively parametrise the wind properties by their ratios to the respective dynamical values, ensuring improved consistency with NR simulations.
While the resulting fits to the data remain acceptable, the posteriors rail against the prior bounds.
Moreover, this NR-informed model is strongly disfavoured against the agnostic runs discussed in Sec.~\ref{sec:NICER_fiesta} with a Bayes factor of $\ln\mathcal{B}_{\rm{NR}}^{\rm{agnostic}}\approx 17.7$.

We may also add data from previously unaccounted messengers to our KN analysis.
A highly desirable messenger would be spectroscopic identifications of heavy nuclei~\citep{Vieira:2024}. 
However, their connection to other intrinsic parameters is far from established and needs to be reserved for future work. 
Here we instead include binary-evolution constraints on the component masses $m_i$ as a messenger that has a well-established link to other model parameters.
To that end, we jointly analyse the KN signal as before alongside the EoS constraints 
used in \cref{app:nuclear} for two different mass priors and for two different pairings between these.
Apart from a uniform prior $\pi(M_{i,s})=\mathcal{U}[1.1, 2.0]$ on the (source-frame) component masses that is agnostic about the formation channel~\citep{Landry:2021}, we also consider a model proposed by \citet{Farrow:2019}.
They suggest distinct mass distributions for the recycled pulsar (typically the more massive star, assuming isolated binary evolution models~\citep{Tauris:2017}) and its slower-spinning companion in NS binaries.
While the former is well matched by the superposition of two Gaussians \begin{align}
    \pi(M_{1,s}) = 0.68\mathcal{N}(1.34, 0.02) + 0.32\mathcal{N}(1.47, 0.15), \label{eq:recycled_prior}
\end{align} they find $\pi(M_{2,s}) = \mathcal{U}[1.16, 1.42]$ sufficient for the latter.

In this configuration, we find a very modest preference for the pulsar-mass informed priors with $\ln\mathcal{B}^{\rm{PSR}}_{\rm{agn.}} \approx 0.9$.
The impact on the inferred NS properties is more significant, though.
The agnostic priors allow low component masses that can generate the required ejecta masses also for softer EoSs as preferred by the NICER posteriors from J0437-4715 \citep{Choudhury:2024} and J0614-3329 \citep{Mauviard:2025}.
Consequently, we observe a strong shift in the inferred $R_{\text{1.4}}$ from \qty{12.6(0.4)}{\kilo\metre} to \qty{12.0(0.4)}{\kilo\metre}, emphasising the impact KN modelling can have even in the absence of GW constraints models.
Conversely, the effective correlation between $L_{\rm sym}$ in our EoS model and the component masses demonstrates how tight nuclear constraints, mediated by multimessenger observations, can inform our modelling of NS binary populations.

Moreover, we study the impact of including a simple likelihood $\mathcal{L}_q\propto q^\beta$ following \citet{Landry:2021}.
We consider the choices $\beta=2$, which would imply a preference for pairing symmetric binaries as seen in binary BH mergers~\citep{Fishbach:2020}, and $\beta=0$ for arbitrary pairing.
Given that the pulsar-informed priors already imply a relatively tight mass range, the inclusion of the pairing likelihood does not significantly alter the results.
For agnostic mass priors we find a marginal preference for the $\beta=2$ model with $\ln\mathcal{B}^{\beta=2}_{\rm{\beta=0}} \approx 0.2$.

\subsection{The heavier component of GW230529}
\label{app:GW230529}

\begin{figure}
    \centering
    \includegraphics[width=0.45\textwidth]{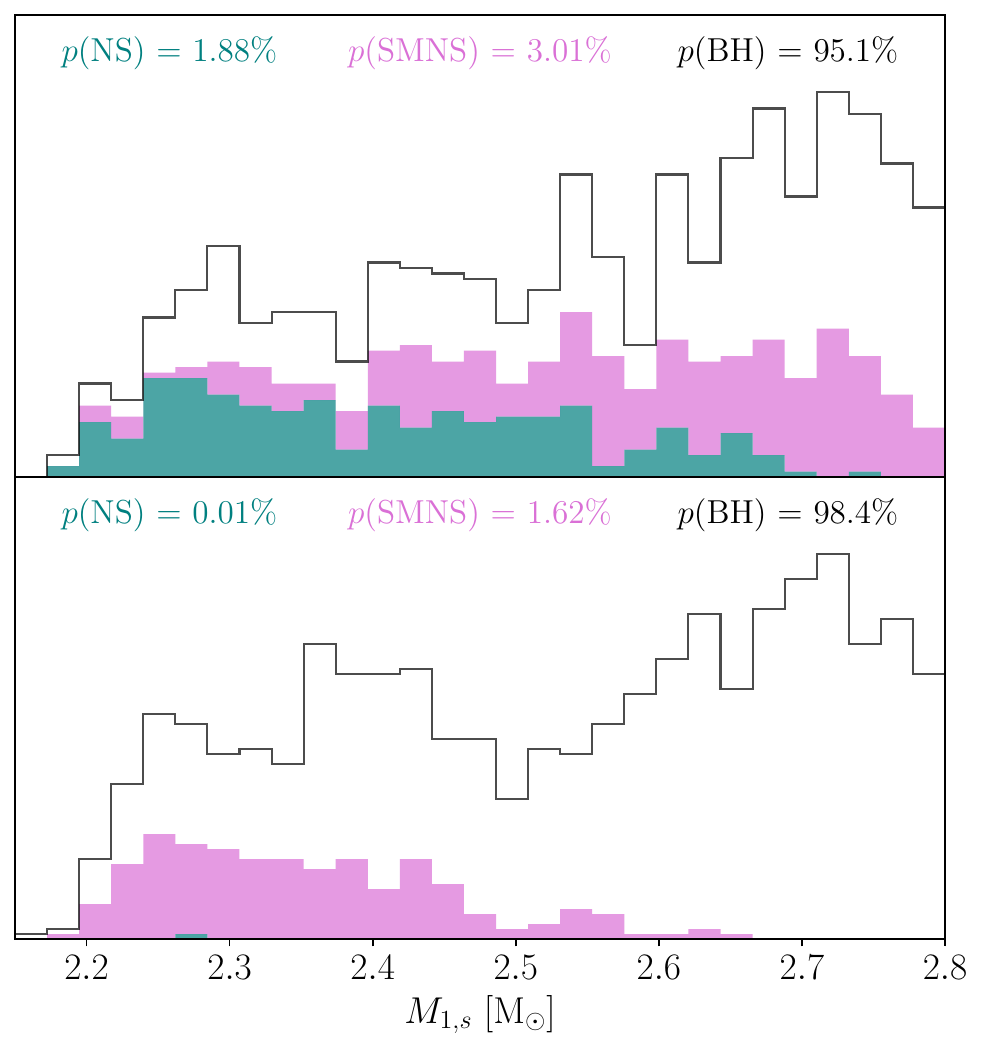}
    \caption{Histogram of the $M_{1,s}$ posterior's lower end for GW230529 when excluding (top) or including (bottom) the upper limit on the TOV mass of~\citet{Rezzolla:2018}. Teal (orchid) bars mark NSs (SMNSs). All other samples (black contour) denote BHs. }
    \label{fig:gw230529}
\end{figure}

In \cref{app:nuclear} we have jointly analysed nuclear and pulsar timing data to quantitatively assess the nature of PSR J0514-4002E's heavier companion. 
This served primarily demonstration purposes because there is no modelling connection between $\mtov$ and $M_{\rm comp}$.
The situation is different for the heavier component in GW230529~\citep{Abac:2024b} due to its potential tidal imprint.
This was a remarkable event that provided the first confident GW detection of a compact object in the lower mass gap.
While its secondary's mass estimate falls into the canonical NS mass range, previous studies have evaluated posterior samples to classify the primary most probably as a BH~\citep{Abac:2024b,Koehn:2024}.
\citet{Wouters:2025c} similarly performed a classification directly from analysing the GW strain data using EoS-informed priors. 
In this example, we instead confirm this result through jointly inferring the GW source properties along with the EoS parameters.

As the primary may have significant spin under an agnostic assumption, we here use \textsc{IMRPhenomXP\_NRTidalv3} as our waveform approximant.
This requires the introduction of additional parameters to account for spin and precession effects that we provide in \cref{tab:example_parameters}.
In particular, we constrain the secondary spin magnitude to remain below \num{0.05}, while we allow values as high as \num{0.99} for the primary in arbitrary direction.
Moreover, we adjust some priors to imitate the state of knowledge after template searches in the detection pipelines~\citep{Adams:2016,Allene:2025,Cannon:2021,Ewing:2024,Dal_Canton:2021,Trevor:2026}.
We set the limits on $\pi(\mathcal{M})$ to \qtyrange{2.02}{2.04}{\msun}, covering the initial chirp mass estimate, and extend the lower bound of $\pi(q)$ to $\frac{1}{8}$ to safely cover a broad mass range. 
Given the initial distance estimates, we increase the upper limit on $\pi(d_{\rm L})$ to \qty{500}{Mpc}, while we assume ignorance on the sky localisation.
We can only expect very weak constraints on the latter because only one detector was available to record the data which moreover had a moderately low signal-to-noise ratio (SNR) of about 11.8~\citep{Abac:2025}.
We further inform the inference similar to Sec.~\ref{ssec:mma_bns2017} by including the lower mass-limit constraints from PST J1614-2230 \citep{Shamohammadi:2023} and J0740+6620 \citep{Fonseca:2021} and optionally the upper mass limit of~\citet{Rezzolla:2018}.

We generally reproduce the results of~\citet{Abac:2024b}, finding $\mathcal{M}=\qty{2.026(0.001)}{\msun}$ at $d_{\rm L}=\qty{210(60)}{\mega\parsec}$ with $q=\num{0.39(0.07:0.11)}$. 
At the \qty{68}{\percent} level, this implies a source-frame primary mass $M_{1,s}$ of \qty{3.7(0.5:0.4)}{\msun}, decisively above the range accessible to NSs under any realistic EoS.
We, too, see non-negligible posterior mass with more symmetric mass ratios, though.
Hence, we obtain a mostly prior-driven $\lambdaT=\num{40(20)}$, but also weak support for $\lambdaT=0$, corresponding to the rather unlikely case that both components were BHs.
This can occur under our EoS assumptions for $q\gtrsim0.65$ and very soft EoS samples. 
If the sampled EoS in contrast is stiff, particularly when we do not take the upper mass-limit constraint into account, these same mass ratios open the possibility that the primary could be an NS.

The identification becomes more intricate when including component spins in our consideration. 
The basic treatment in \nmma\ identifies a binary component as an NS if its mass remains below the obtained $\mtov$.
This is only justified for slow-spinning NSs.
Differentially rotating (hypermassive) NSs may support up to \qty{1.6}{\mtov}, albeit only on a sub-second timescale~\citep{Baumgarte:2000,Siegel:2013}.
A supramassive NS (SMNS) in contrast maintains rigid rotation on a secular timescale and can still reach masses up to $M_{\rm SMNS} \lesssim 1.2 \mtov$ at its Kepler limit where mass-shedding occurs.

Due to a degeneracy~\citep{Cutler:1994,Baird:2013} between $q$ and the leading-order spin parameter \begin{align}
    \chi_{\rm eff} = \frac{M_1\chi_{1,z} +M_2\chi_{2,z}}{M_1+M_2},
\end{align}
where $\chi_{i,z}$ denotes the dimensionless spin component perpendicular to the orbital plane, the posterior samples with very low primary mass actually require a strong anti-aligned spin $\chi_{1,z}\leq \num{-0.4}$.
Such high spins naturally affect other structural quantities, too, increasing both $R$ and $\Lambda$~\citep{Stergioulas:2003}.
An exact treatment would require numerical solutions of the perturbed NS structure equations. 
Software like \textsc{rns}~\citep{Stergioulas:1995} is in principle available to address this task, but its computational requirements stand in contrast to the intent of this work.

Given the approximate nature of our approach, we use the relation of~\citet{Breu:2016} instead to express $M_{\rm max}(\chi)$ up to the maximum $\chi_K$ as \begin{align}
    \frac{M_{\rm max}(\chi)}{\mtov} = 1+0.1316\left(\frac{\chi}{\chi_K}\right)^2 + 0.07111\left(\frac{\chi}{\chi_K}\right)^4.
\end{align}
$\chi_K$ denotes the Kepler limit above which mass shedding occurs.
While the rotational frequency of the Kepler limit varies significantly with mass and EoS, matching the intuition that more compact stars should be capable to sustain higher spins, \citet{Lo:2011,Cipolletta:2015} have found $\chi_K\approx 0.7$ to be nearly independent of the EoS and the NS mass.
We therefore interpret posterior samples as characterising an NS if $M_{1,s}\leq\mtov$ and $\chi_1 \leq 0.7$, an SMNS if $\mtov<M_{1,s}<M_{\rm max}(\chi_1)$ and $\chi_1 \leq 0.7$ and a BH otherwise.

\Cref{fig:gw230529} shows the resulting shares of the $M_{1,s}$ posterior with and without upper $\mtov$ limit. 
When including the constraint of \citet{Rezzolla:2018}, an NS interpretation is fully negligible.
Even in the more permissive case, a BH interpretation is favoured at almost \qty{95}{\percent} with a mere \qty{1.9}{\percent} NS probability.
But even in that case, an SMNS or NS interpretation is effectively ruled out when considering our current understanding of compact binary formation.
The detection of a pulsar with $\chi\approx0.4$~\citep{Hessels:2006} confirms that higher NS spin magnitudes are in principle found in binaries, but there is some evidence that $|\chi|\approx0.4$ represents an upper-limit to spin-up in isolated binaries~\citep{Chakrabarty:2008,Gusakov:2014}. 
Moreover, it is currently unclear whether or how this spin could be maintained until merger, given that spins in the observed BNS population with short orbital periods, leading to mergers within a Hubble time, is bounded by $\chi<0.05$~\citep{Burgay:2003}.

\end{document}